\documentclass[preprint, bj]{imsart}

\usepackage{textcomp,marvosym}

\usepackage{fixltx2e}

\usepackage{amsmath,amssymb}
\usepackage{cite}

\usepackage{nameref,hyperref}

\usepackage[right]{lineno}

\usepackage{microtype}

\usepackage{rotating}

\usepackage{tikz}
\usepackage{subfig}
\usepackage{algorithm}
\usepackage{algpseudocode}
\usepackage{calc}
\usepackage{siunitx}
\usepackage[aboveskip=1pt,labelfont=bf,labelsep=period,justification=raggedright,singlelinecheck=off]{caption}

\startlocaldefs
\newcommand{\R}{\mathbb{R}} 

\newcommand{\I}{\mathbb{I}} 

\newcommand{\by}{\boldsymbol{y}}
\newcommand{\bc}{\boldsymbol{c}}
\newcommand{\bd}{\boldsymbol{d}}
\newcommand{\bP}{\boldsymbol{\Phi}}
\newcommand{\bZ}{Z}
\newcommand{\bV}{V}

\newcommand{\bt}{\boldsymbol{t}}
\newcommand{\btheta}{\boldsymbol{\vartheta}}

\newcommand{\bz}{\boldsymbol{z}}
\newcommand{\bx}{{\boldsymbol{x}}}

\newcommand{\bw}{\boldsymbol{w}}

\newcommand{\bepsilon}{\boldsymbol{\varepsilon}}

\endlocaldefs

\begin{document}

\begin{frontmatter}

\title{Separating timing, movement conditions and individual differences in the analysis of human movement}

\runtitle{Timing and movement path separation}

\begin{aug}

\author{\fnms{Lars Lau} \snm{Raket}\corref{}\thanksref{a}\ead[label=e1]{larslau@math.ku.dk}},
\author{\fnms{Britta} \snm{Grimme}\thanksref{b}},
\author{\fnms{Gregor} \snm{Sch\"oner}\thanksref{b}},
\author{\fnms{Christian} \snm{Igel}\thanksref{c}},
\and
\author{\fnms{Bo} \snm{Markussen}\thanksref{a}}

\address[a]{Department of Mathematical Sciences, University of Copenhagen, Universitetsparken 5, 2100 Copenhagen, Denmark.}

\address[b]{Institut f\"ur Neuroinformatik, Ruhr-Universit\"at Bochum, Universit\"atsstra\ss{}e 150, D-44801 Bochum, Germany.}

\address[c]{Department of Computer Science, University of Copenhagen, Sigurdsgade 41
2200 Copenhagen, Denmark.}

\runauthor{Raket et al.}

\affiliation{University of Copenhagen}

\end{aug}

\begin{abstract}
\paragraph{Abstract:}
A central task in the analysis of human movement behavior is to determine systematic patterns and differences across experimental conditions, participants and repetitions. This is possible because human movement is highly regular, being constrained by invariance principles. Movement timing and movement path, in particular, are linked through scaling laws. Separating variations of movement timing from the spatial variations of movements is a well-known challenge that is addressed in current approaches only through forms of preprocessing that bias analysis. Here we propose a novel nonlinear mixed-effects  model for analyzing temporally continuous signals that contain systematic effects in both timing and path. Identifiability issues of path relative to timing are overcome by using maximum likelihood estimation in which the most likely separation of space and time is chosen given the variation found in data. The model is applied to analyze experimental data of human arm movements in which participants move a hand-held object to a target location while avoiding an obstacle. The model is used to classify movement data according to participant. Comparison to alternative approaches establishes nonlinear mixed-effects models as viable alternatives to conventional analysis frameworks.  The model is then combined with a novel factor-analysis model that estimates the low-dimensional subspace within which movements vary when the task demands vary. Our framework enables us to visualize different dimensions of movement variation and to test hypotheses about the effect of  obstacle placement and height on the movement path. We demonstrate that the approach can be used to uncover new properties of human movement.
\end{abstract}

\end{frontmatter}




\section*{Author Summary}
When you move a cup to a new location on a table, the movement of lifting, transporting, and setting down the cup  appears to be completely automatic.  Although the hand could take continuously many different paths and move on any temporal trajectory, real movements are highly regular and reproducible. From repetition to repetition movemens vary, and the pattern of variance reflects movement conditions and movement timing. If another person performs the same task, the movement will be similar.  When we look more closely, however, there are systematic individual differences. Some people will overcompensate when avoiding an obstacle and some people will systematically move slower than others. When we want to understand human movement, all these aspects are important. We want to know which parts of a movement are common across people and we want to quantify the different types of variability. Thus, the models we use to analyze movement data should contain all the mentioned effects. In this work, we developed a framework for statistical analysis of movement data that respects these structures of movements. We showed how this framework modeled the individual characteristics of participants better than other state-of-the-art modeling approaches. We combined the timing-and-path-separating model with a novel factor analysis model for analyzing the effect of obstacles on spatial movement paths. This combination allowed for an unprecedented ability to quantify and display different sources of variation in the data. We analyzed data from a designed experiment of arm movements under various obstacle avoidance conditions. Using the proposed statistical models, we  documented three findings: a linearly amplified deviation in mean path related to increase in obstacle height; a consistent asymmetric pattern of variation along the movement path related to obstacle placement; and the existence of obstacle-distance invariant focal points where mean trajectories intersect in the frontal and vertical planes.

\section*{Introduction}
When humans move and manipulate objects, their hand paths are smooth, but also highly flexible. Humans do not move in a jerky, robot-like way that is sometimes humorously invoked to illustrated ``unnatural'' movement behavior. In fact, humans have a hard time making ``arbitrary'' movements. Even when they scribble freely in three dimensions, their hand moves in a regular way that is typically piecewise planar \cite{Morasso83,soechting1987organization}. Movement generation by the nervous systems, the neuro-muscular systems, and the body is constrained by implied laws of motion signatures which are found empirically through invariances of movement trajectories and movement paths. Among these, laws decoupling space and time are of particular importance. For instance, the fact that the trajectories of the hand have approximately bell-shaped velocity profiles across varying movement amplitudes \cite{morasso1981spatial} implies a scaling of the time dependence of velocity. The 2/3 power law \cite{lacquaniti1983law} establishes an analogous scaling of time with the spatial path of the hand's movement. Similarly, the isochrony principle \cite{viviani1983relation} captures that the same spatial segment of a movement takes up the same proportion of movement time as movement amplitude is rescaled. {Several of these invariances can be linked to geometrical invariance principles \cite{bennequin2009movement}.}

These invariances imply that movements as a whole have a reproducible temporal form, which can be characterized by movement parameters. Their values are specified before a movement begins, so that one may predict the movement's time course and path based on just an initial portion of the trajectory \cite{Erlhagen2002a}. Movement parameters are assumed to reside at the level of end-effector trajectories in space and their neural encoding begins to be known \cite{Georgopoulos95,Schwartz2007, HarpazFlashDinstein2015}. The set of possible movements can thus be spanned by a limited number of such parameters. {Moreover, the choices of these movement parameters are constrained. For instance, in sequences of movements, earlier segments predict later segments \cite{zhang2008planning}.} 

A key source of variance of kinematic variables is, of course, the time course of the movement itself. The invariance principles suggest that this source of variance can be disentangled from the variation induced when the movement task varies. In this paper, we will first address time as a source of variance, focussing on a fixed movement task, and then use the methods developed to address how movements vary when the task is varied. 

Given a fixed movement task, movement trajectories also vary across individuals. {Individual differences in movement, a personal movement style, are reproducible and stable over time, as witnessed, for instance, by the possibility to identify individuals or individual characteristics such as gender by movement information alone \cite{CuttingKozlowski77, pollick2005gender, lu2014human}. }

A third source of variation are fluctuations in how movements are performed from trial to trial or across movement cycles in rhythmic movements. Such fluctuations are of particular interest to movement scientists, because they reflect not only sources of random variability such as neural or muscular noise, but also the extent to which the mechanisms of movement generation stabilize movement against such noise. Instabilities in patterns of coordination have been detected by an increase of fluctuations \cite{SchonerKelso88} and differences in variance among different degrees of freedom have been used to establish priorities of neural control \cite{ScholzSchoner99,LatashScholzSchoner2007}.

A systematic method to disentangle these three sources of movement variation, time, individual differences, and fluctuations, would be a very helpful research tool. Such a method would decompose sets of observed kinematic time series into a common trajectory (that may be specific to the task), participant-specific movement traits, and random effects.  Given the observed decoupling of space and time, such a decomposition would also separate the rescaling of time across these three factors from the variation of the spatial characteristics of movement. 

The statistical subfield that deals with analysis of temporal trajectories is the field of functional data analysis. In the literature on functional data, the typical approach for handling continuous signals with time-warping effects is to pre-align samples under an oversimplified noise model in the hope of eliminating the effects of movement timing \cite{Ramsay}. In contrast, we propose an analytic framework in which the decomposition of the signal is done simultaneously with the estimation of movement timing effects, so that samples are continually aligned under an estimated noise model. Furthermore, we account for both the task-dependent variation of movement and for individual differences (a brief review of warping in the modeling of biological motion is provided in the Methods section). 

Decomposition of time series into a common effect (the time course of the movement given a fixed task), an individual effect, and random variation naturally leads to a mixed-effects formulation \cite{pinheiro2000mixed}. The addition of nonlinear timing effects
gives the model the structure of a hierarchical nonlinear mixed-effects model \cite{LindstromBates}. We present a framework for
maximum-likelihood estimation in the model 
and demonstrate that the method leads to high-quality templates that foster subsequent analysis (e.g., classification). 
We then show that the results of this analysis can be combined with other models to test hypotheses about the invariance of movement patterns across participants and task conditions. We demonstrate this by using the individual warping functions in a novel factor analysis model that captures variation of movement trajectories with task conditions. 

We use as of yet unpublished data from a study of naturalistic movement that extends published work  \cite{grimme2012naturalistic}. In the study, human participants transport a wooden cylinder from a starting to a target location while avoiding obstacles at different spatial positions along the path. Earlier work has shown that movement paths and trajectories in this relatively unconstrained, naturalistic movement task  clearly reflect typical invariances of movement generation, including the planar nature of movement paths,  spatiotemporal invariance of velocity profiles, and a local isochrony principle that reflects the decoupling of space and time \cite{grimme2012naturalistic}. By varying the obstacle configuration, the data include significant and non-trivial task-level variation.  We begin by modeling a one-dimensional projection of the time courses of acceleration of the hand in space,  which we decompose into a common pattern and the deviations from it that characterize each participant.  The timing of the acceleration profiles is determined by individual time warping functions which are of higher quality than conventional estimates, since timing and movement noise are modeled simultaneously.  The high quality of the estimates is demonstrated by classifying movements according to participant. Finally, the results of the nonlinear mixed model are analyzed using a novel factor analysis model that estimates a low-dimensional subspace within which movement paths change when the task conditions are varied. This combination of statistical models makes it possible to separate and visualize the variation caused by experimental conditions, participants and repetition. Furthermore, we can formulate and test hypotheses about the effects of experimental conditions on movement paths. Using the proposed statistical models, we  document three findings: a linearly amplified deviation in mean path related to increase in obstacle height; a consistent asymmetric pattern of variation along the movement path related to obstacle placement; and the existence of obstacle-distance invariant focal points where mean trajectories intersect in the frontal and vertical plane.

Software for performing the described types of simultaneous analyses of timing and movement effects  are publicly available through the \texttt{pavpop} R package \cite{pavpop}. A short guide on model building and fitting in the proposed framework is available in Supporting Information, along with an application to handwritten signature data. 


\section*{Methods}

\subsection*{Experimental data set}
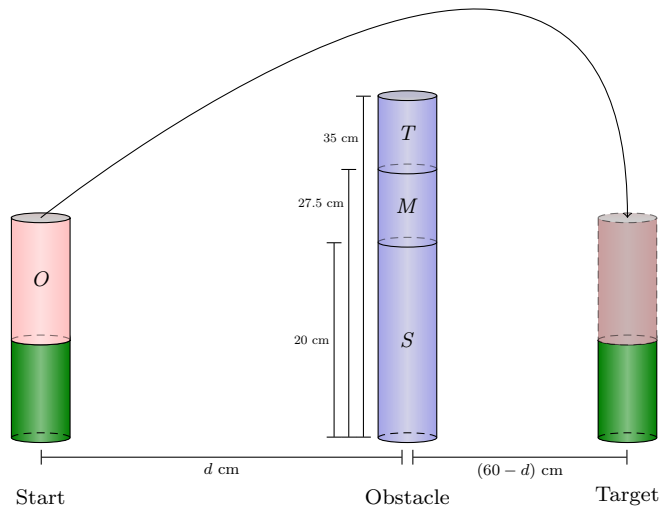
\begin{figure}[!h]
\centering
\vspace{-6em}
\begin{tikzpicture}[scale=0.13]
\fill[left color=green!50!black,right color=green!50!black,middle color=green!50!black!50,shading=axis,opacity=1] (3,0) -- (3,10) arc (360:180:3cm and 0.5cm) -- (-3,0) arc (180:360:3cm and 0.5cm);
\draw (-3,10) -- (-3,0) arc (180:360:3cm and 0.5cm) -- (3,10) arc (180:360:-3cm and 0.5cm);
\draw[densely dashed] (-3,0) arc (180:0:3cm and 0.5cm);
\node[align=center] (obj) at (0,-6) {Start};

\fill[left color=green!50!black,right color=green!50!black,middle color=green!50!black!50,shading=axis,opacity=1] (63,0) -- (63,10) arc (360:180:3cm and 0.5cm) -- (57,0) arc (180:360:3cm and 0.5cm);
\draw (57,10) -- (57,0) arc (180:360:3cm and 0.5cm) -- (63,10) arc (180:360:-3cm and 0.5cm);
\draw[densely dashed] (57,0) arc (180:0:3cm and 0.5cm);
\node[align=center] (obj) at (60,-6) {Target};

\fill[left color=pink,right color=pink,middle color=pink!50,shading=axis,opacity=1]  (3,10) -- (3,22.5) arc (360:180:3cm and 0.5cm) -- (-3,10) arc (180:360:3cm and 0.5cm);
\fill[top color=pink!90!,bottom color=pink!2,middle color=pink!30,shading=axis,opacity=0.25] (0,22.5) circle (3cm and 0.5cm);
\draw (-3,22.6) -- (-3,10) arc (180:360:3cm and 0.5cm) -- (3,22.5) ++ (-3,0) circle (3cm and 0.5cm);
\draw[densely dashed, opacity = 0.5] (-3,10) arc (180:0:3cm and 0.5cm);
\node[align=center] (obj) at (0,16.25) {\emph{O}};

\fill[left color=pink,right color=pink,middle color=pink!50,shading=axis,opacity=0.7]  (63,10) -- (63,22.5) arc (360:180:3cm and 0.5cm) -- (57,10) arc (180:360:3cm and 0.5cm);
\fill[top color=pink!90!,bottom color=pink!2,middle color=pink!30,shading=axis,opacity=0.25] (60,22.5) circle (3cm and 0.5cm);
\draw[densely dashed, opacity = 0.5] (57,22.5) -- (57,10) arc (180:360:3cm and 0.5cm) -- (63,22.5) ++ (-3,0) circle (3cm and 0.5cm);
\draw[densely dashed, opacity = 0.5] (57,10) arc (180:0:3cm and 0.5cm);

\fill[fill=white!50,shading=axis,opacity=0.25] (37.5,0) circle (3cm and 0.5cm);
\fill[left color=blue!30,right color=blue!30,middle color=blue!10,shading=axis,opacity=0.9] (40.5,0) -- (40.5,35) arc (360:180:3cm and 0.5cm) -- (34.5,0) arc (180:360:3cm and 0.5cm);
\fill[top color=blue!30!,bottom color=blue!2,middle color=blue!30,shading=axis,opacity=0.25] (37.5,35) circle (3cm and 0.5cm);
\draw (34.5,35) -- (34.5,0) arc (180:360:3cm and 0.5cm) -- (40.5,35) ++ (-3,0) circle (3cm and 0.5cm);
\draw[densely dashed] (34.5,0) arc (180:0:3cm and 0.5cm);

\draw[] (34.5,20) arc (180:360:3cm and 0.5cm);
\draw[densely dashed,opacity=0.5] (40.5,20) arc (0:180:3cm and 0.5cm);
\draw[] (34.5,27.5) arc (180:360:3cm and 0.5cm);
\draw[densely dashed,opacity=0.5] (40.5,27.5) arc (0:180:3cm and 0.5cm);

\node[align=center] (obj) at (37.5, 10) {\emph{S}};
\node[align=center] (obj) at (37.5, 23.75) {\emph{M}};
\node[align=center] (obj) at (37.5, 31.25) {\emph{T}};
\node[align=center] (obj) at (37.5,-6) {Obstacle};

\draw[|-|] (30,0) --++ (0,20) node [midway,left,scale=0.6] {20 cm};
\draw[|-|] (31.5,0) --++ (0,27.5) node [very near end,left,scale=0.6] {27.5 cm};
\draw[|-|] (33,0) --++ (0,35) node [very near end,left,scale=0.6] {35 cm};

\draw[|-|] (0,-2) --++ (37,0) node [midway,below,scale=0.75] {$d$ cm};
\draw[|-|] (38,-2) --++ (22,0) node [midway,below,scale=0.75] {$(60-d)$ cm};

\draw[->] (0,22.5) .. controls +(left:0cm) and +(up:48cm) .. (60,22.5);

\end{tikzpicture}

\caption{Obstacle avoidance paradigm. Participants move the cylindrical object \emph{O} from the starting platform (green) to the target platform by lifting it over an obstacle. Obstacles of three different heights, small (\emph{S}), medium (\emph{M}), and tall (\emph{T}), were used in the experiment, and the distance from starting position to obstacle $d$ was varied.}  \label{fig:setup}
\end{figure}

Ten participants performed a series of simple, naturalistic motor acts in which they moved a wooden cylinder
from a starting to a target position while avoiding a cylindric obstacle. The obstacle's height and positition along the movement path were varied across experiments (Figure~\ref{fig:setup}.)  

\begin{figure}[!ht]
\centering
\subfloat[Experiment $d=15$, \emph{S}]{\includegraphics[scale = 0.35, trim = 100 0 120 200, clip]{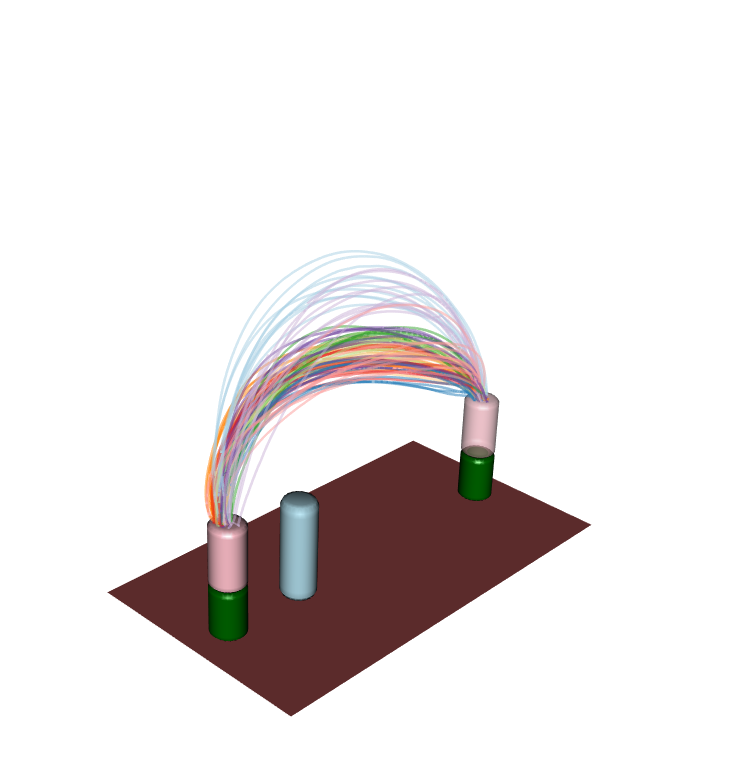}}
\subfloat[Experiment  $d=30$, \emph{T}]{\includegraphics[scale = 0.35, trim = 100 0 120 200, clip]{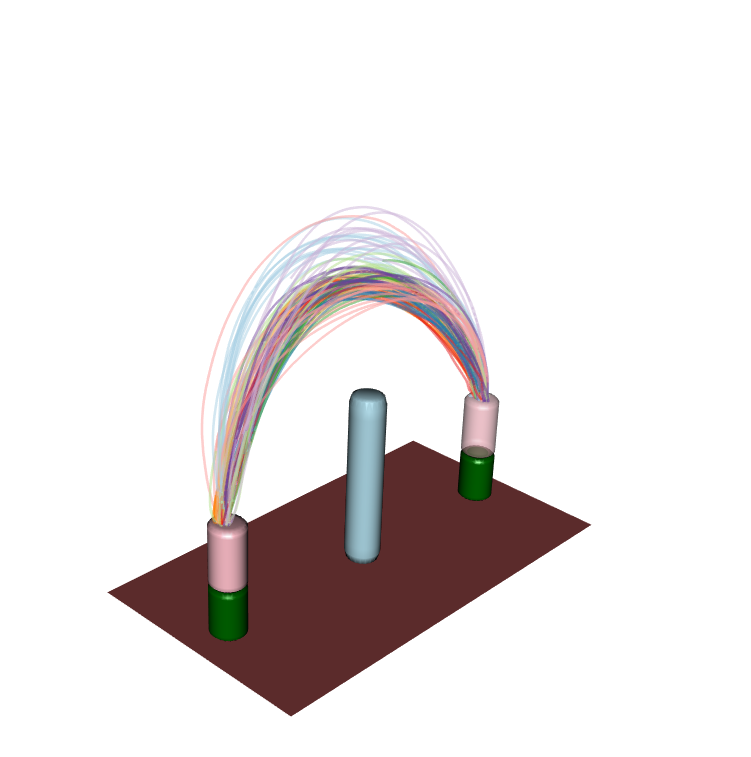}} 

\vspace{1em}\includegraphics[scale = 0.7]{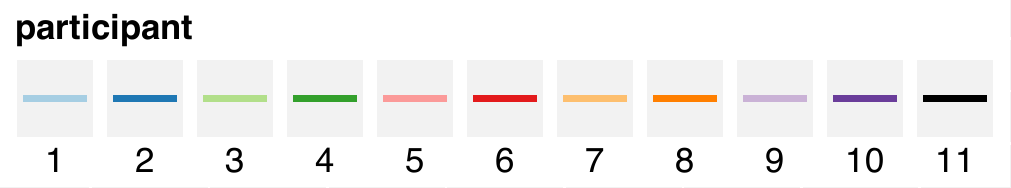}

\caption{Recorded paths of the hand-held object in space for two experimental conditions.}\label{fig:samples}
\end{figure}

\begin{figure}[!ht]
\centering
\subfloat[Acceleration]{\includegraphics[scale = 0.45, trim = 60 160 40 140, clip]{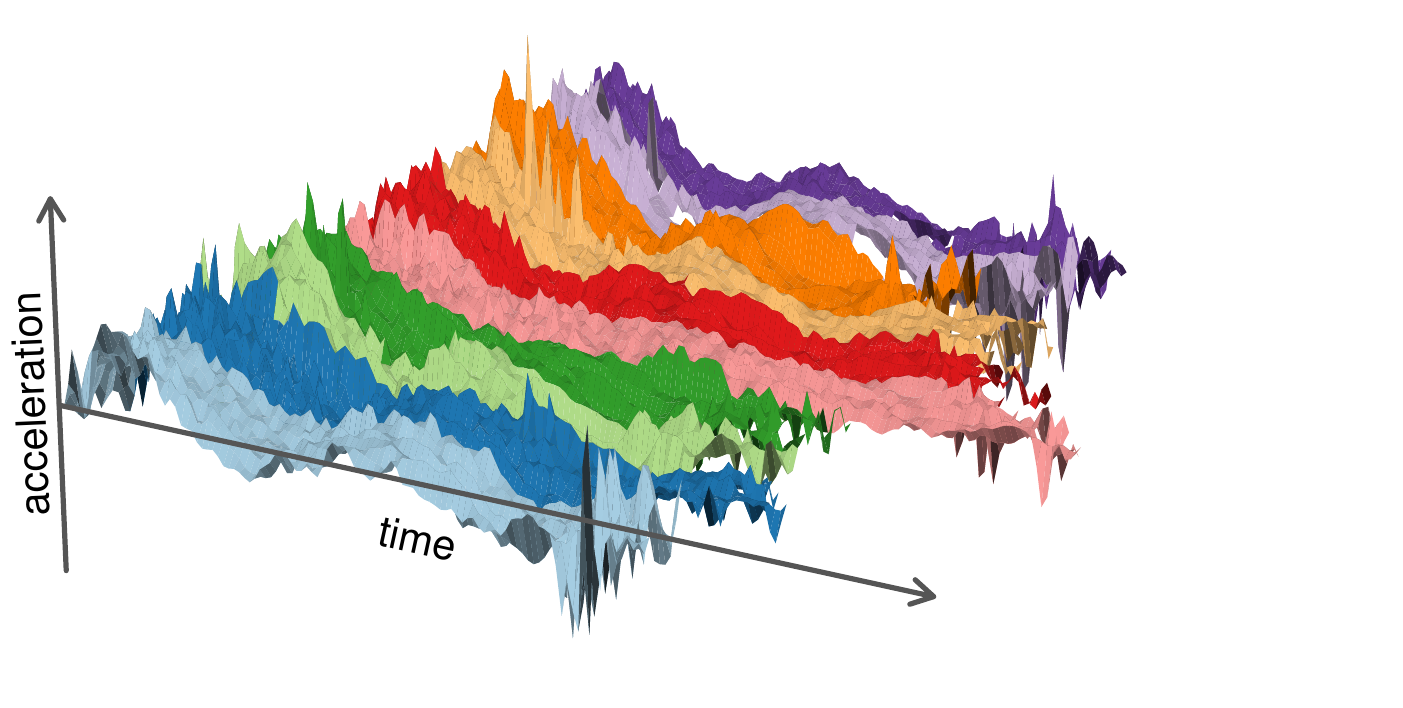}}
\subfloat[Acceleration, percentual time]{\includegraphics[scale = 0.45, trim = 60 160 40 140, clip]{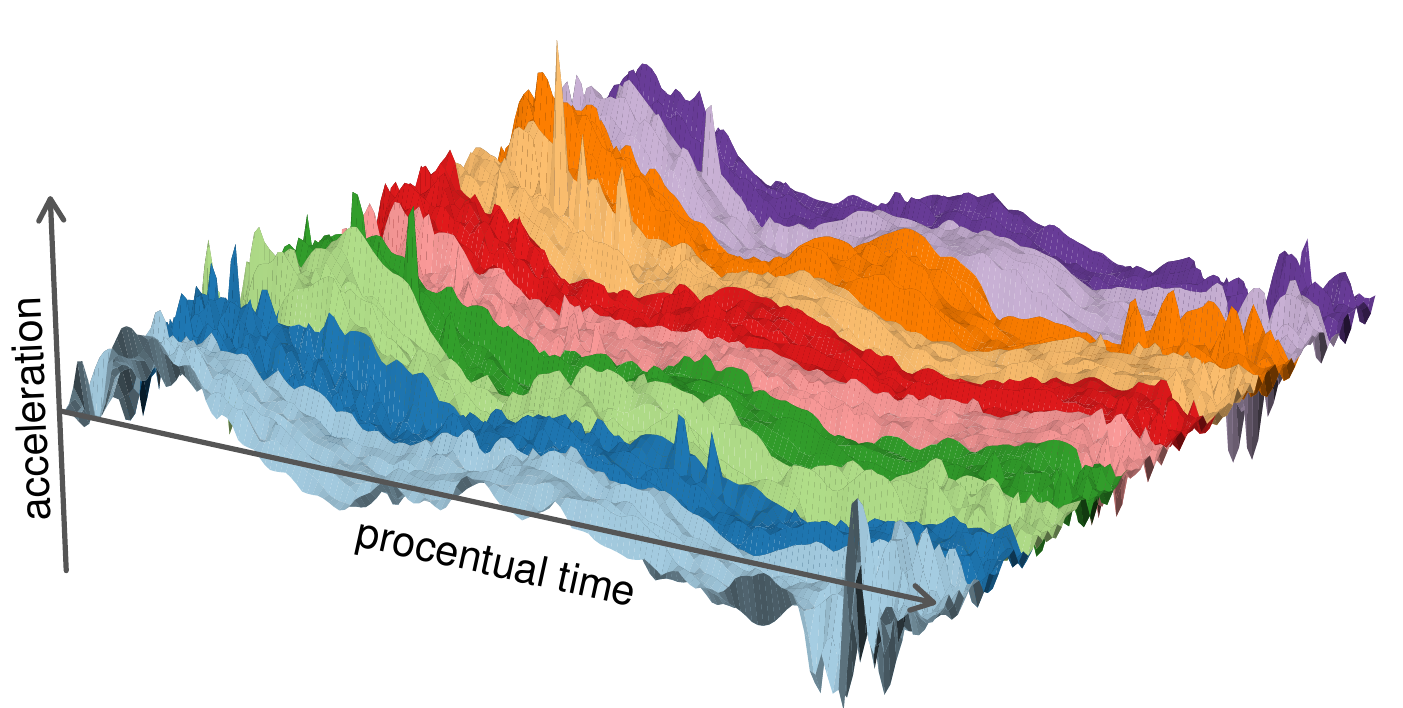}}

\subfloat[Smoothed acceleration]{\includegraphics[scale = 0.45, trim = 60 160 40 140, clip]{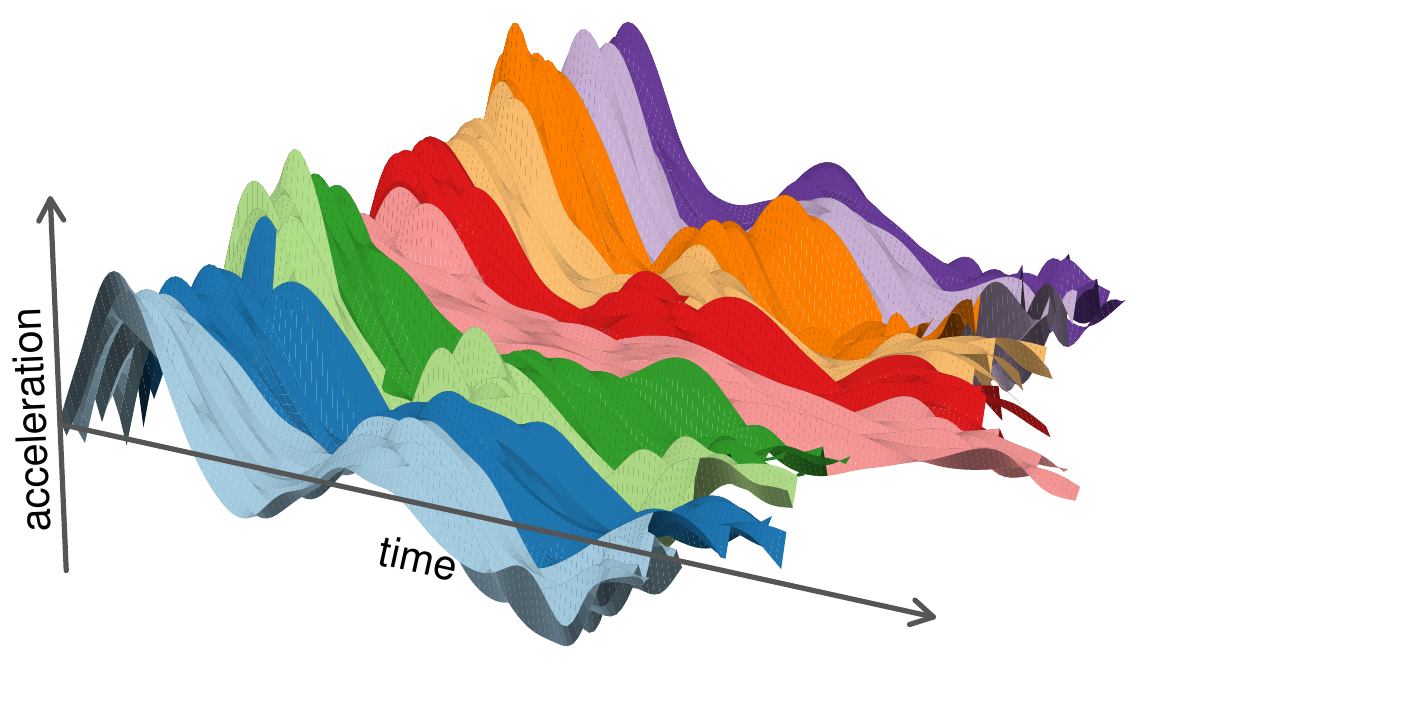}}
\subfloat[Smoothed acceleration, percentual time]{\includegraphics[scale = 0.45, trim = 60 160 40 140, clip]{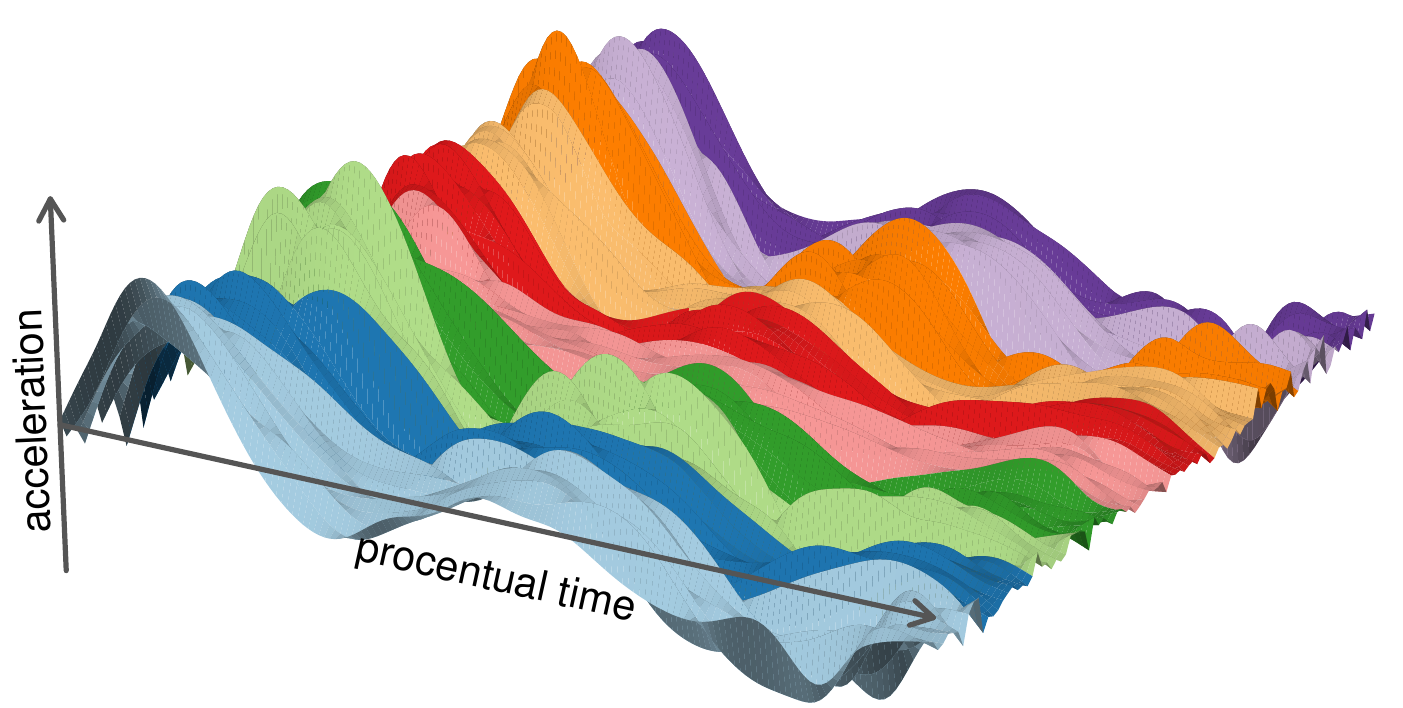}}

\vspace{1em}\includegraphics[scale = 0.7]{img/person.pdf}

\caption{Surface plots of acceleration profiles ordered by repetition ($y$-axis) in the experiment with  $d=30$, \emph{T}. The plots allow visualization of the variation across participants and repetitions.}\label{fig:acceleration}
\end{figure}

The movements were recorded with the Visualeyez (Phoenix Technologies Inc.) motion capture system VZ 4000. Two trackers, each equipped with three cameras, were mounted on the wall 1.5 m above the working surface, so that both systems had an excellent view of the table. A wireless infrared light-emitting diode (IRED) was attached to the wooden cylinder. The trajectories of markers were recorded in three Cartesian dimensions at a sampling rate of 110 Hz based on a reference frame anchored on the table. The starting position projected to the table was taken as the origin of each trajectory in three-dimensional Cartesian space. Recorded movement paths for two experimental conditions are shown in Figure~\ref{fig:samples}. The acceleration profiles considered in the following sections were obtained by using finite difference approximations of the raw velocity magnitude data, see Figure~\ref{fig:acceleration}. 

Obstacle avoidance was performed in 15 different conditions that combined three obstacle heights \emph{S}, \emph{M}, or \emph{T} with five distances of the obstacle from the starting position $d\in \{15, 22.5, 30, 37.5, 45\}$. A control condition had no obstacle.  The participants performed each condition 10 times. Each \emph{experimental condition} provided $n=100$ functional samples for a total of $n_f =1600$ functional samples in the dataset, leading to a total data size of $m=175,535$ observed time points.


The present data set is described in detail in \cite{Grimme2014}. The experiment is a refined version of the experiment described in \cite{grimme2012naturalistic}.

\subsection*{Time warping of functional data}

Not every movement has the exact same duration. Comparisons across movement conditions, participants, and repetitions are hampered by the resulting lack of alignment of the movement trajectories.  For a single condition, this is illustrated in Figure~\ref{fig:acceleration} (a) and (c), in which the duration of the movement clearly varies from participant to participant but also from repetition to repetition. Without alignment, it is difficult to compare movements. In an experiment such as the present, in which start and target positions are fixed, the standard solution for aligning samples is to use \emph{percentual time}; the onset of the individual movement corresponds to time 0\% and the end of the movement to time 100\%.  Such linear warping is based on detecting movement onset and offset through threshold criteria.  As can be seen in Figure~\ref{fig:acceleration} (b) and (c), linear warping does not align the characteristic features of the acceleration signals very well, however, as there is still considerable variation in the times at which acceleration peaks. There is, in other words, a nonlinear component to the variation of timing. 

\begin{figure}[!tp]
\centering
\subfloat[Recorded $z$-coordinate, observed time]{\includegraphics[width = 0.4\textwidth]{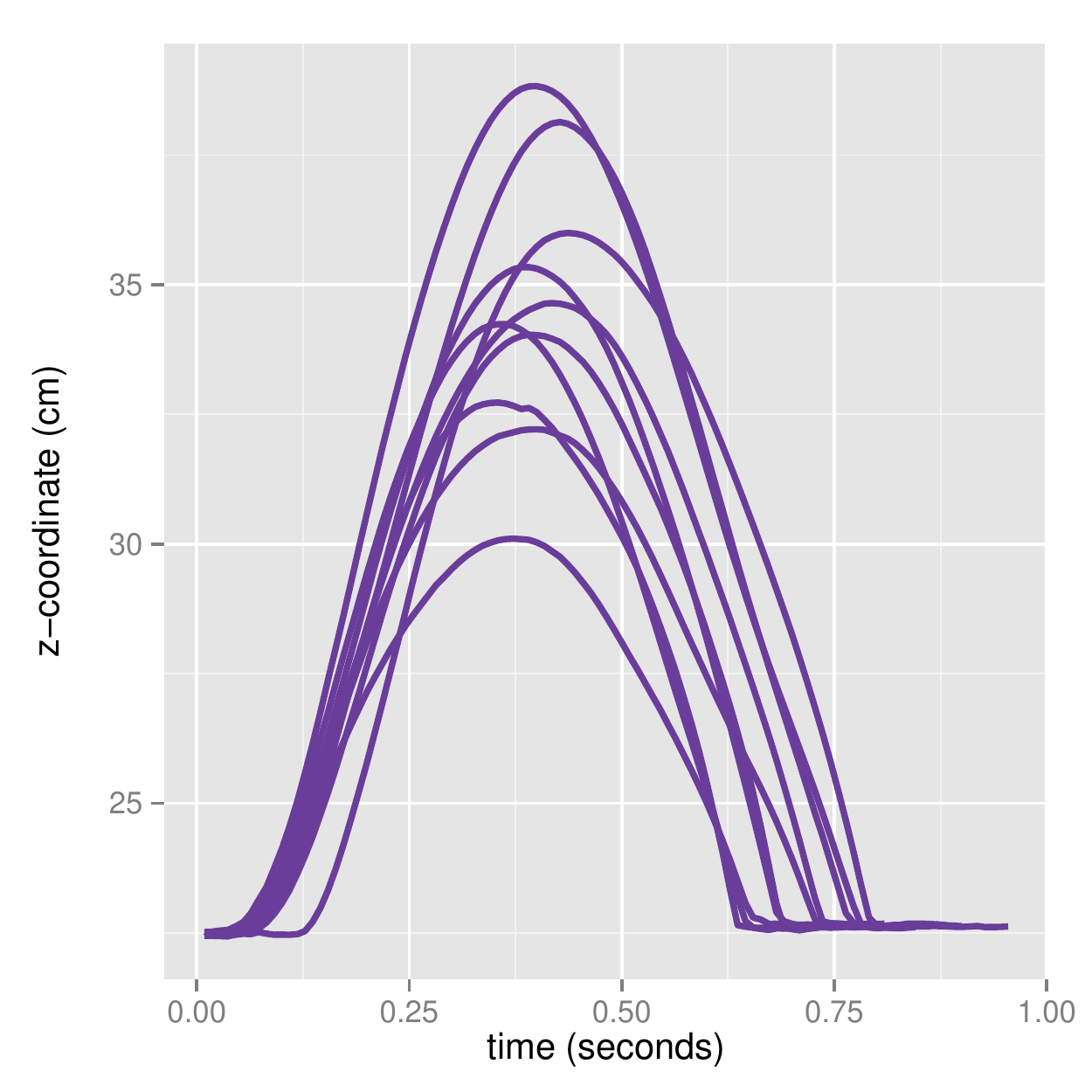}}\quad
\subfloat[Recorded $z$-coordinate, percentual time]{\includegraphics[width = 0.4\textwidth]{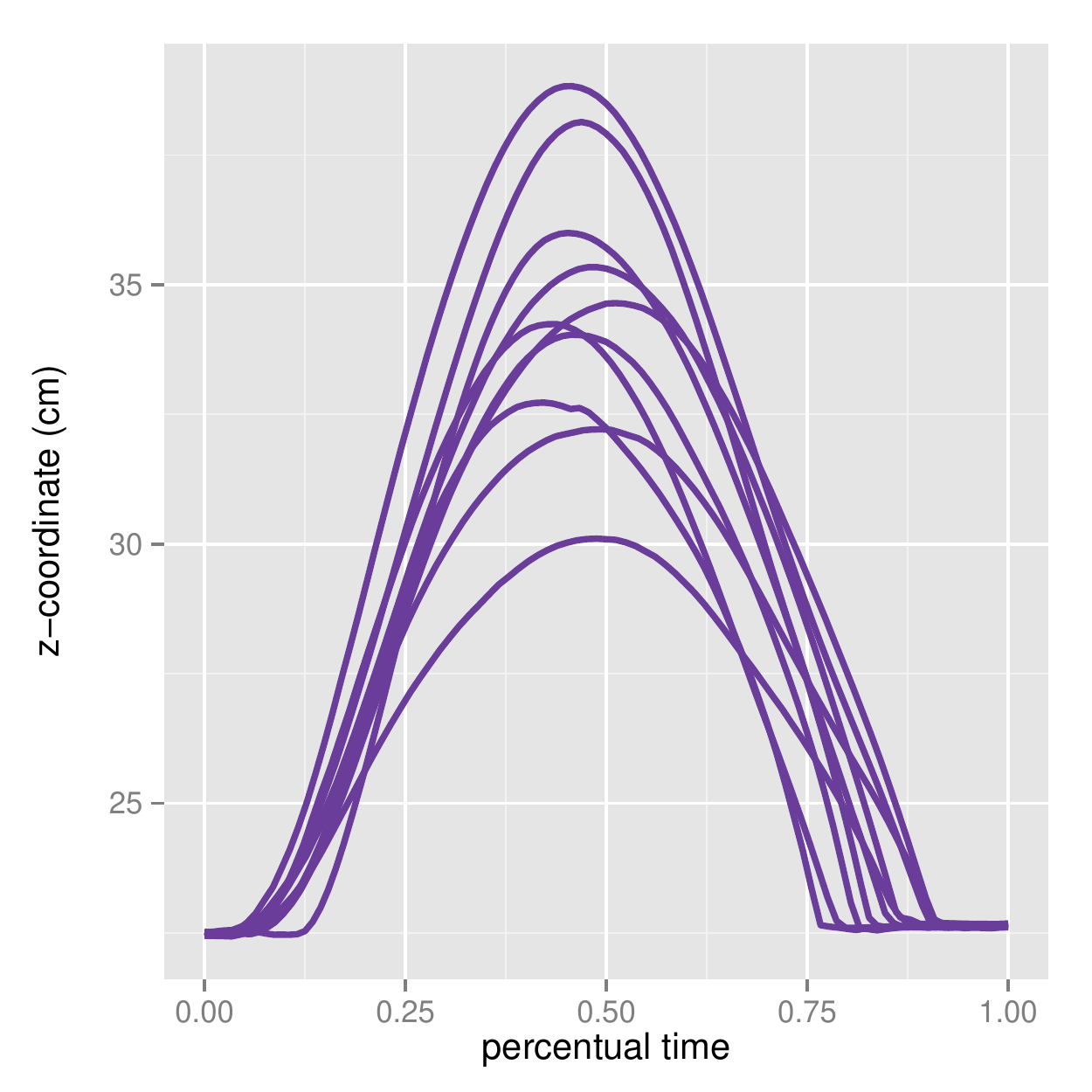}}\quad
\caption{Elevation ($z$-dimension) for one participant's (no.10) repetitions of the control condition without obstacle plotted on different time scales.  } \label{fig:warp_z}
\end{figure}

\begin{figure}[!tp]
\centering
\subfloat[Recorded $z$-coordinate, warped percentual time, DTW symmetric step pattern]{\includegraphics[width = 0.4\textwidth]{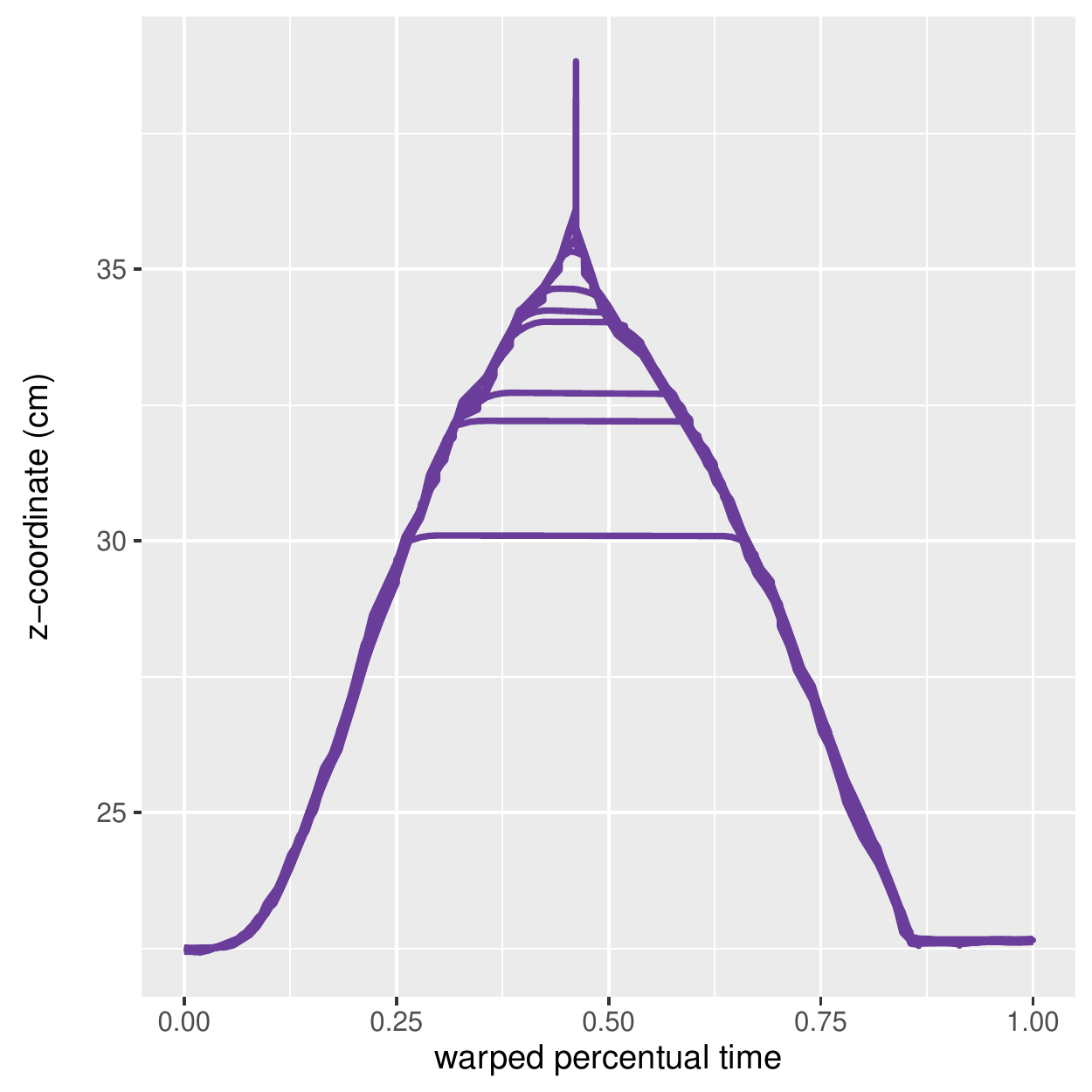}}\quad
\subfloat[Warping functions, white dashed line shows the identity, DTW symmetric step pattern]{\includegraphics[width = 0.4\textwidth]{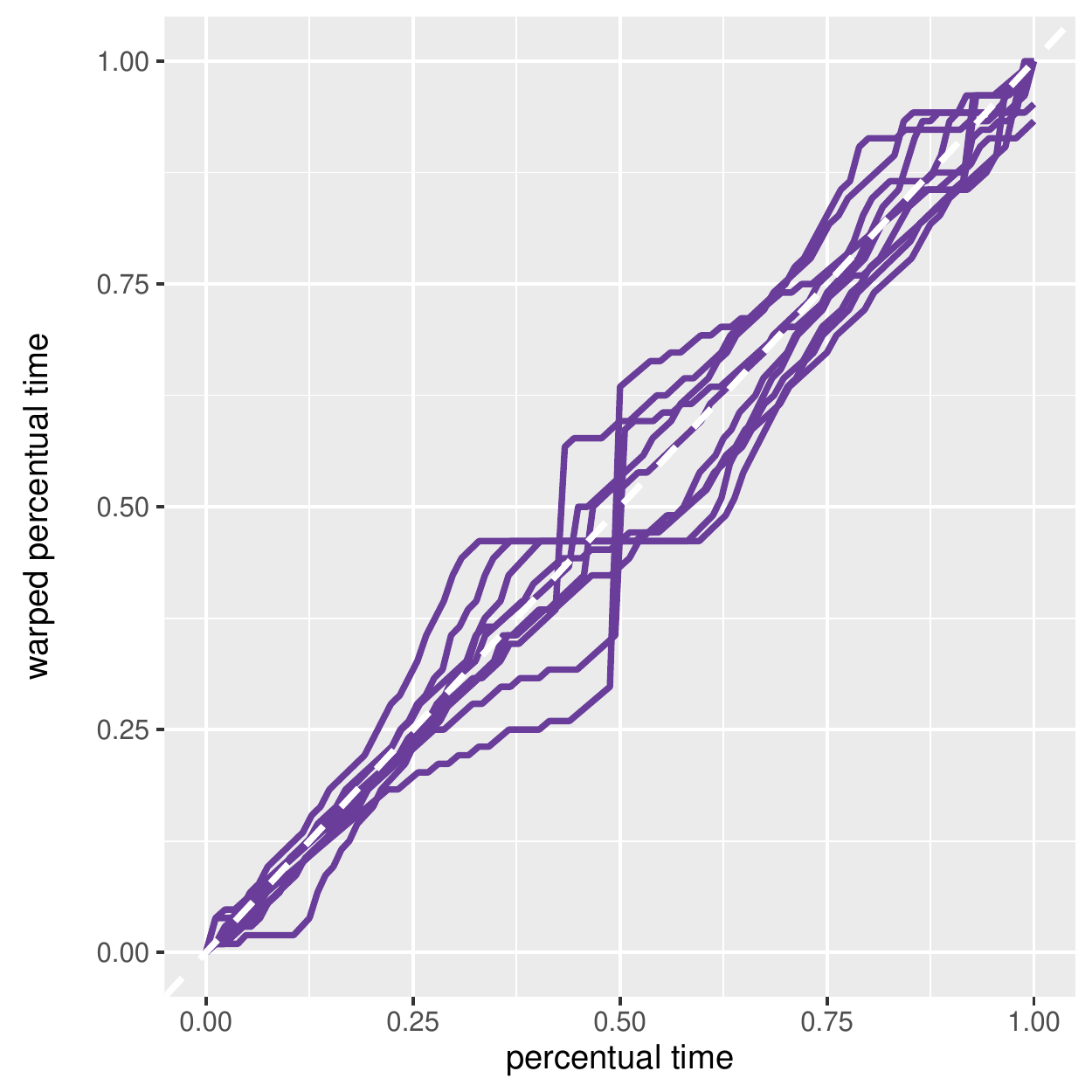}}\quad
\subfloat[Recorded $z$-coordinate, warped percentual time, DTW asymmetric step pattern]{\includegraphics[width = 0.4\textwidth]{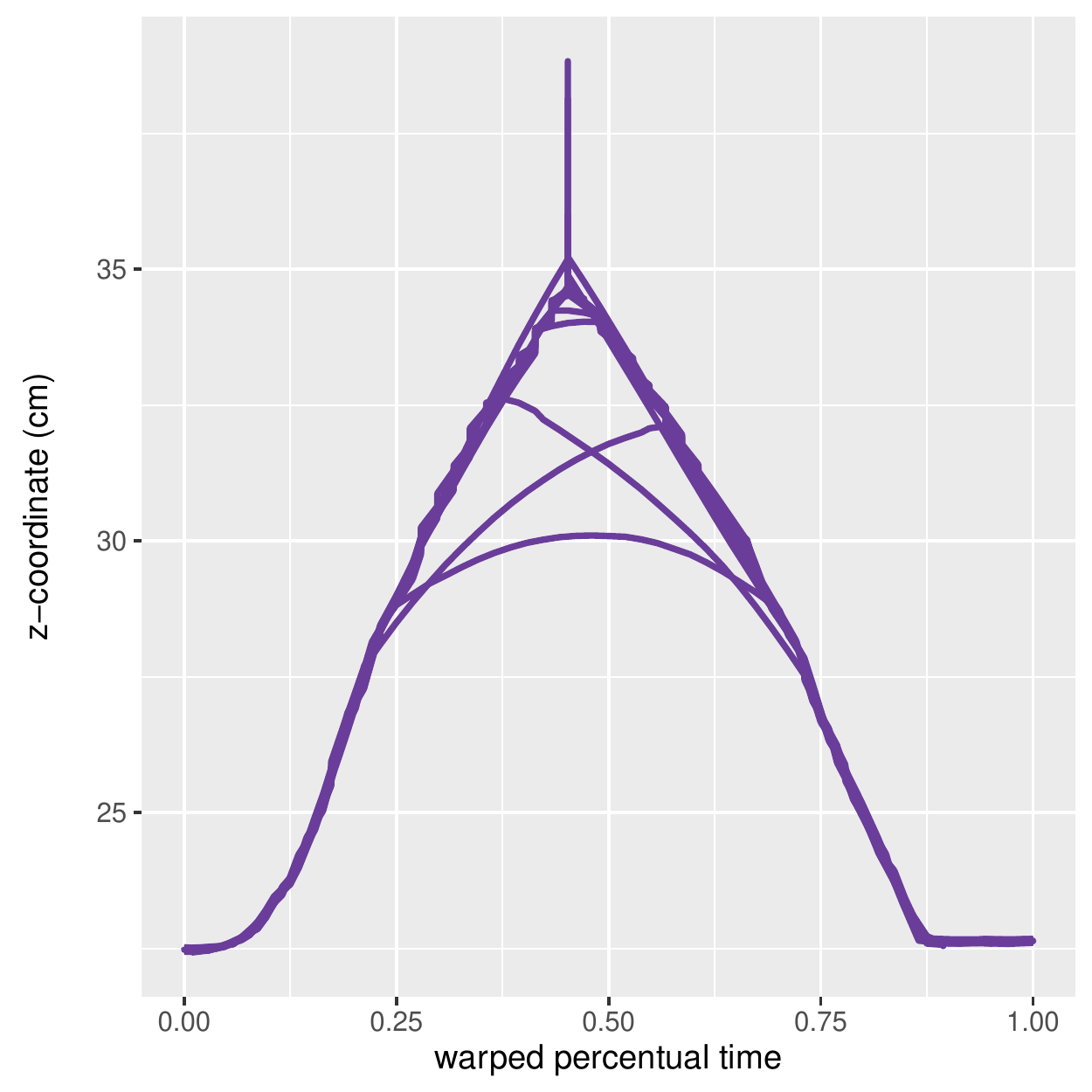}}\quad
\subfloat[Warping functions, white dashed line shows the identity, DTW asymmetric step pattern]{\includegraphics[width = 0.4\textwidth]{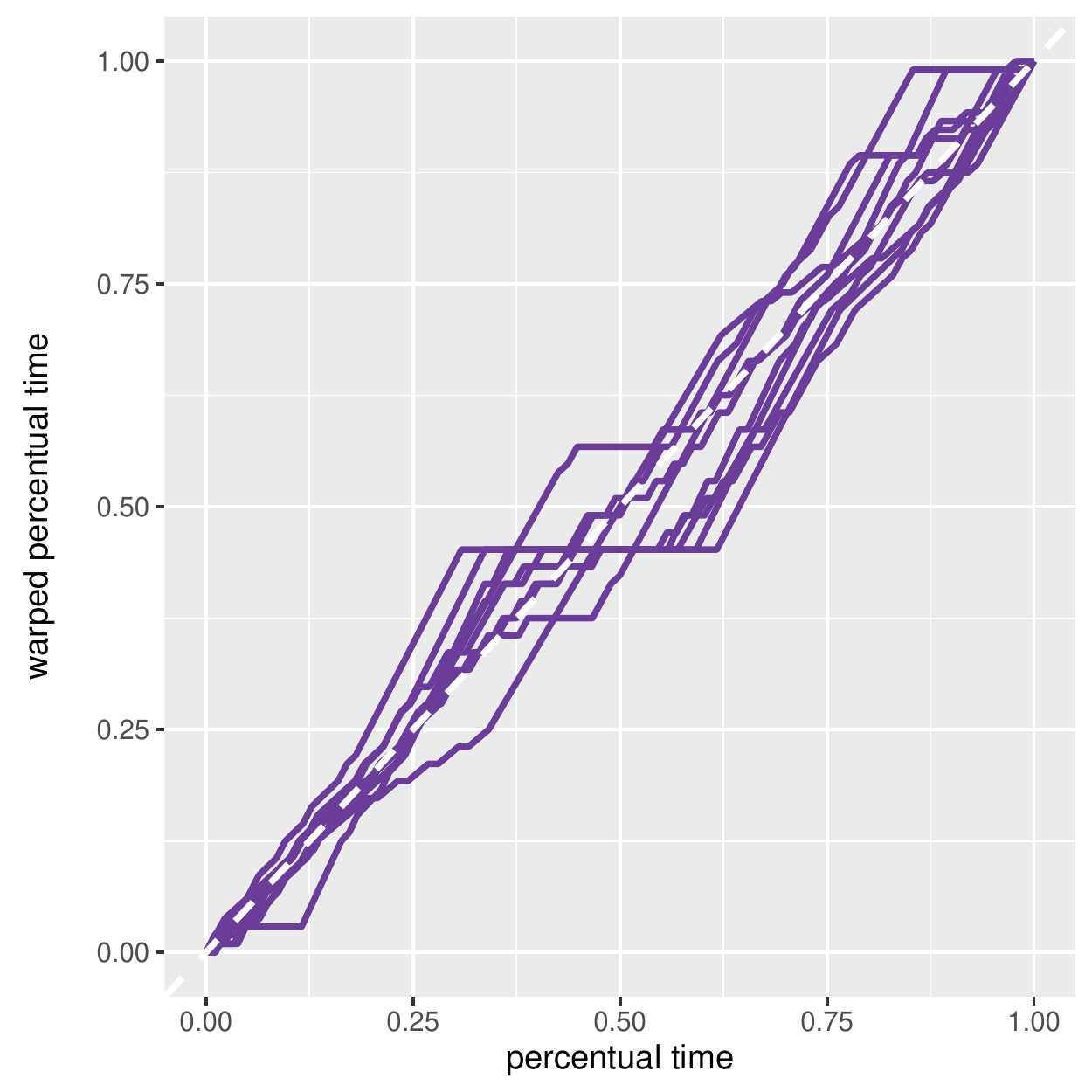}}\quad
\subfloat[Recorded $z$-coordinate, warped percentual time, DTW Sakoe-Chiba step pattern]{\includegraphics[width = 0.4\textwidth]{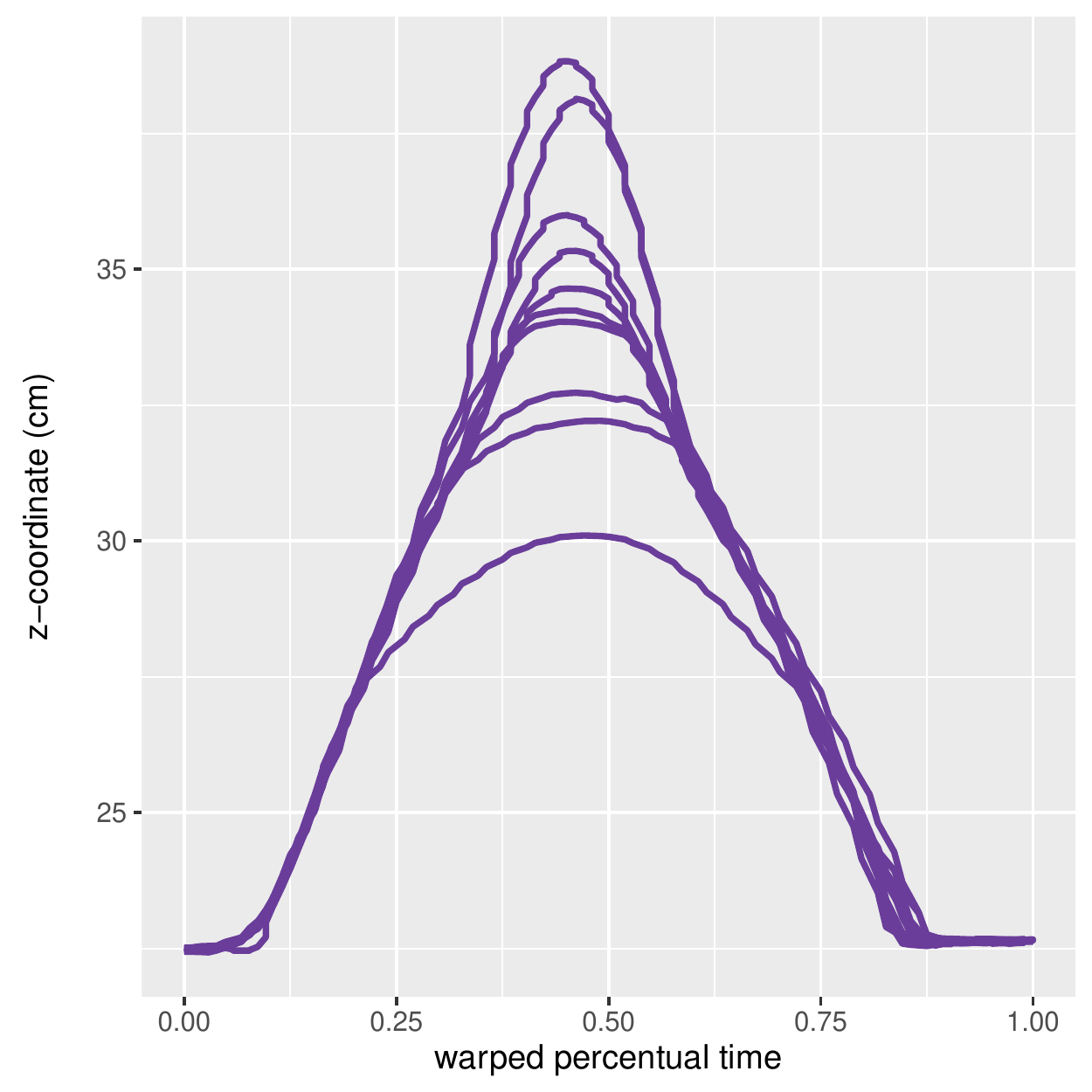}}\quad
\subfloat[Warping functions, white dashed line shows the identity, DTW Sakoe-Chiba step pattern]{\includegraphics[width = 0.4\textwidth]{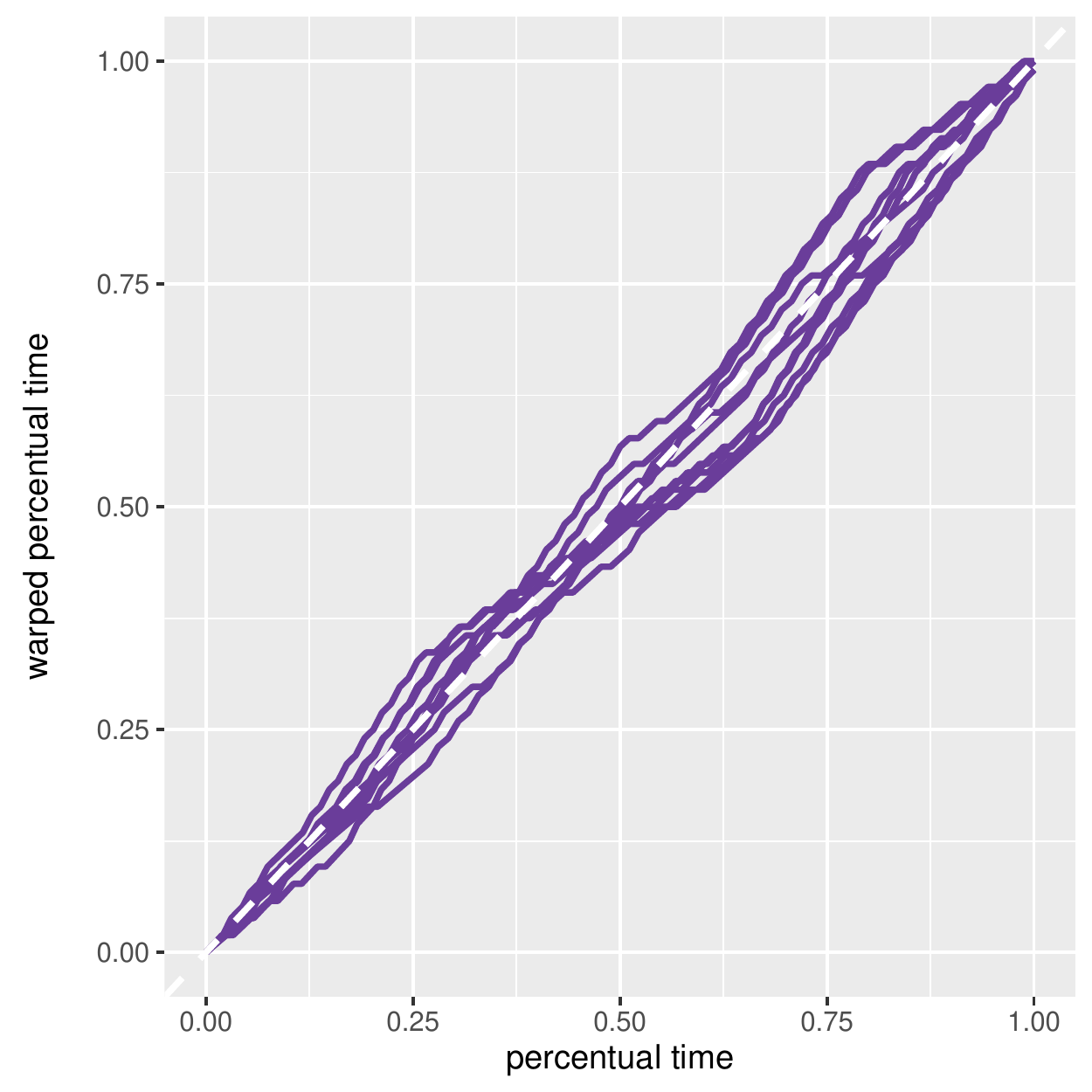}}\quad
\caption{The elevation trajectories from Figure~\ref{fig:warp_z} warped with dynamic time warping using different step patterns. \emph{Symmetric} step pattern denotes   the so-called White-Neely step pattern that has no local constraints, \emph{asymmetric} step pattern denotes  a slope-constrained step pattern where local slopes are required to be between 0 and 2, and \emph{Sakoe-Chiba} step pattern denotes the asymmetric step pattern proposed in \cite[Table I]{sakoe1978dynamic} with a slope constraint of 2.} \label{fig:warp_z_dtw}
\end{figure}

\begin{figure}[!tp]
\centering
\subfloat[Recorded $z$-coordinate, warped percentual time, proposed model]{\includegraphics[width = 0.4\textwidth]{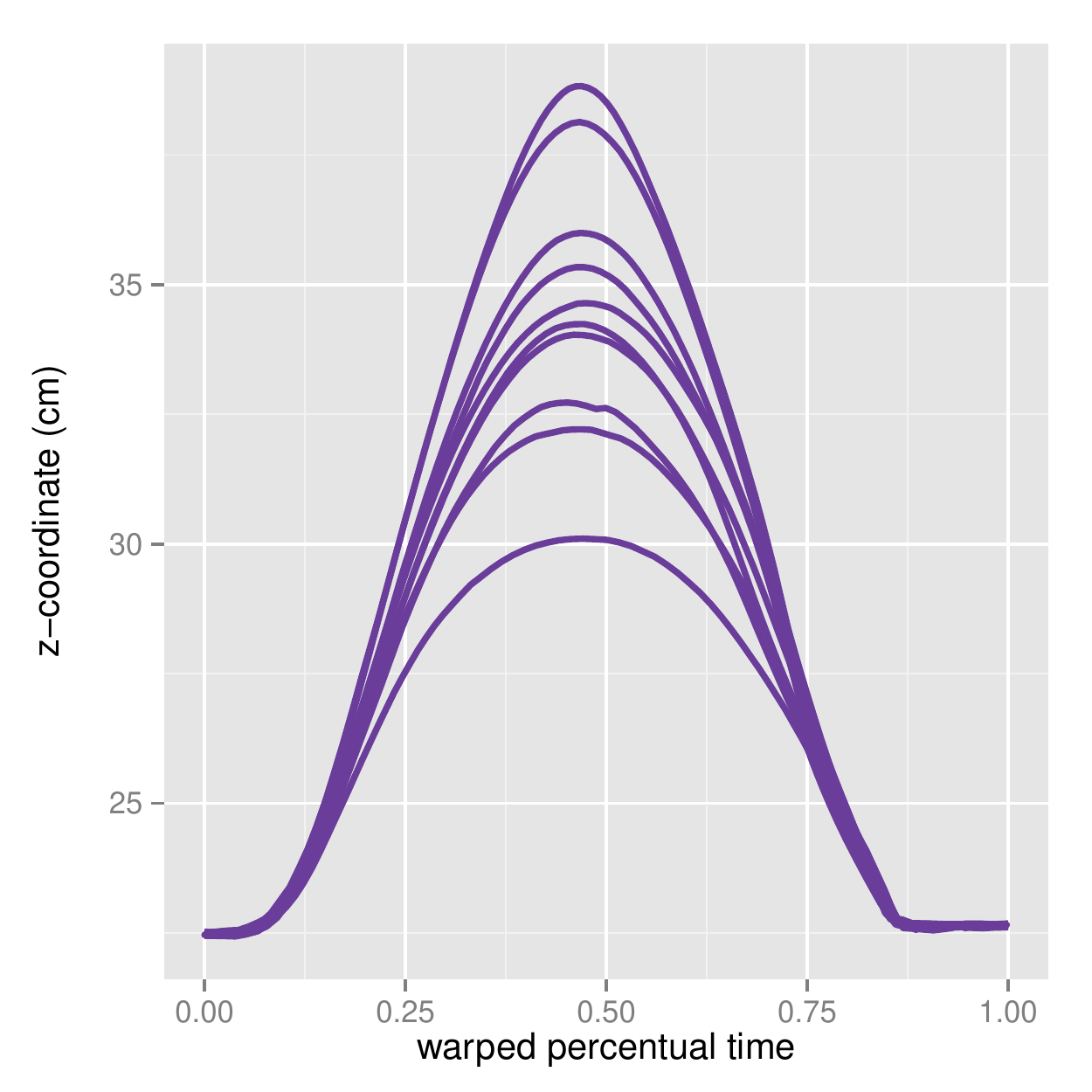}}\quad
\subfloat[Warping functions, white dashed line shows the identity, proposed model]{\includegraphics[width = 0.4\textwidth]{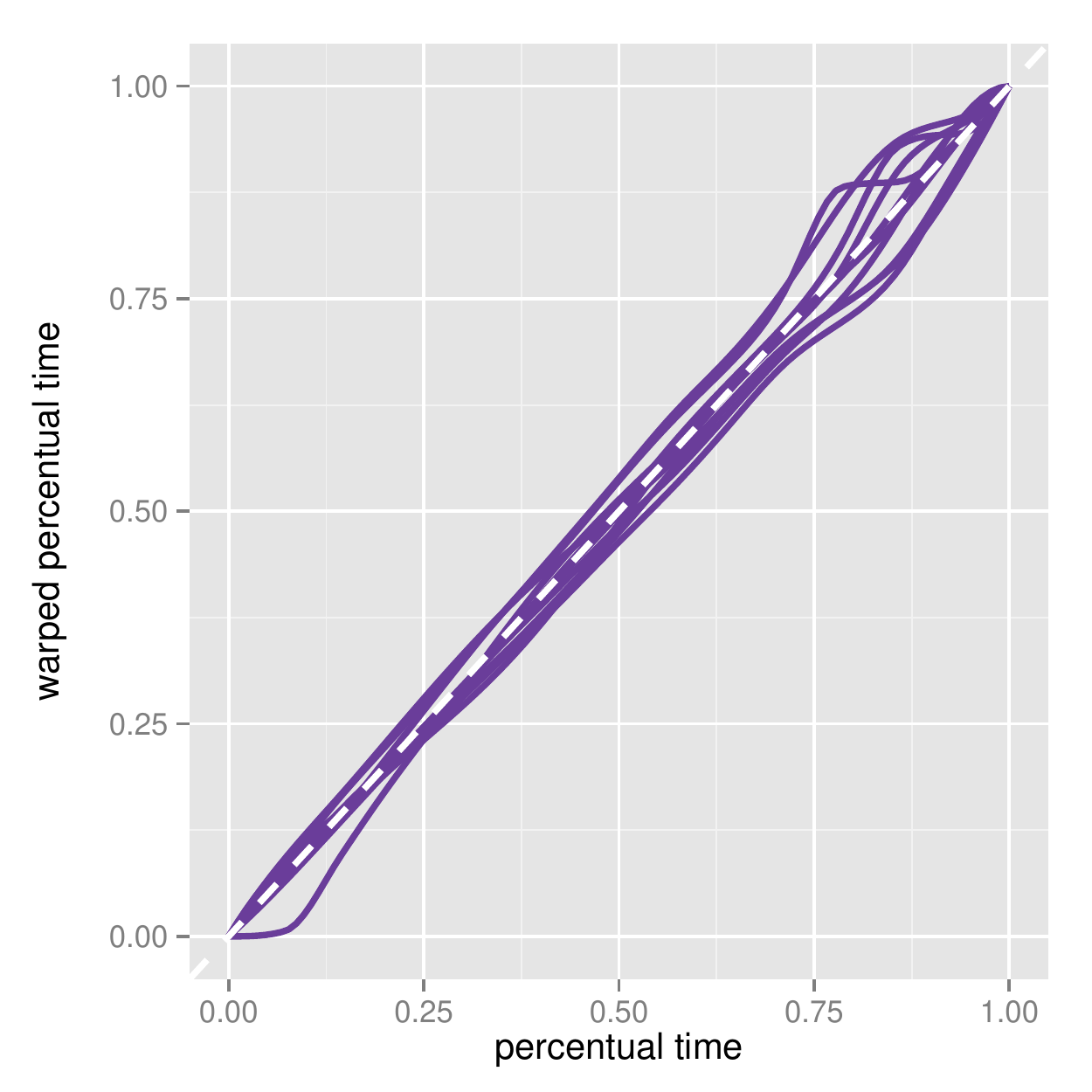}}\quad
\caption{The elevation trajectories from Figure~\ref{fig:warp_z} warped with the proprosed model.  } \label{fig:warp_z_pavpop}
\end{figure}

To handle nonlinear variation of timing,  the signal must be time warped based on an estimated, continuous, and monotonically increasing function that maps percentual time to warped percentual time, such that the functional profiles of the signals are best aligned with each other. {Such warping has traditionally been achieved by using the dynamic-time warping (DTW) algorithm \cite{sakoe1978dynamic} which offers a fast approach for globally optimal alignment under a prespecified distance measure(for reviews of time warping in the domains of biological movement modeling see \cite{bruderlin1995motion,troje2decomposing}). DTW is both simple and elegant, but while it will often produce much better results than cross-sectional comparison of time-warped curves \cite{berndt1994using, gavrila1995towards, giese2000morphable}, it does suffer from some problems. In particular, DTW requires a pointwise distance measure such as Euclidean distance. Therefore, the algorithm cannot take serially correlated noise effects in a signal into account. As a result, basic unconstrained DTW will overfit in the sense of producing perfect fits whenever possible, and for areas that cannot be perfectly matched, either stretch them or compress them to a single point. In other words, DTW cannot model curves with systematic amplitude differences, and using DTW to naively computing time warped mean curves is in general problematic \cite{niennattrakul2007inaccuracies, petitjean2011global}. This lacking ability to model serial correlated effects can be somewhat mitigated by restricting the DTW step pattern, in particular through a reduced search window for the warping function and constraints on the maximal step sizes. These are, however, hard model choices, they are restrictions on the set of possible warping functions, and they are a difficult to interpret since they seek to fix a problem in amplitude variation by penalizing warping variation. Instead, a much more natural approach would be to use a data term that models the amplitude variation encountered in data, and to impose warp regularization by using a cost function that puts high cost on undesired warping functions. In the following sections, we will propose a model with these properties, which, in addition, allows for estimating the data term, warp regularization and their relative weights from the data. 

To illustrate the difference between DTW and the proposed method, consider the example displayed in Figure~\ref{fig:warp_z} where the recorded $z$-coordinates (elevation) of one participant's 10 movements in the control condition (without obstacle) are plotted in recorded time (a) and percentual time (b). These samples have been aligned using DTW by iteratively estimating a pointwise mean function and aligning the samples to the mean function (10 iterations). The three rows of  Figure~\ref{fig:warp_z_dtw} display the results of the procedure using three different step patterns. We first note the strong overfitting of the symmetric and asymmetric step patterns, where the sequences with highest elevation are collapsed to minimize the residual. Secondly, we note the jagged warping functions that are results of the discrete nature of the DTW procedure. For comparison, we fitted a variant of the proposed model with a continuous model for the warping function controlled by 13 basis functions. We modeled both amplitude and warping effects as random Gaussian processes using simple, but versatile classes of covariance functions (see Supporting Information), and estimated the internal weighting of the effect directly from the samples. The results are displayed in Figure~\ref{fig:warp_z_pavpop}. We see that the warped elevation trajectories seem perfectly aligned, and that the corresponding warping functions are relatively simple, with the majority of variation being near the end of the movement. It is evident from the figures that both the alignment is much more reasonable, and the warping functions are much simpler than the warping functions found using DTW. 
}

Two questions naturally arise. Firstly, are the warping functions unique? Other warping functions could perhaps have produced similarly well-aligned data.  Secondly, do we want perfectly aligned trajectories? There is still considerable variation of the amplitude  in the warped $z$-coordinates of the movement trajectories in Figure~\ref{fig:warp_z} (c), for instance.  Some of the  variation visible in the unwarped variant in Figure~\ref{fig:warp_z} (b)  could be due to random variations in amplitude rather than timing.  

\begin{figure}[!tp]
\centering
\subfloat[Recorded $x$-coordinate, warped percentual time]{\includegraphics[width = 0.4\textwidth]{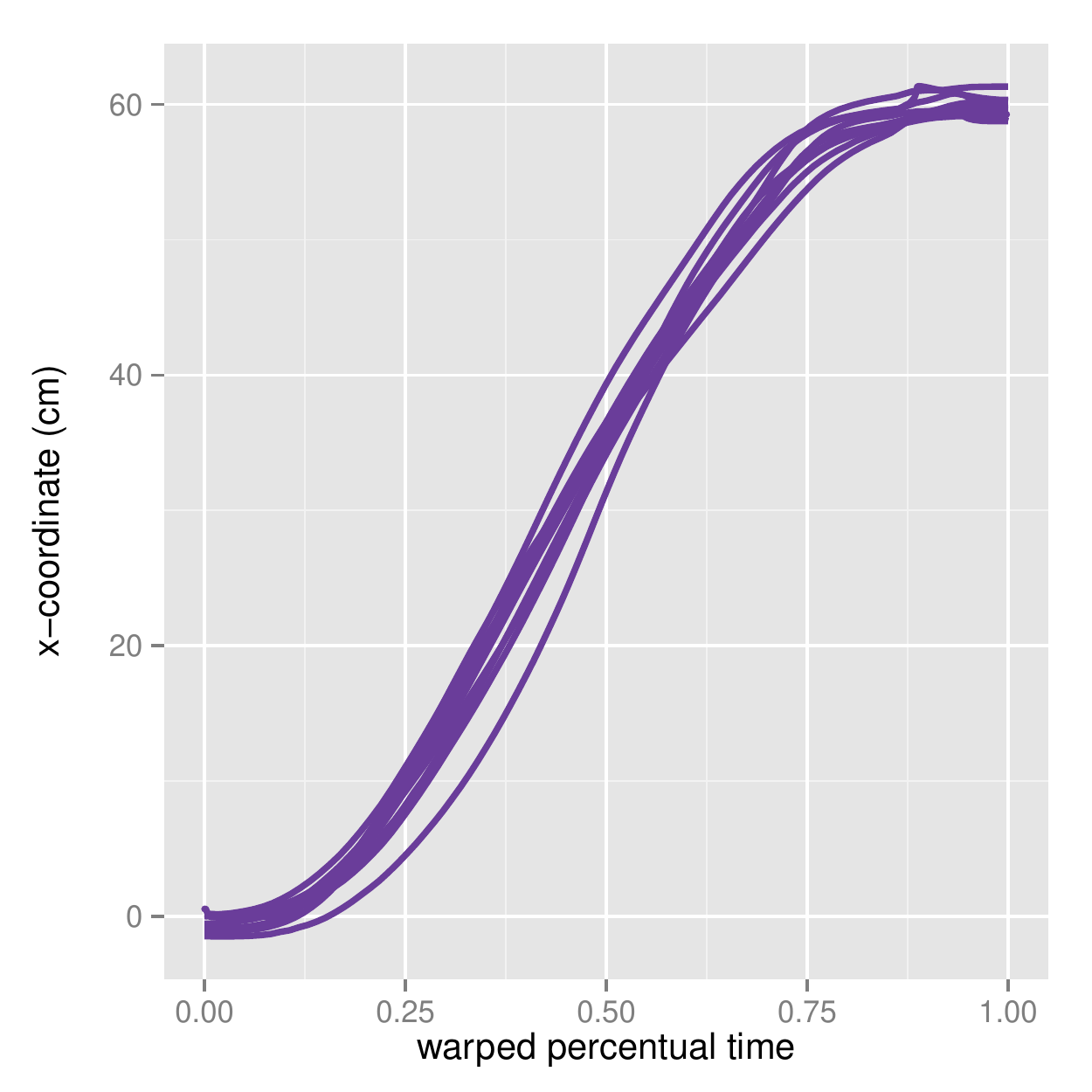}}\quad
\subfloat[Recorded $y$-coordinate, warped percentual time]{\includegraphics[width = 0.4\textwidth]{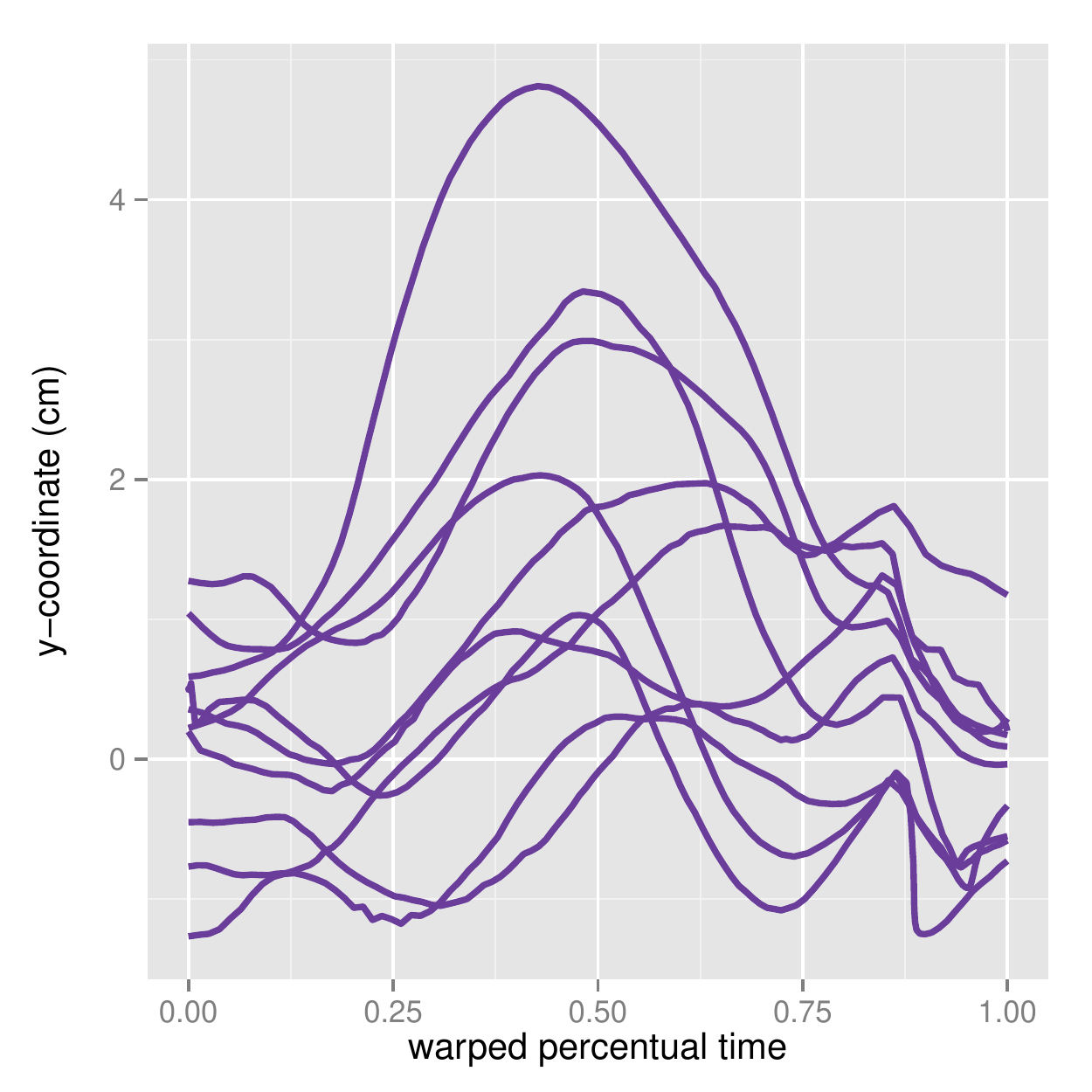}}\quad
\caption{The other two spatial coordinates, $x$, and $y$, of the movements from Figure~\ref{fig:warp_z} (a) warped with the warping functions from Figure~\ref{fig:warp_z_pavpop}, that was estimated from the elevation component $z$.} \label{fig:warp_xy}
\end{figure}

Using the time-warping functions that were determined by aligning the z-coordinates of the movements to now warp the trajectories for the $x$- and $y$-coordinates (Figure~\ref{fig:warp_xy}), we see that the alignment obtained is not perfect. Conversely, were we to use warping functions determined for the $x$- or $y$-coordinate to warp the z-coordinate we would encounter similarly  imperfect alignment. Thus, we need a method that avoids over-aligning, and represents a trade-off between the complexity of the warping functions and of the amplitude variation in data. In the next section, we introduce a statistical model that handle this trade-off in a data-driven fashion by using the patterns of variation in the data to find the most likely separation of amplitude and timing variation.

\subsubsection*{Statistical modeling of movement data to achieve time warping}
In the following, we describe inference for a single experimental condition.  
For a given experimental condition---an object that needs to be moved to
a target and an obstacle that needs to be avoided---we assume there is a common underlying pattern in all acceleration profiles; all
$n_p=10$ participants will lift the object and move it toward the target, lifting it over the obstacle at some point.  This assumption is supported by the pattern in the data that Figure~\ref{fig:acceleration} visualizes.  We denote the hypothesized underlying  acceleration profile shared across participants and repetitions  by $\theta$. In addition to this fixed acceleration profile, we assume that each participant,  $i$, has a typical deviation $\varphi_{i}$ from $\theta$, so that the acceleration profile that is characteristic of that participant is $\theta + \varphi_{i}$. Such a systematic pattern characteristic of each participant is apparent in Figure~\ref{fig:acceleration} (c) and (d).  The individual trials (repetitions) of the movement deviate from this characteristic profile of the individual. We model these deviations as additive random effects with serial correlations so that for each repetition, $j$, of the experimental condition we have an additive random effect $x_{ ij}$ that causes deviation from the ideal profile. Finally we assume that the data contains observation noise $\varepsilon_{ ij}$ tied to the tracking system and data processing.

Time was implicit, up to this point, and the observed acceleration profile was decomposed into additive, linear contributions. We now assume, in addition, that each participant, $i$, has a consistent timing of the movement across repetitions, that is reflected in the temporal deformation of the acceleration profile (Figure~\ref{fig:acceleration} (a) and (c)) and is captured by the time warping function $\nu_{i}$. On each repetition, $j$, of the condition, the timing of participant, $i$, contains a random variation of timing  around $\nu_{ i}$ captured by a random warping function $v_{ ij}$ (see Figure~\ref{fig:acceleration}). 

Altogether, we have described the following statistical model of the observed acceleration profiles across participants: 
\begin{align}
y_{ ij}(t) = (\theta+ \varphi_{ i})\circ (\nu_{ i} + v_{ ij})(t) + x_{ ij}(t)+\varepsilon_{ ij}(t)\label{model}
\end{align}
where $\circ$ denotes functional composition, $t$ denotes time, $\theta, \,\varphi_{ i}, \,\nu_{ i}\,:\,\R \rightarrow \R$ are fixed effects and $v_{ ij}$, $x_{ i j}$ and $\varepsilon_{ i j}$ are random effects. The serially correlated effect $x_{ ij}$ is assumed to be a  zero-mean Gaussian process with a parametric covariance function $\mathcal{S}\,:\, \R\times \R \rightarrow \R$; the randomness of the warping function $v_{ij}$ is assumed to be completely characterized by a latent vector of $n_w$ zero-mean Gaussian random variables $\bw_{ ij}$ with covariance matrix $\sigma^2 C$; and $\varepsilon_{ij}$ is Gaussian white noise with variance $\sigma^2$.


{
Compared to conventional methods for achieving time warping, the proposed model~\eqref{model} models amplitude and warping variation between repetition as random effects, which enables separation of the effects from the joint likelihood. Conventional approaches for warping model warping functions as fixed effects and do not contain amplitude effects \cite{Ramsay}. The idea of modeling warping functions as random effects have previously been considered by \cite{ronn, Gervini, RonnSkovgaard} where warps were modeled as random shifts or random polynomials. None of these works however included amplitude variation. Recently, some works have considered models with random affine transformations for warps and amplitude variation in relation to growth curve analysis \cite{beath2007infant, cole2010sitar}. A generalization that does not require affine transformations for warp and amplitude variation is presented in \cite{RaketSommerMarkussen}. The presented model~\eqref{model} is a hierarchical generalization of the model presented in \cite{RaketSommerMarkussen}. In the context of aligning image sequences in human movement analysis, morphable models \cite{giese2000morphable, ilg2003hierarchical} model an observed movement pattern as a linear combination of prototypical patterns using both nonlinear warping functions (estimated using DTW) and spatial shifts. Thus morphable models are similar to the warping approaches that model both warp and amplitude effects as fixed \cite{marron2015functional}.
}

\subsubsection*{Maximum likelihood estimation of parameters}

Model \eqref{model} has a considerable number of parameters,  both for linear and nonlinear dependencies on the underlying state variable acceleration. The model also has effects that interact. This renders direct simultaneous likelihood
estimation intractable. Instead we propose a scheme in which fixed effects and parameters are estimated and random effects are predicted iteratively on three different levels of modeling. 
\begin{description}
\item[Nonlinear model] At the nonlinear level, we consider the original model \eqref{model}, and simultaneously perform conditional likelihood estimation of the participant-specific warping functions and predict the random warping functions from the negative log posterior. All other parameters remain fixed.

\item[Fixed warp model] At the fixed warp level, we fix the participant-specific warping effect $\nu_i$ at the conditional maximum likelihood estimate, and the random warping function $v_{ij}$, at the predicted values. The resulting model is an approximate linear mixed-effects model with  Gaussian random effects $x_{ ij}$ and $\varepsilon_{ ij}$, that allows direct  maximum-likelihood estimation of the remaining fixed effects, $\theta$ and $\varphi_{ i}$.

\item[Linearized model] At the linearized level, we consider the first-order Taylor approximation of model \eqref{model} in the random warp $v_{ij}$. This linearization is done around the estimate of $\nu_{ i}$ plus the given prediction of $v_{ij}$ from the nonlinear model. The result is again a linear mixed-effects model, for which one can compute the likelihood explicitly, while taking the uncertainty of all random effects---including the nonlinear effect $v_{ij}$---into account. At this level all variance parameters are estimated using maximum-likelihood estimation.
\end{description}

The estimation/prediction procedure is inspired by the algorithmic framework proposed in  \cite{RaketSommerMarkussen}. The estimation procedure is, however, adapted to the hierarchical structure of data and refined in several respects. On the linearized model level, the nonlinear Gaussian random effects are approximated by linear combinations of correlated Gaussian variables around the mode of the nonlinear density. The linearization step thus corresponds to a Laplace approximation of the likelihood, and the quality of this approximation is approximately second order \cite{wolfinger1993laplace}.

Let $\by_{ ij}$ be the vector of the $m_{ij}$ observations for participant $i$'s $j$th replication of the given experimental condition,  and let $\by_i$ denote the concatenation of all functional observations of participant $i$ in the experimental condition,  and  $\by$ the concatenation all these observations across participants. We denote the lengths of these vectors by $m_i$ and $m$. Furthermore, let $\sigma^2 S_{ ij}$, $\sigma^2 S_{ i}$ and $\sigma^2  S$ denote the covariance matrices of $\bx_{ ij}=(x_{ ij}(t_k))_k$, $\bx_{ i}=(\bx_{ ij})_{j}$, and $\bx=(\bx_{ i})_{i}$ respectively. We note that the index set for $k$ depends on $i$ and $j$ since the covariance matrices $S_{ ij}$ vary in size due to the different durations of the movements and because of possible missing values when markers are occluded. 

We note that all random effects are scaled by the noise standard deviation $\sigma$. This parametrization is chosen because it simplifies the likelihood computations, as we shall see.
Finally, we denote the norm induced by a full-rank covariance matrix $A$ by $\|\bz\|^2_{A} = \bz^\top A^{-1} \bz $. 

\paragraph{Fixed warp level}
We model the underlying profile, $\theta$, and the participant-specific
variation around this trajectory, $\varphi_i$, in the common (warped)
functional basis {$\bP\in \R^{m\times K}$}, with weights $\bc= (c_1,\dots, c_{K})$ for
$\theta$ and $\bd_i = (d_{i1},\dots, d_{iK})$ for $\varphi_i$.
We assume that the participant-specific  variations, $\varphi_i$, are centered around
$\theta$  and thus $\sum_i\bd_i = \boldsymbol{0}\in \R^{K}$. Furthermore, 
the square magnitude of the weights, $\bd_i$, is penalized with a weighting
factor $\eta$. This penalization helps guiding the alignment process in
the direction of the highest possible level of detail in the common
profile $\theta$ when the initial alignment is poor. 

For fixed warping functions $\nu_i$ and $v_{ij}$, the negative log likelihood function in $\theta=\bP\bc$ is proportional to	
\[
\ell (\bc) =\|{\by}-\bP\bc\|_{\I_{n}+S}^2
\]
where $\I_{n}$ denotes the $n\times n$ identity matrix. This yields the estimate
\[
\hat \bc = (\bP^\top (\I_{m} + S)^{-1} \bP )^{-1}\bP^\top (\I_{m} + S)^{-1} \by.
\]
The penalized negative profile log likelihood for the weights $\bd_i$ for $\varphi_i$ is proportional to	
\[
\ell (\bd_i ) = \|{\by}_{i}-\bP_{i}(\hat\bc + \bd_i)\|^2_{I_{m_{i}}+S_{i}}  
+ \eta \bd_i^\top \bd_i,
\]
which gives the maximum likelihood estimator
\[
\hat\bd_i = (\bP_{i}^\top (\I_{m_i} + S_i)^{-1} \bP_{i} + \eta\I_{K} )^{-1}\bP_{i}^\top (\I_{m_i} + S_i)^{-1} (\by_i-\bP_i \hat\bc).
\]

\paragraph{Nonlinear level}
Similarly to the linear mixed-effects setting \cite{Robinson}, it is natural to predict nonlinear random effects from the posterior \cite{LindstromBates}, since these predictions correspond to the most likely values of the random effects given the observed data. Recall that the Gaussian variables, $\bw_{ ij}$, parametrize the randomness of the repetition-specific warping function $v_{ij}$. Since the conditional negative (profile) log likelihood function in $\nu_{ i}$ given the random warping function $v_{i j}$ and the negative (profile) log posterior for $\bw_{ i j}$ coincide, we propose to simultaneously estimate the fixed warping effects $\nu_{ i}$ and predict the random warping effects $v_{ij}$  from the joint conditional negative log likelihood/negative log posterior  which is proportional to
\begin{align}
p(\nu_{ i}, \bw_{ ij})& = \sum_{j} \|\by_{ ij}-(\hat\theta+ \hat\varphi_{ i})\circ (\nu_{ i} + v_{ij})(t_k)_k\|^2_{\I_{n_{ ij}}+ S_{ ij}}
 + \sum_{j}\|\bw_{ ij}\|_C^2.
 \label{posterior}
\end{align}
Since the variables $\bw_{ij}$ can be arbitrarily transformed through the choice of warping function $v_{ij}$, the assumption that variables are Gaussian is merely one of computational convenience. 

\paragraph{Linearized level}

We can write the local linearization of model \eqref{model} in the random warping parameters $\bw_{ij}$ around a given prediction $\bw_{ij}^0$ as a vectorized linear mixed-effects model
\begin{align}
\by \approx \btheta+\bZ (\bw-\bw^0)+\bx+\bepsilon\label{model:linear}
\end{align}
with effects given by
\[
\btheta=\{(\theta + \varphi_{i})\circ(\nu_i + v_{ij}^0)(t_k)\}_{ijk}\in \R^{m},
\]
\[
\bZ = \mathrm{diag}(\bZ_{ij})_{ij},\qquad
\bZ_{ij} = \{\partial_{\bt} (\theta + \varphi_{i})\circ(\nu_i + v_{ij}^0)(t_k)(\nabla_{\bw}v^0_{ij}(t_k))^\top\}_{k}\in \R^{m_i\times n_w},
\]
\[
\bw=(\bw_{ij})_{ij}\sim\mathcal{N}_{n_p n_{\bw}}(0, \sigma^2 \I_{n_p}\otimes C),\qquad
\bx \sim\mathcal{N}_{m}(0, \sigma^2 S), \qquad\bepsilon\sim \mathcal{N}_{m}(0,\sigma^2\I_{m}),
\]
where $v_{ij}^0$ indicates that the warping function is evaluated at the prediction $\bw_{ij}^{0}$, $\mathrm{diag}(\bZ_{ij})_{ij}$ is the block diagonal matrix with the $\bZ_{ij}$ matrices along its diagonal, and $\otimes$ denotes the Kronecker product.

Altogether, twice the negative profile log likelihood function for the linearized model \eqref{model:linear} is
\begin{align}
\ell(\sigma^2, C, S)=m\log\sigma^2 +\log\det \bV+\sigma^{-2} \|\by - \hat\btheta+\bZ\bw^0\|_{V}^2\label{likelihood}
\end{align}
where $\bV = S+\bZ (\I_n\otimes C) \bZ^\top + \I_{m}$.

\subsubsection*{Modeling of effects and algorithmic approach}\label{sec:fit_results}

So far, the model \eqref{model} has only been presented in a general sense. We now consider the specific modeling choices. The acceleration data has been rescaled using a common scaling for all experimental conditions, { such that the span of data values has length 1} and the global timespan is the interval $[0,1]$.

To model the amplitude effects, we use a cubic B-spline basis $\bP$ with $K$ knots \cite{deBoor2001practical}.

We require that the fixed warping function $\nu_i$ is an increasing piecewise linear homeomorphism parametrized by $n_{w}$ equidistant anchor points in $(0,1)$, and assume that $v_{ij}$ is of the form
\[
v_{ij}(t)=t +\mathcal{E}_{ij}(t),
\]
where $\mathcal{E}_{ij}(t)$ is the linear interpolation at $t$ of the values $\bw_{ij}$ placed at the $n_w$ anchor points in $(0,1)$. In the given experimental setting, the movement path is fixed at the onset and the end of the movement. The movement starts when the cylindrical object is lifted and ends when it is placed at its target position. Thus, we would expect the biggest variation in timing to be in the middle of the movement (in percentual time). These properties can be modeled by assuming that $\bw_{ij}$ is a discretely observed zero-drift Brownian bridge with scale $\sigma^2\gamma^2$ \cite[Chapters 8-9]{billingsley2009convergence}, which means that the covariance matrix $\sigma^2C$ is given by point evaluation of the covariance function
\[
\mathcal{C}(t, t')=\sigma^2\gamma^2\, t(1 - t')
\]
for $t \le t'$. When predicting the warps from the negative log posterior we restrict the search space to warps $\nu_i$ and $\nu_i + v_{ij}$ that are increasing homeomorphic maps of the domain $[0,1]$ onto itself. The conditional distribution of $\nu_{ij}$ given this restriction is slightly changed. For the used numbers of anchor points $n_w$ and the estimated variance parameters the difference is however minuscule, and we use the original Brownian model as a high-quality approximation of the true distribution.

We assume that the sample paths of the serially correlated effects $x_{ij}$ are continuous and that the process is stationary \cite{flash1985coordination}. A natural choice of covariance is then the Mat\'ern covariance with smoothness parameter $\mu$, scale $\sigma^2\tau^2$ and range $1/\alpha$ \cite{stein1999interpolation}, since it offers a broad class of stationarity covariance functions. 

Finally, in order to consistently penalize the participant-specific spline across experimental conditions with varying variance parameters, we will use penalization weights that are normalized with the variance of the amplitude effects, $\eta=\lambda/(1+\tau^2)$.

The algorithm for doing inference in model~\eqref{model} is outlined in  Algorithm~\ref{alg}. We have found that $i_{\max}=j_{\max}=5$ outer and inner loops are sufficient for convergence. A wide variety of these types of models can be fitted using the \texttt{pavpop} R package \cite{pavpop}. {A short guide on model building and fitting is available in Supporting Information. }

\begin{algorithm}
\caption{Maximum likelihood estimation for model \eqref{model}}\label{alg}
\begin{algorithmic}[1]
\Procedure{MLE}{$\by$, $\eta$, $\tau^2$, $\alpha$, $\gamma^2$}
   \State Compute $\hat\theta$ and $\hat\varphi_1,\dots,\hat\varphi_{m}$ assuming an identity warp\Comment{Initialize}
  \For{$i=1,\dots,i_{\max}$} \Comment{Outer iterations}
    \For{$j=1,\dots,j_{\max}$}  \Comment{Inner iterations}
    \State Estimate and predict warping functions by minimizing the posterior \eqref{posterior}
    \If{Estimates and predictions do not change}
    \textbf{break}
    \EndIf
    \State Recompute $\hat\theta$ and $\hat\varphi_1,\dots,\hat\varphi_{m}$
    \EndFor
    \State Estimate variance parameters by minimizing the linearized likelihood \eqref{likelihood}
  \EndFor
   \State \textbf{return} $\hat\theta$, $\hat\varphi_1,\dots,\hat\varphi_{m}$, $\hat\nu_1,\dots,\hat\nu_{m}$, $\hat\sigma^2$, $\hat\eta$, $\hat\tau^2$, $\hat\alpha$, $\hat\gamma^2$\Comment{Maximum likelihood estimates}
\EndProcedure
\end{algorithmic}
\end{algorithm}

In the following we consider two approaches,  with (1) samples parametrized by recorded time (Figure~\ref{fig:acceleration}a) and (2) samples parametrized by percentual time (Figure~\ref{fig:acceleration}b). Parameters for the latter case will be denoted by a subscript $\text{p}$.

\begin{figure}[!h]
\centering
\begin{tabular}{m{1em}p{2.8\textwidth/10}p{2.8\textwidth/10}p{2.8\textwidth/10}}
& \centering\emph{S} & \centering\emph{M} & \centering\emph{T} \cr
\vspace{-10em}\rotatebox{90}{\scriptsize 15.0 cm} & \mbox{\hspace{-1.0em}\includegraphics[scale = 0.6]{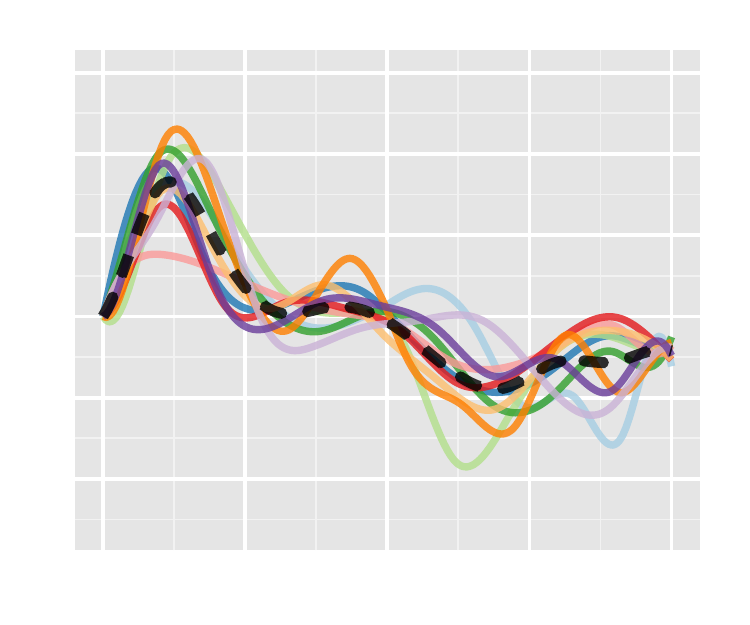}} &
\mbox{\hspace{-1.0em}\includegraphics[scale = 0.6]{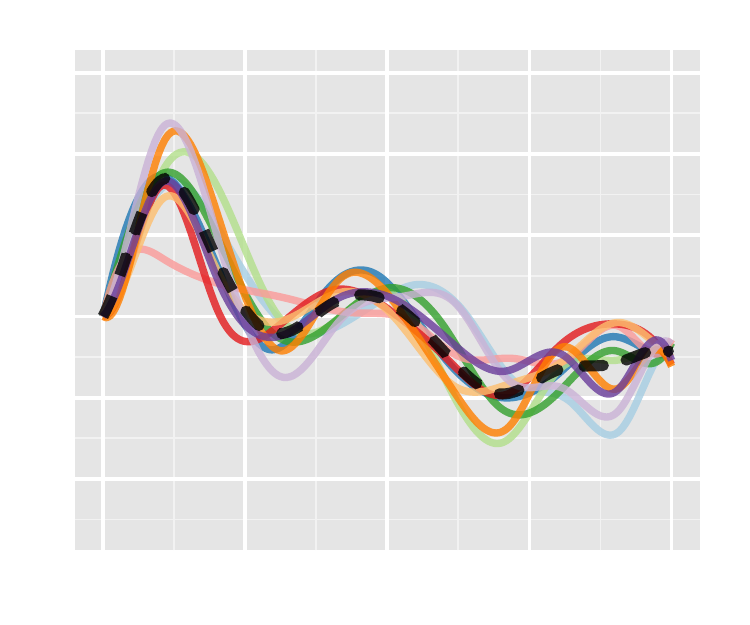}} &
\mbox{\hspace{-1.0em}\includegraphics[scale = 0.6]{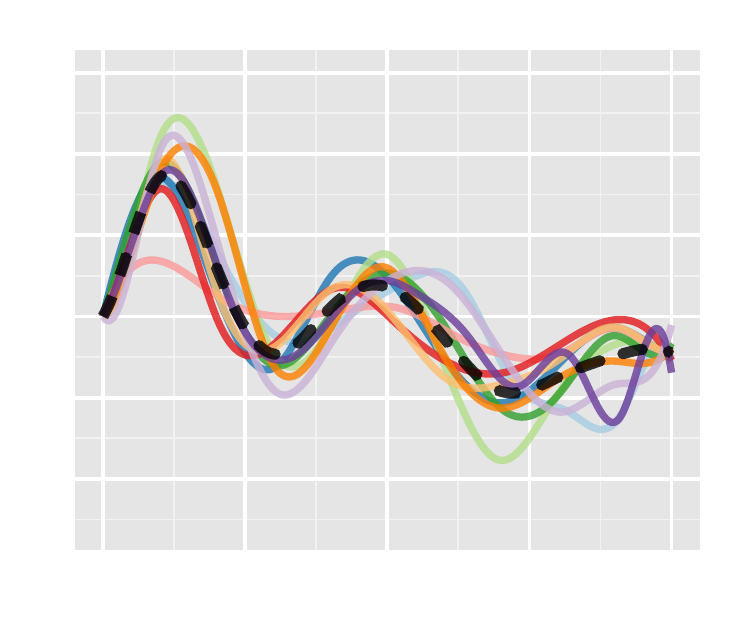}}\\[-1.5em]
\vspace{-10em}\rotatebox{90}{\scriptsize 22.5 cm} & \mbox{\hspace{-1.0em}\includegraphics[scale = 0.6]{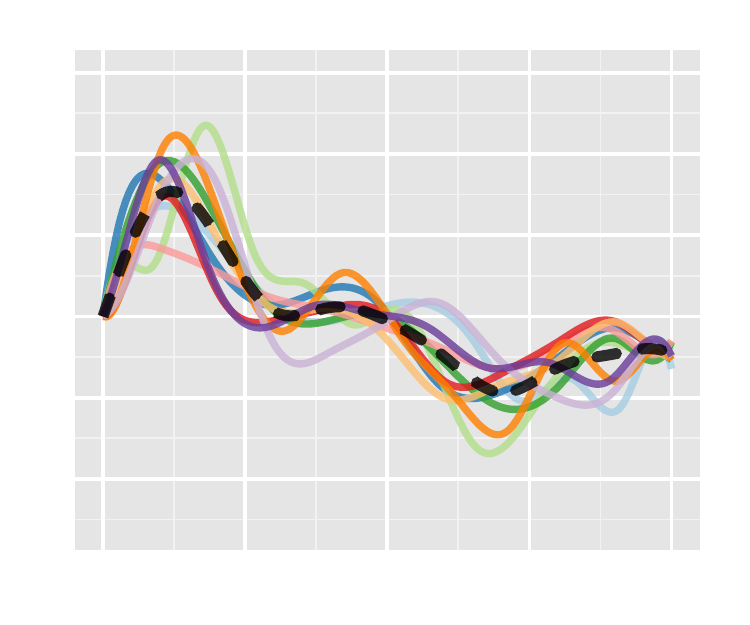}} &
\mbox{\hspace{-1.0em}\includegraphics[scale = 0.6]{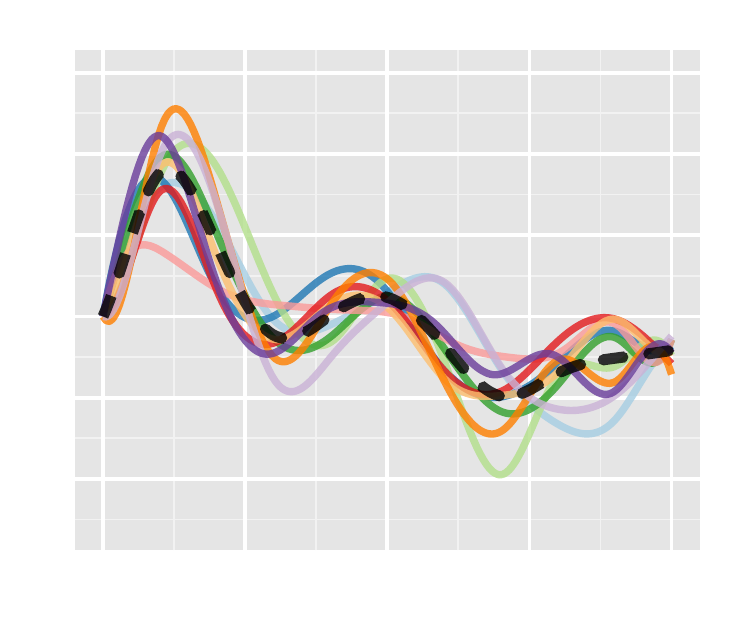}} &
\mbox{\hspace{-1.0em}\includegraphics[scale = 0.6]{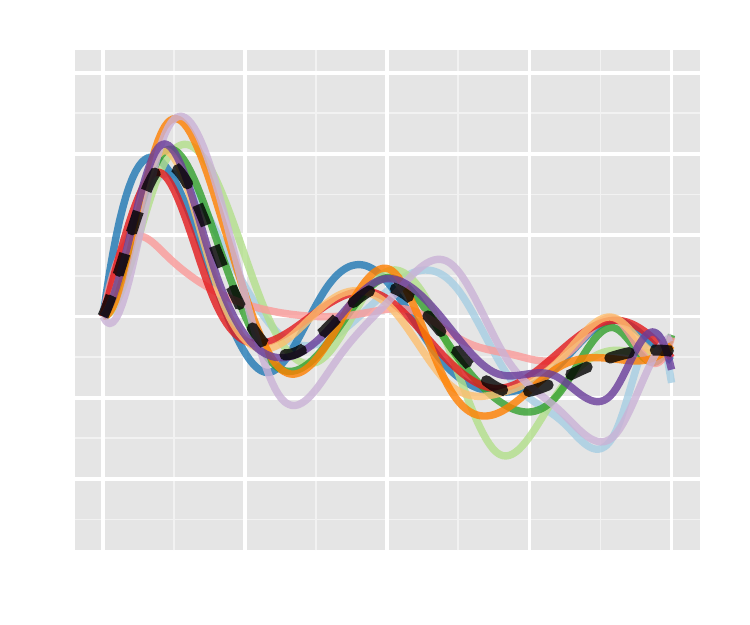}}\\[-1.5em]
\vspace{-10em}\rotatebox{90}{\scriptsize 30.0 cm} & \mbox{\hspace{-1.0em}\includegraphics[scale = 0.6]{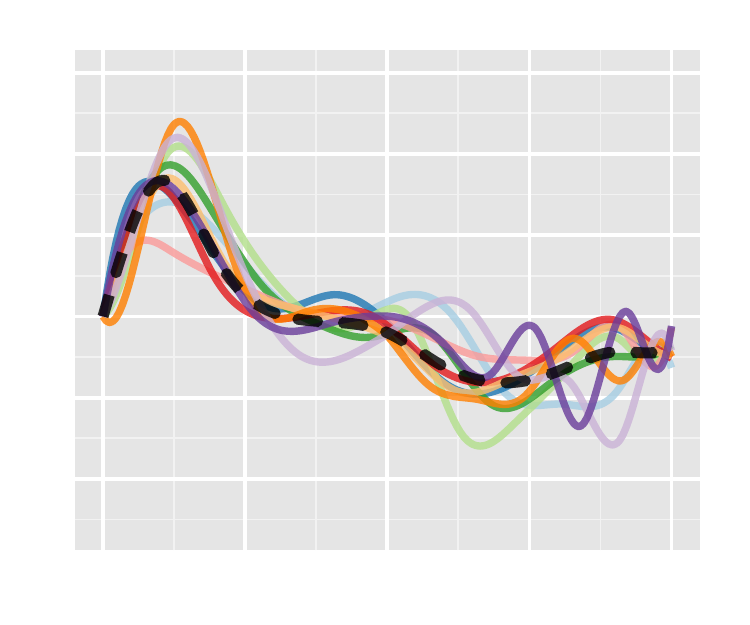}} &
\mbox{\hspace{-1.0em}\includegraphics[scale = 0.6]{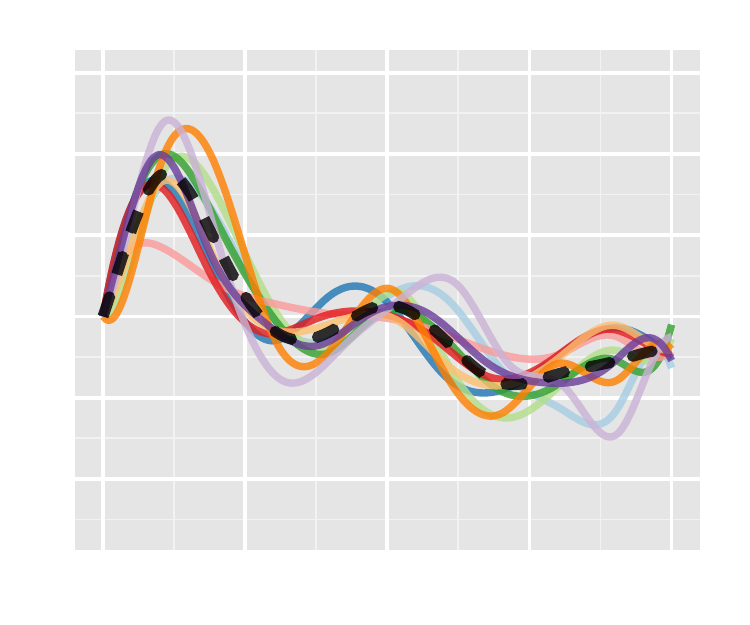}} &
\mbox{\hspace{-1.0em}\includegraphics[scale = 0.6]{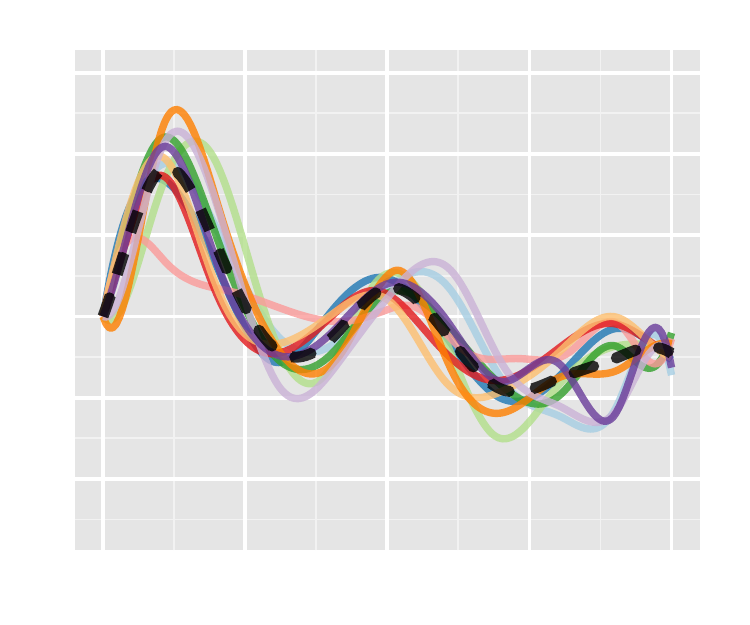}}\\[-1.5em]
\vspace{-10em}\rotatebox{90}{\scriptsize 37.5 cm} & \mbox{\hspace{-1.0em}\includegraphics[scale = 0.6]{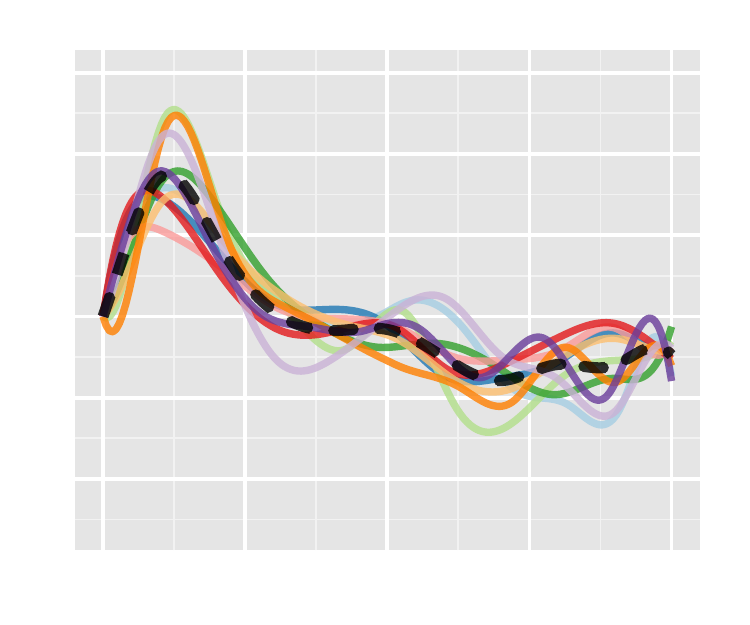}} &
\mbox{\hspace{-1.0em}\includegraphics[scale = 0.6]{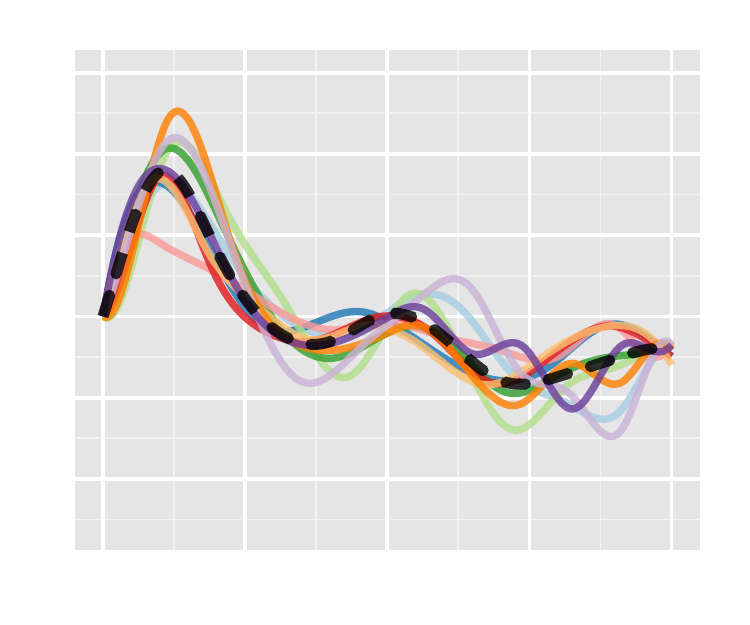}} &
\mbox{\hspace{-1.0em}\includegraphics[scale = 0.6]{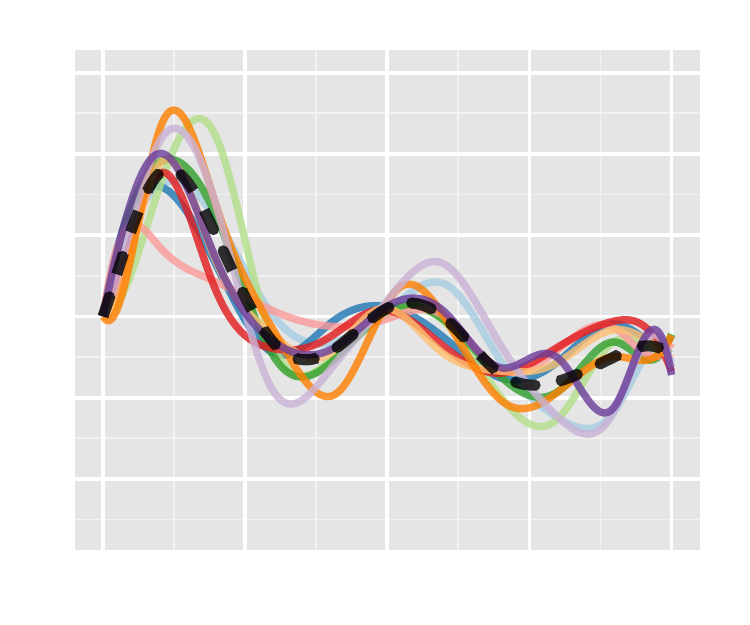}}\\[-1.5em]
\vspace{-10em}\rotatebox{90}{\scriptsize 45.0 cm} & \mbox{\hspace{-1.0em}\includegraphics[scale = 0.6]{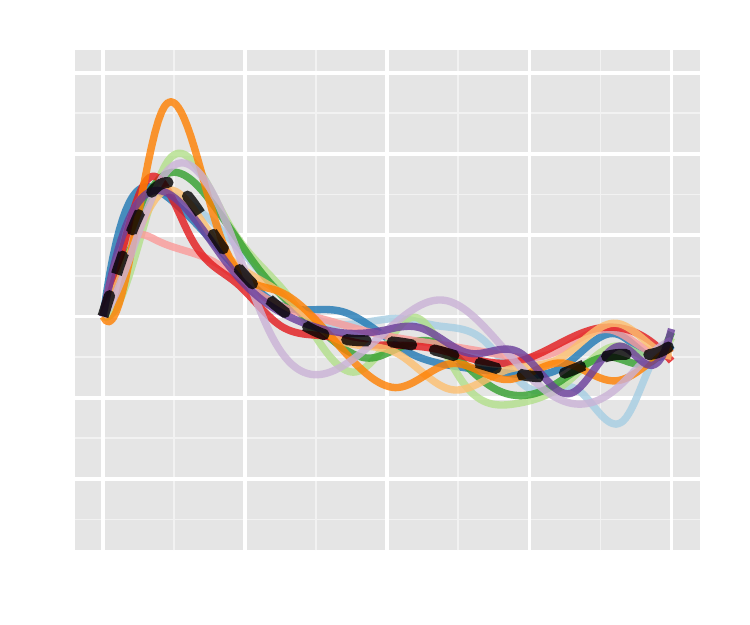}} &
\mbox{\hspace{-1.0em}\includegraphics[scale = 0.6]{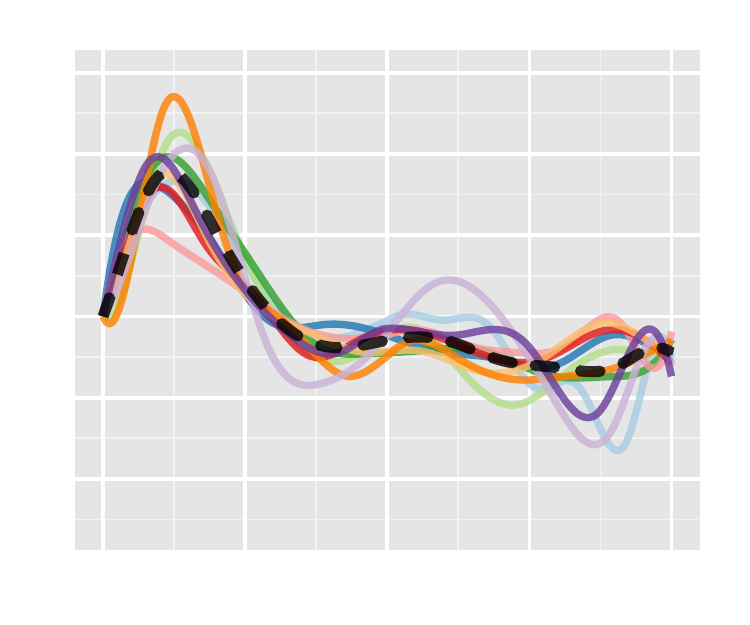}} &
\mbox{\hspace{-1.0em}\includegraphics[scale = 0.6]{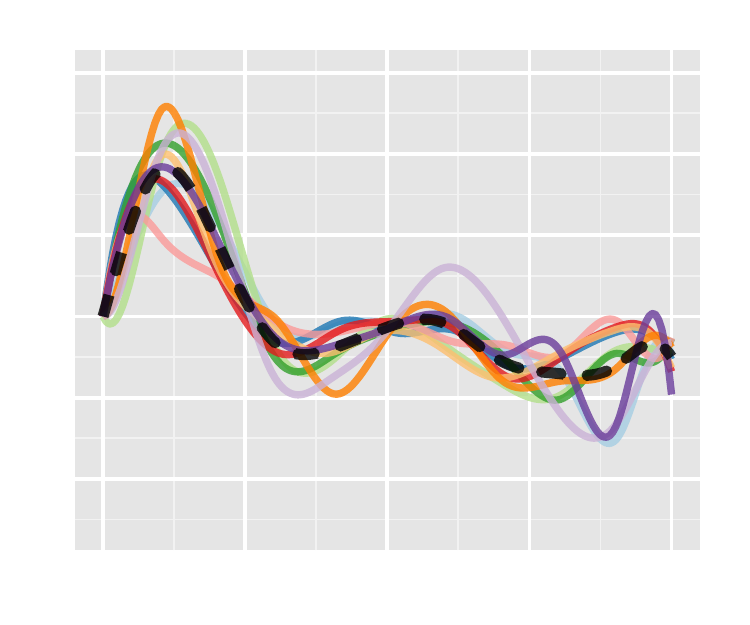}}\\[-1em]
\end{tabular}
\vspace{0.5em}

\centering \hspace{1.4em}\includegraphics[scale = 0.7]{img/person.pdf}
\caption{Estimated fixed effects $(\theta + \varphi_i)\circ \nu_i$ in the 15 obstacle avoidance experiments using percentual time. The dashed trajectory shows the estimate for $\theta$. The average percentual warped time between two white vertical bars corresponds to 0.2 seconds.}\label{fig:fixed}
\end{figure}

The number of basis functions $K$, the number of warping anchor points $n_w$, and the regularization parameter $\lambda$ were determined by the average 5-fold cross-validation score on each of three experimental conditions ($d=30$ cm and obstacle heights \emph{S}, \emph{M}, and \emph{T}). The models were fitted using the method described in the previous section, and the quality of the models was evaluated through the accuracy of classifying the participant from a given movement in the test set, using posterior distance between the sample and the combined estimates for the fixed effects $(\theta + \varphi_i)\circ \nu_i$. The cross-validation was done over a grid of the following values $\mu, \mu_{\text{p}}\in \{0.5, 1, 2\}$, $K \in \{8, 13, 18, 23, 28, 33\}$, $K_{\text{p}} \in \{8, 9, \dots, 18\}$, $n_w, n_{w\text{p}}\in \{0, 1, 2, 3, 5, 10\}$, $\lambda, \lambda_{\text{p}} \in \{0, 1, 2, 3\}$. The best values were found to be $\mu = 2$, $K = 23$, $n_w = 2$, $\lambda = 2$ and $\mu_{\text{p}}=1$ $K_{\text{p}} = 12$, $n_{w\text{p}} = 1$, $\lambda_{\text{p}} = 0$. {We note that the smoothness parameter $\mu = 2$ is on the boundary of the cross-validation grid. The qualitative difference between second order smoothness $\mu=2$ (corresponding to twice differentiable sample paths of amplitude effects) and higher order is so small, however, that we chose to ignore higher order smoothness.} Furthermore, $\lambda_{\text{p}}=0$ indicates that we do not need to penalize participant-specific amplitude effects when working with percentual time. The reason is most likely that the samples have better initial alignment in percentual time. The estimated participant-specific acceleration profiles using percentual time can be seen in Figure~\ref{fig:fixed}. 

{A simulation study that validates the method and implementation on data simulated using the maximum likelihood estimates of the central experimental condition ($d=30.0$ cm, medium obstacle) is available in Supporting Information.}

\section*{Results}

\subsection*{Identification of individual differences}

A first assessment of the strength of the statistical model~\eqref{model} is to examine the extent to which the model captures individual differences. Proper modeling of systematic individual differences is not only of scientific interest per se, but also provides perspective for interpreting any observed experimental effects. To validate the capability of the model to capture systematic individual differences, we use the model to identify an individual from the estimated individual templates.  Such identification of individuals is becoming increasingly relevant also in a practical sense with the recent technological advances in motion tracking systems, and the growing array of digital sensors in handheld consumer electronics. 
Consistent with the framing of model~\eqref{model}, we perform identification of individual participants on the basis of the data from a single experimental condition. This is, in a sense, a conservative approach. Combining data across the different conditions of the experimental tasks would likely provide more discriminative power given that personal movement styles tend to be reproducible. 


The classification of the movement data is based on the characteristic acceleration profile computed for each participant. For this to work it is important that individual movement differences are not smoothed away.  The hyperparameters of the model were chosen with this requirement in mind. In the following, we describe alternative methods we considered. For all approaches, the stated parameters have been chosen by 5-fold cross-validation on the experimental conditions with obstacle distance $d=30.0$ cm. The  grids used for cross-validation are given in Supporting Information section. Recall that subscript $\mathrm{p}$ indicates the use of percentuall time. 

 \begin{description}
 \item[Nearest Participant (NP)] {NP classification classifies using the minimum combined pointwise $L^2$ distance to all samples for every individual in the training set. }

 \item[Modified Band Median (MBM)] MBM classification estimates templates using the modified band median proposed in \cite{arribas2012robust}, which under mild conditions is a consistent estimator of the underlying fixed amplitude effects warped according to the modified band medians of the warping functions. Classification is done using $L^2$ distance to the estimated templates. In the computations we count the number of bands defined by $J=4$ curves \cite[Section 2.2]{arribas2012robust}.  MBM$_{\text{p}}$ used $J_{\text{p}}=2$.
 \item[Robust Manifold Embedding (RME)] RME classification estimates templates using the robust manifold embedding algorithm proposed in \cite{dimeglio2014robust}, which, assuming that data lies on a low-dimensional smooth manifold, approximates the geodesic distance and computes the empirical Fr\'echet median function.  Classification is done using $L^2$ distance to the estimated templates. 
  \item[Dynamic Time Warping (DTW)] {DTW classification estimates templates by iteratively time warping samples to the current estimated personal template (5 iterations per template) using an asymmetric step pattern (slopes between $0$ and $2$). The template is modeled by a B-spline with $33$ degrees of freedom. DTW$_{\text{p}}$ used $16$ degrees of freedom.}
 \item[Fisher-Rao (FR)] { FR classification estimates templates as Karcher means under the Fisher-Rao Riemannian metric \cite{Kurtek} of the data represented using a single principal component \cite{fdasrvf}. \v{C}encov's theorem states that the Fisher-Rao distance is the only distance that is preserved under warping \cite{cencov2000statistical}, and in practice the distance is computed by using a dynamic time warping algorithm on the square-root slope functions of the data.  Classification is done using $L^2$ distance to the estimated templates. }
  
 \item[Elastic Fisher-Rao (FR$_{\text{E}}$)] FR$_{\text{E}}$ classification estimates templates analogously to FR, but classifies using the weighted sum of elastic amplitude and phase distances \cite[Definition 1 and Section 3.1]{tucker2013generative}. The phase distance was weighted by a factor 1.5. FR$_{\text{Ep}}$ uses two principal components and a phase distance weight of 1. 
 
 \item[Timing and Motion Separation (TMS)] The proposed TMS classification estimates templates of the fixed effects $(\theta + \varphi_i)\circ \nu_i$ using Algorithm~\ref{alg}. Classification is done using least distance measured in the negative log posterior \eqref{posterior} as a function of the test samples. The parameters were set as described in the previous section.
 \end{description}

We evaluate classification accuracy using 5-fold cross-validation, which means that eight samples are available in the training set for every participant. The folds of the cross-validation are chosen chronologically, such that the first fold contains replications 1 and 2, the second contains 3 and 4 and so on. The results are available in Table~\ref{table:combined}. We see that TMS and TMS$_{\text{p}}$ achieve the highest classification rates, followed by FR$_{\text{Ep}}$, FR$_{\text{p}}$  and RME$_{\text{p}}$. Thus, the model enables identification of individual movement style. 
Furthermore, we note that there is little effect of using percentual time for the proposed method, which for all other methods gives a considerable boost in accuracy. This suggest that the TMS methods align data well without the initial linear warping and the endpoint constraints of percentual time.

\begin{table}[!ht]
\centering
\resizebox{\textwidth}{!}{
\begin{tabular}{cc|cccccccccccccc}
\hline
$d$& obstacle & NP & NP$_{\text{p}}$ & MBM & MBM$_{\text{p}}$ & RME & RME$_{\text{p}}$ & DTW & DTW$_{\text{p}}$ & FR & FR$_{\text{p}}$ & FR$_{\text{E}}$ & FR$_{\text{Ep}}$ &
TMS & TMS$_{\text{p}}$\\
\hline\hline
& \emph{S} & 0.36 &  0.48 & 0.53 & 0.43 & 0.55 & 0.57 & 0.52 & 0.52 & 0.47 & 0.54 & 0.62 & 0.51  & 0.70 & \textbf{0.76}\\
15.0 cm & \emph{M} & 0.36 & 0.46 & 0.38 & 0.45& 0.41 & 0.43 & 0.49 & 0.56 & 0.36 & 0.49  & 0.47 & 0.46  & \textbf{0.69} & 0.66\\
& \emph{T} & 0.41 & 0.47 & 0.41& 0.46 & 0.49 & 0.50 & 0.43 & 0.43 & 0.32 & 0.56 & 0.49 & 0.49  & \textbf{0.64} & 0.62\\
\hline
& \emph{S} & 0.36 & 0.49 & 0.34& 0.46& 0.37 & 0.50 & 0.45 & 0.44 & 0.44 & 0.51  & 0.50 & 0.51  & \textbf{0.70} & 0.68\\
22.5 cm & \emph{M} & 0.38 & 0.44 & 0.42 &0.53& 0.46 & 0.55 & 0.38 & 0.42 & 0.32 & 0.45 & 0.42 & 0.55  & 0.62 & \textbf{0.74}\\
& \emph{T} & 0.36 & 0.49 & 0.45 &0.54& 0.46 & 0.57 & 0.40 & 0.53 &  0.48 & 0.57 & 0.53 & \textbf{0.64}  & 0.61 & \textbf{0.64}\\
\hline
& \emph{S} & 0.27 & 0.29 & \emph{0.37} & \emph{0.47} & 0.41 & 0.45 & \emph{0.43} & \emph{0.44} & \emph{0.40} & \emph{0.43} & \emph{0.46} & \emph{0.55} & \emph{{0.63}} & \emph{\textbf{0.65}}\\
30.0 cm  & \emph{M} & 0.30 & 0.42 & \emph{0.38}& \emph{0.46} & 0.40 & 0.45 & \emph{0.34} & \emph{0.49} & \emph{0.36} & \emph{0.48} & \emph{0.46} & \emph{0.47} & \emph{\textbf{0.65}} & \emph{\textbf{0.65}}\\
& \emph{T} & 0.37 & 0.44 & \emph{0.42} & \emph{0.50} & {0.50} & 0.45  & \emph{0.44} & \emph{0.44} & \emph{0.37} & \emph{0.50} & \emph{0.39} &\emph{0.43}&  \emph{\textbf{0.74}} &\emph{{0.69}}\\
\hline
& \emph{S} & 0.28 & 0.45 & 0.41& 0.49  & 0.42 & 0.51 & 0.45 & 0.50 & 0.36  & 0.51 & 0.39 & 0.56 & 0.69 & \textbf{0.74}\\
37.5 cm  & \emph{M} & 0.26 & 0.33 & 0.33 & 0.37& 0.35 & 0.41& 0.40 & 0.49 & 0.35 & 0.37 & 0.32  & 0.53 & 0.57 & \textbf{0.62}\\
& \emph{T} & 0.31 & 0.43 & 0.38 & 0.43 & 0.40 & 0.46 & 0.37 & 0.29 & 0.50 & 0.48 & 0.49  & 0.55 & 0.63 & \textbf{0.65} \\
\hline
& \emph{S} & 0.25 & 0.38 & 0.33 & 0.45 & 0.32 & 0.42 & 0.34 & 0.51 & 0.32 & 0.45 & 0.37 &0.41 & \textbf{0.68} & 0.65 \\
45.0 cm  & \emph{M} & 0.29 & 0.31 & 0.29  & 0.38& 0.38 & 0.39 & 0.43 & 0.43 & 0.36 & 0.48 & 0.38 &0.45  & 0.53 & \textbf{0.57}\\
& \emph{T} & 0.29 & 0.39 & 0.45 & 0.48 & 0.48 & 0.57 & 0.38  & 0.45 & 0.39 & 0.44 & 0.44 &0.47  & \textbf{0.66} & 0.58\\
\hline\hline
average & & 0.323 & 0.418  & 0.393 & 0.460& 0.427 & 0.482 & 0.417 & 0.463 & 0.387 & 0.484 & 0.449 & 0.487  &  0.649 & \bf0.660\\
\hline
\end{tabular} 
}
\caption{Classification accuracies of various methods. {\bf Bold} indicates best result(s), \emph{italic} indicates that the given experiments were used for training.}\label{table:combined}
\end{table}

\subsection*{Factor analysis of spatial movement paths}


In the previous section, the proposed modeling framework was shown to give unequalled accuracy of modeling the time series data for acceleration. In this section we use the warping functions obtained to analyze the spatial movement paths and their dependence on task conditions. Temporal alignment of the spatial positions along the path for different repetitions and participants is necessary to avoid spurious spatial variance of the paths. The natural alignment of two movement paths  is the one that matches their acceleration signatures. In other words, spatial positions along the paths at which similar accelerations are experienced should correspond to the same times.  Thus, each individual spatial trajectory was aligned using the  time warping predicted from the TMS$_{\text{p}}$-results of the previous section. Every sample path was represented by 30 equidistant sample points in time at which values were obtained by  fitting a three-dimensional B-splines with 10 equidistantly spaced knots to each trajectory. 

As an exemplary study, we analyze how spatial paths depend on obstacle height.  We do this separately for each distance, so that we perform five separate analyses, one for each obstacle placement. In each analysis we have 10 participants with 10 repetitions for each of 3 obstacle heights.  The three-dimensional spatial positions along the movement path, $\boldsymbol{y}_{ijh}\in \R^{30\times 3}$, depend on participant $i=1, \dots, 10$, repetition $j=1,\dots, 10$ and height $h = 1,2, 3$.

We are interested in understanding how the space-time structure of movement captured the 30 by 3 dimensions of  the trajectories, $\boldsymbol{y}_{ijh}$, varies when obstacle height is varied.  We the seek to find a low-dimensional affine subspace of the space-time representation of the movements within which movements vary, once properly aligned. That subspace provides a  low-dimensional model of movement paths on the basis of which we can analyze the data. 

We identify the low-dimensional subspace based on a novel factor analysis model. In analogy to principal component analysis (PCA), 
 $q$ so-called \emph{loadings} are estimated that represent dominant patterns of variation along movement trajectories. In contrast to PCA, the factor analysis model does not only model the residual-variance of independent paths around the mean, but also allows one to include covariates from the experimental design, for example by taking the repetition structure of participants and systematic effects of obstacle height into account. {In other words, the proposed factor analysis model is a generalization of PCA suitable for addressing the question at hand while obeying the study design.}  

The idea is to use the mean movement trajectory, $\boldsymbol{\theta}\in \R^{30\times 3}$, of one condition, the lowest obstacle height, as a reference. The movement trajectories $\boldsymbol{y}_{ijh}$ (30 time steps and 3 cartesian coordinates; participant $i$;  repetition $j$, and experimental condition with height $h$) are then represented through their deviation from the reference path. We estimate the hypothesized low-dimensional affine subspace in which these deviations lie. That subspace is spanned by the $q$ orthonormal $(30\times 3)$-dimensional columns of the loadings matrix $W\in \R^{(30\times 3)\times q}$. We assume a mixed-effect structure on the weights for the loadings that takes into account both the categorical effect of obstacle-height and random effects of participant and repetition. This amounts to a statistical model
\begin{align}
\boldsymbol{y}_{ijh} = \boldsymbol{\theta} + (X_{h}\boldsymbol{\beta} + \sum_{l=1}^3 Z_{i, g_l(j, h), l})W^\top + \boldsymbol{\varepsilon}_{ij},\label{eq:factor_model}
\end{align}
 where $X_h\in \R^{1\times 2}$ represents the covariate design that indicates obstacle heights: \emph{S}: $X_1 = (0, 0)$; \emph{M}: $X_2 = (1, 0)$; and \emph{T}: $X_3 = (0, 1)$. The parameters, $\boldsymbol\beta \in \R^{2\times q}$, are the weights for the loadings that account for the systematic deviation of obstacle heights, \emph{M} and \emph{T} from the reference height, \emph{S}. $g_l$ is the factor that describes the $l$th level random effects design (participant, participants' reaction to obstacle-height change, and  repetition). $Z_{i, g_l(j, h), l}\in \R^{1\times q}$ are independent latent Gaussian variables with zero-mean and a covariance structure modeled with three $q\times q$ covariance matrices, each describing the covariance between loadings within a level of the random-effect design.  $\boldsymbol{\varepsilon}_{ij}\in \R^{30\times 3}$ is zero-mean Gaussian noise with diagonal covariance matrix $\Lambda$ with one variance parameter per dimension. 
 
 { The loading matrix $W$ is identifiable in a similar way as for usual PCA. Firstly, the scaling of $W$ is identified by the assumption that $W^\top W$ is the $q$-dimensional identity matrix. Secondly, the rotation of $W$ is identified by the assumption that the total variance of the latent variables for a single curve
\begin{equation*}
\sum_{l=1}^3 \text{Var}(Z_{i,g_l(j,h),l}),
\end{equation*}
is a diagonal matrix. This identifies the loading matrix $W$ with probability 1.}


The models were fitted using maximum likelihood estimation by using an ECM algorithm \cite{meng1993maximum} that had been accelerated using the SQUAREM method \cite{varadhan2008simple}.

\begin{figure}[!thp]
\centering
\includegraphics[width = 1.0\textwidth]{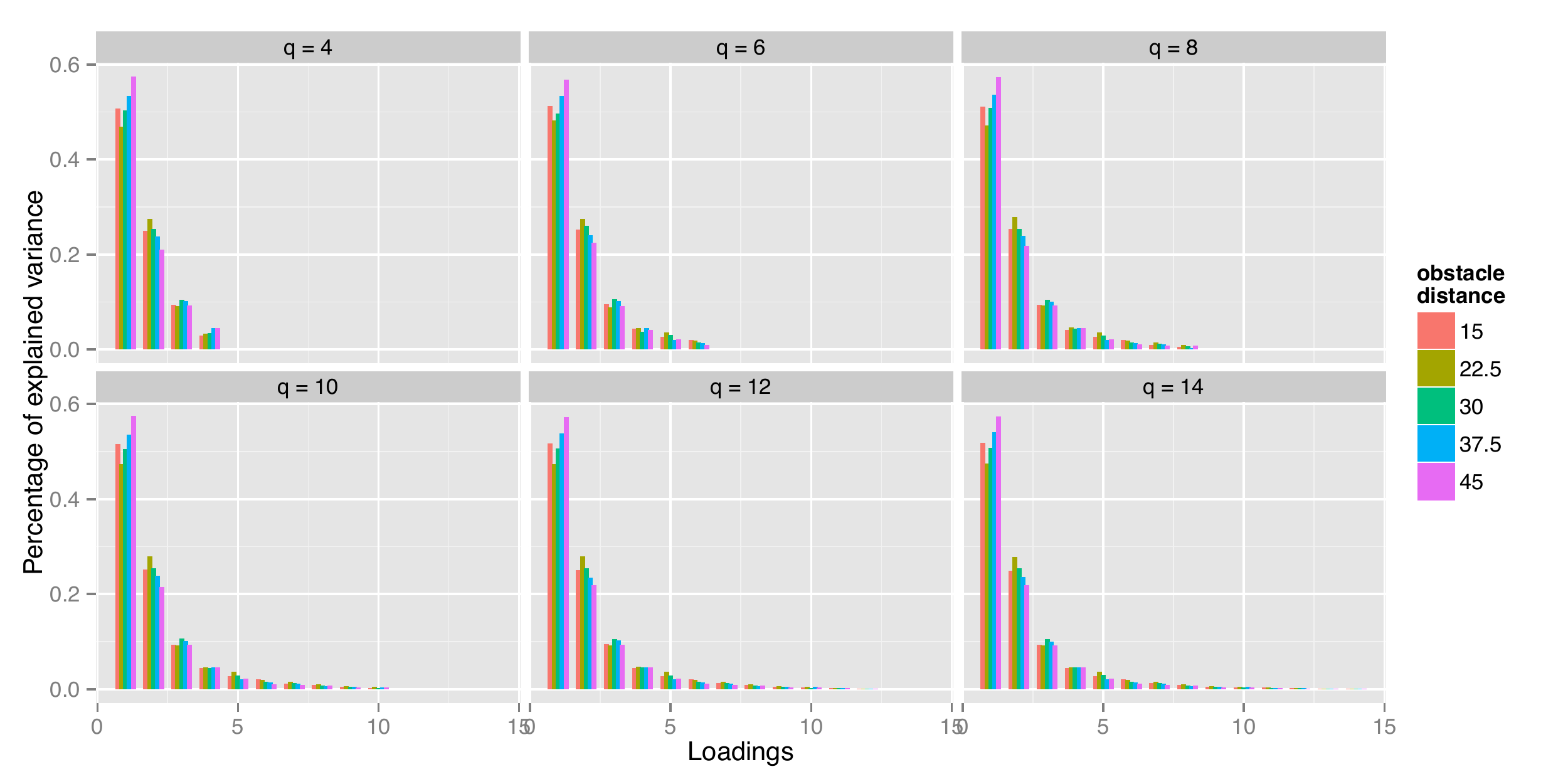}
\caption{Percentage of variance explained by individual loadings under different total number of loadings $q$ and obstacle distance.} \label{fig:scree}
\end{figure}

Figure~\ref{fig:scree} shows the percentage of explained variance for various values of $q$ across different obstacle distances.  For $q > 8$, the average percentage of variance explained by the ninth loading ranges from 1\% to 2\% and the combined percentage of variance explained by the loadings beyond number eight remains under 3\%. From this perspective, $q=8$ seems like a reasonable choice, with the loadings explaining 97.1\% of the variance, meaning that the error term $\boldsymbol{\varepsilon}_{ij}$ should account for the remaining 2.9\%. In the following, all results are based on the model with $q=8$.

\begin{figure}[!thp]
\centering
\includegraphics[width = 0.45\textwidth, trim = 100 80 100 50, clip]{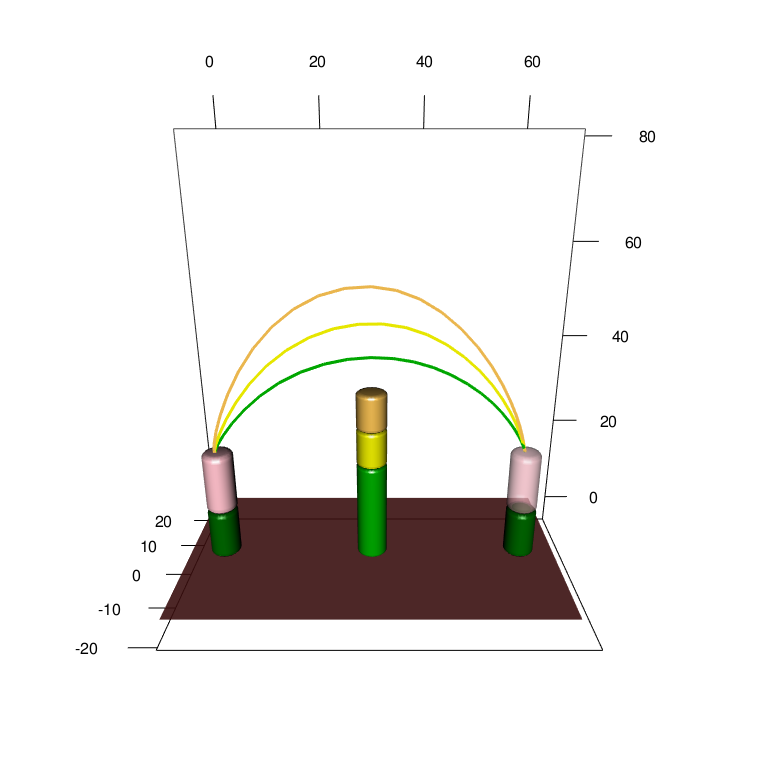}
\includegraphics[width = 0.45\textwidth, trim = 30 210 90 210, clip]{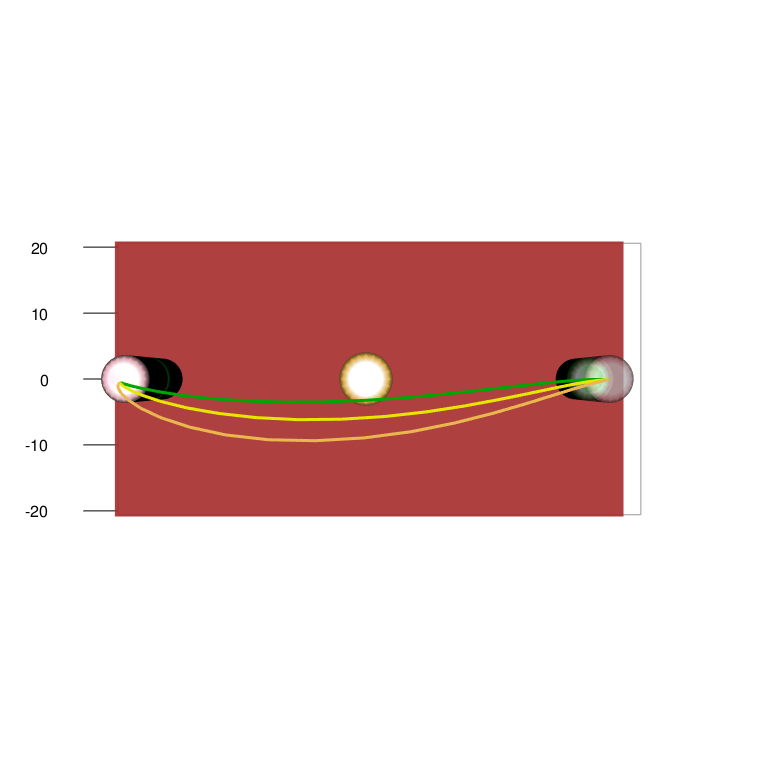}
\caption{Mean paths for the three obstacle heights (green: small; yellow: middle, orange: tall) at obstacle distance $d = 30.0$ cm. } \label{dist30mean}
\end{figure}

\paragraph{Modeling obstacle height}
The fitted mean trajectories for the three different obstacle heights at distance $d = 30.0$ cm can be found in Figure~\ref{dist30mean}. A striking feature of the mean paths is the apparent linear scaling of elevation, but also of the lateral excursion with height. (The difference in the frontal plane, not shown, was very small, but follows a similar pattern.) This leads to the hypothesis that  the scaling of the mean trajectory with obstacle height can be described by a one-parameter regression model in height increase ($X_1=0$, $X_2=7.5$, $X_3=15$) rather than a more generic two-parameter ANOVA model. 

We fitted both models for every obstacle distance and performed likelihood-ratio tests. The p-values can be found in Table~\ref{table:p_val}. They were obtained by evaluating twice the difference in log likelihood for the two models using a $\chi^2$-distribution at $q=8$ degrees of freedom. We see that no p-values are significant, so there is no significant loss in the descriptive power of the linear scaling model compared to the ANOVA model. The remaining results in the paper are all based on the model with a regression design.

\begin{table}[!thp]
\caption{p-values for the hypothesis of linearly amplified path changes in obstacle height increase factor.}\label{table:p_val}
\centering
\begin{tabular}{c|ccccc}
\hline
Obstacle distance & $15.0$ cm &  $22.5$ cm &  $30.0$ cm &  $37.5$ cm &  $45.0$ cm\\
\hline\hline
p-value & 0.478 & 0.573 & 0.093 & 0.764 & 0.362\\
\hline
\end{tabular}
\end{table}

\begin{figure}[!pth]
\centering
\includegraphics[width = 0.3\textwidth , trim = 100 80 100 100, clip]{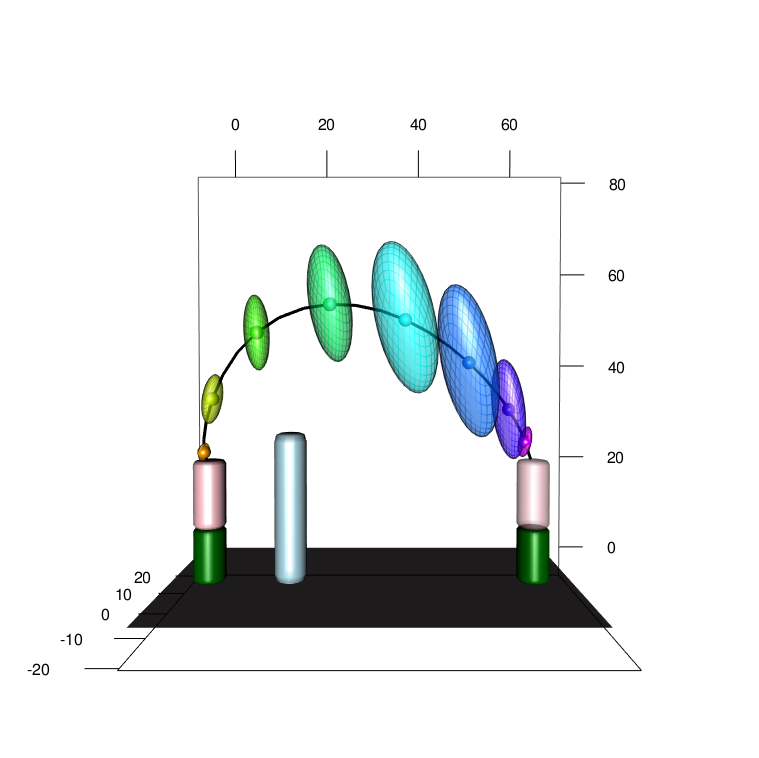}
\includegraphics[width = 0.3\textwidth , trim = 100 80 100 100, clip]{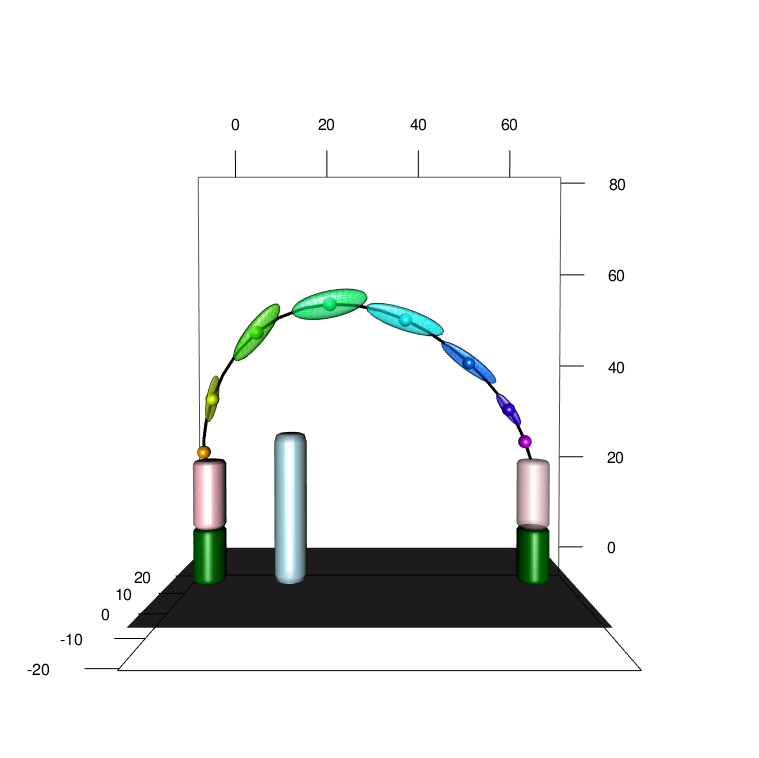}
\includegraphics[width = 0.3\textwidth , trim = 100 80 100 100, clip]{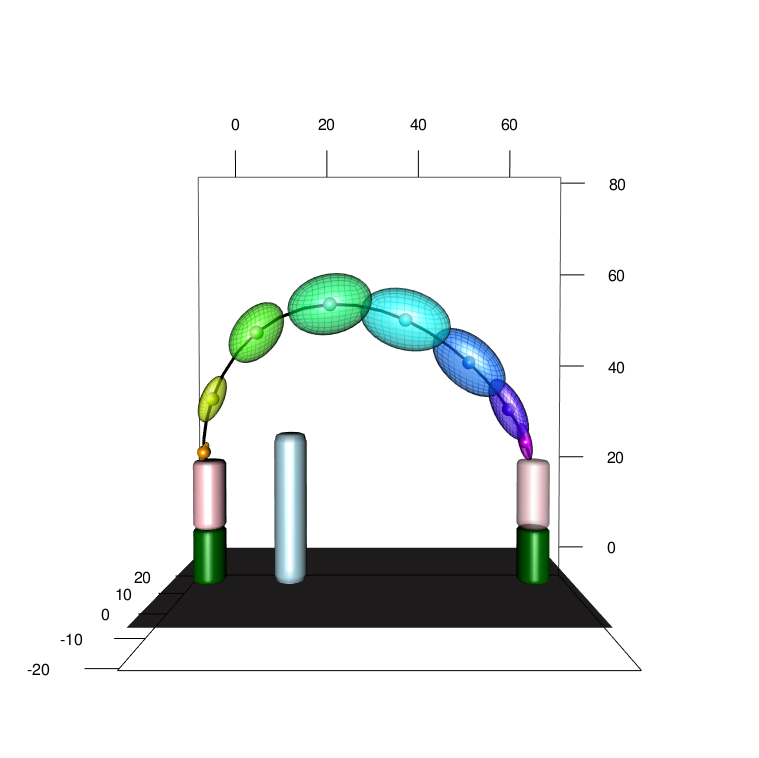}
\includegraphics[width = 0.3\textwidth , trim = 100 80 100 100, clip]{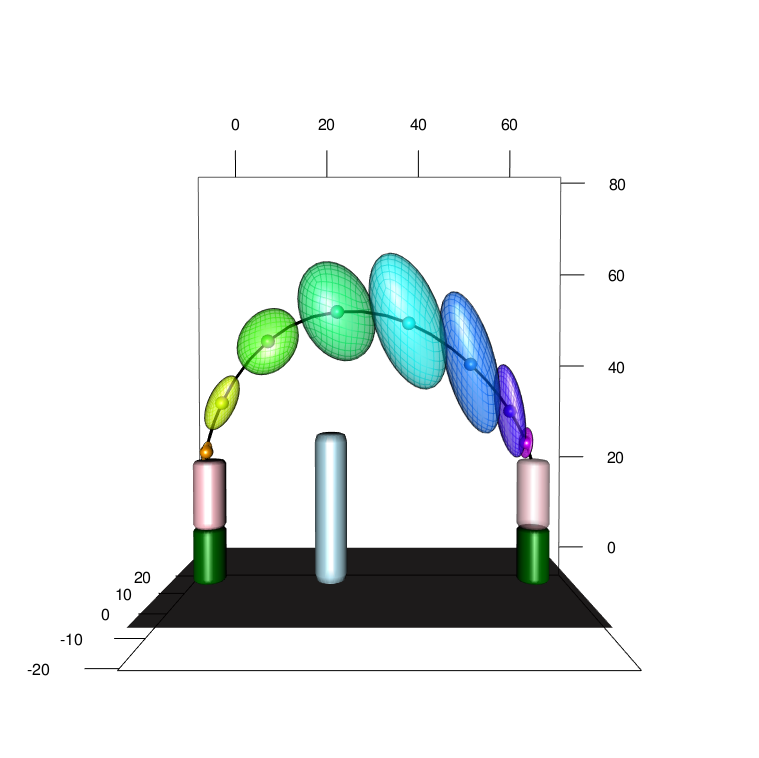}
\includegraphics[width = 0.3\textwidth , trim = 100 80 100 100, clip]{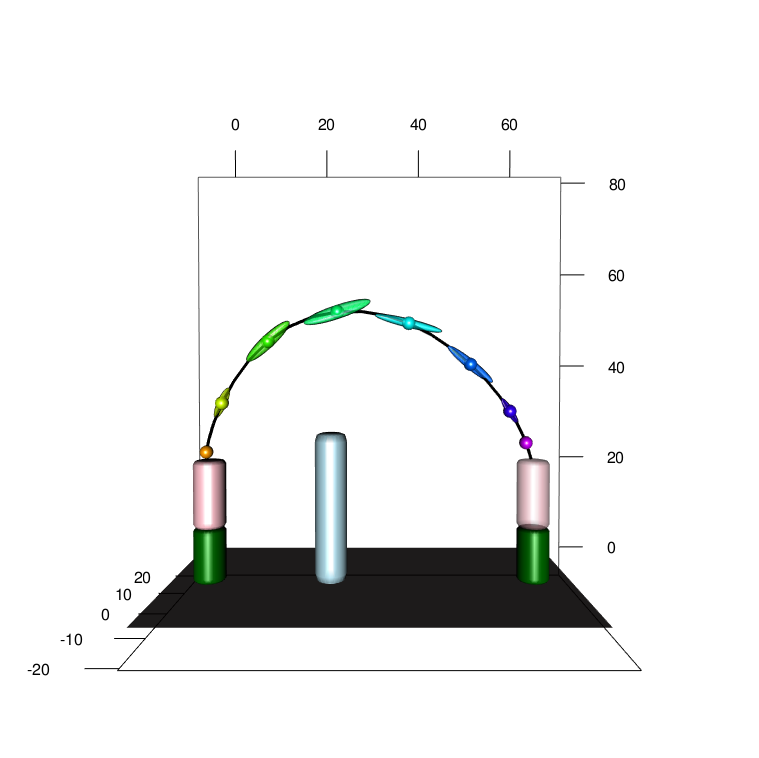}
\includegraphics[width = 0.3\textwidth , trim = 100 80 100 100, clip]{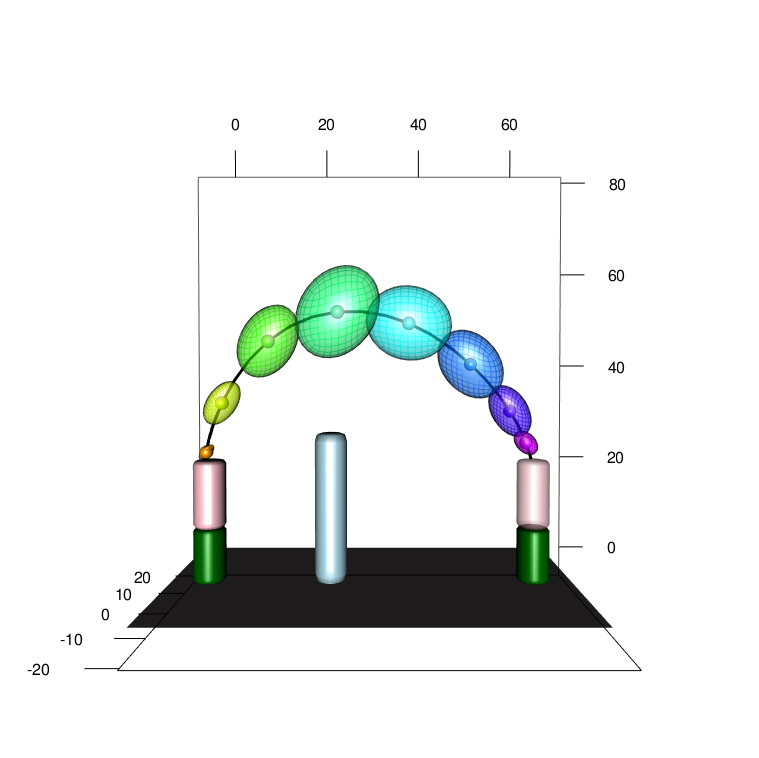}
\includegraphics[width = 0.3\textwidth , trim = 100 80 100 100, clip]{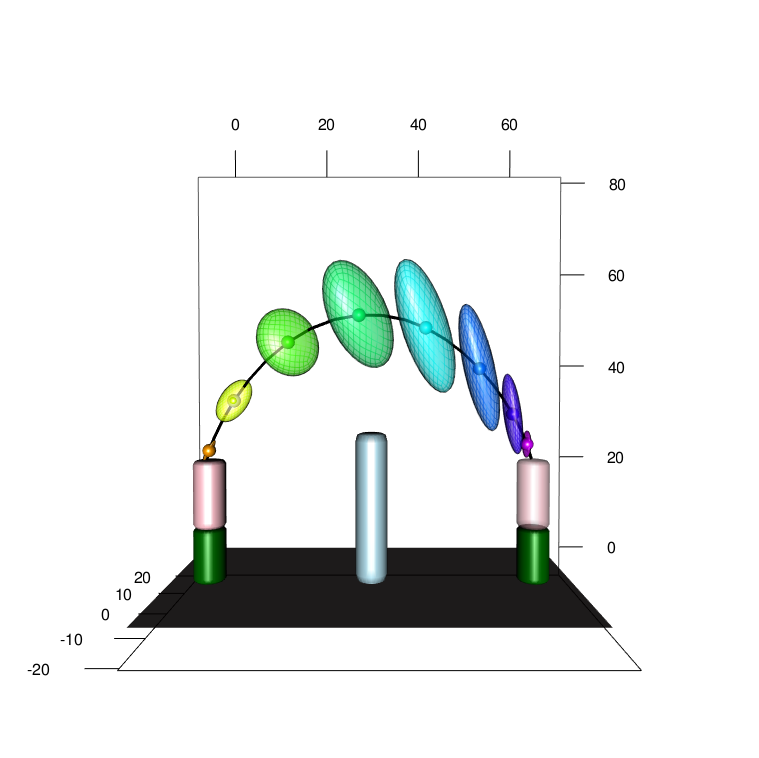}
\includegraphics[width = 0.3\textwidth , trim = 100 80 100 100, clip]{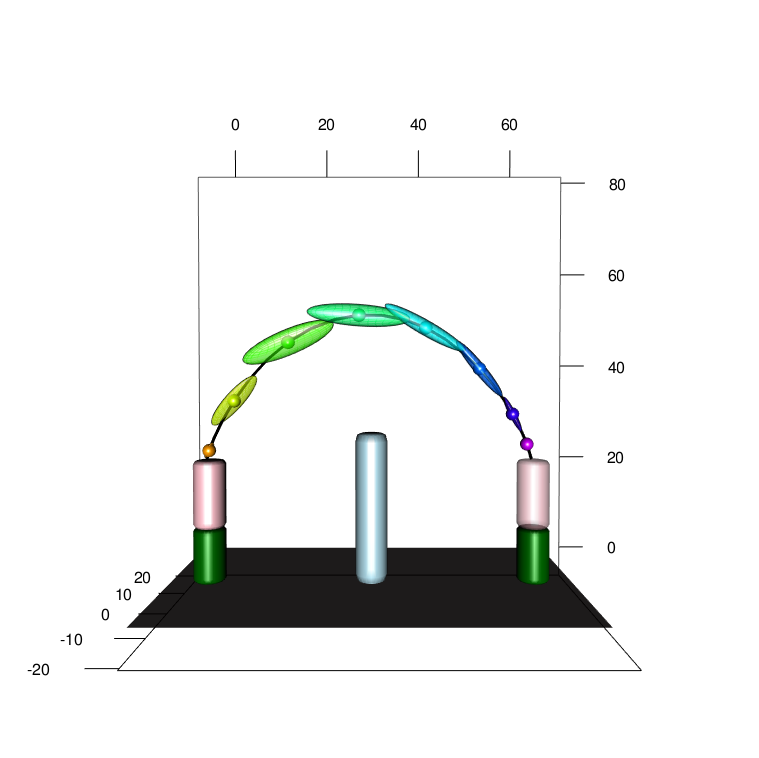}
\includegraphics[width = 0.3\textwidth , trim = 100 80 100 100, clip]{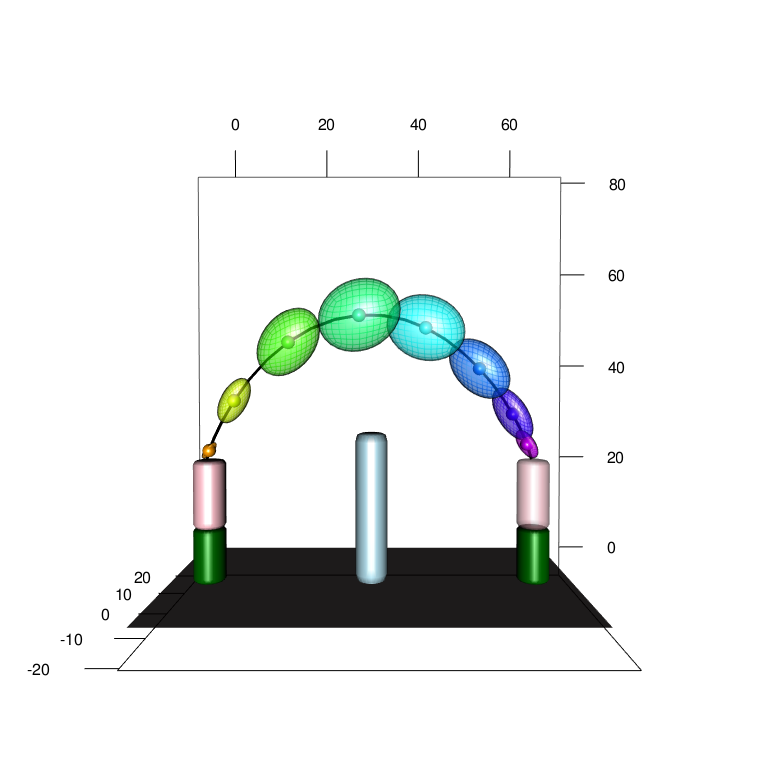}
\includegraphics[width = 0.3\textwidth , trim = 100 80 100 100, clip]{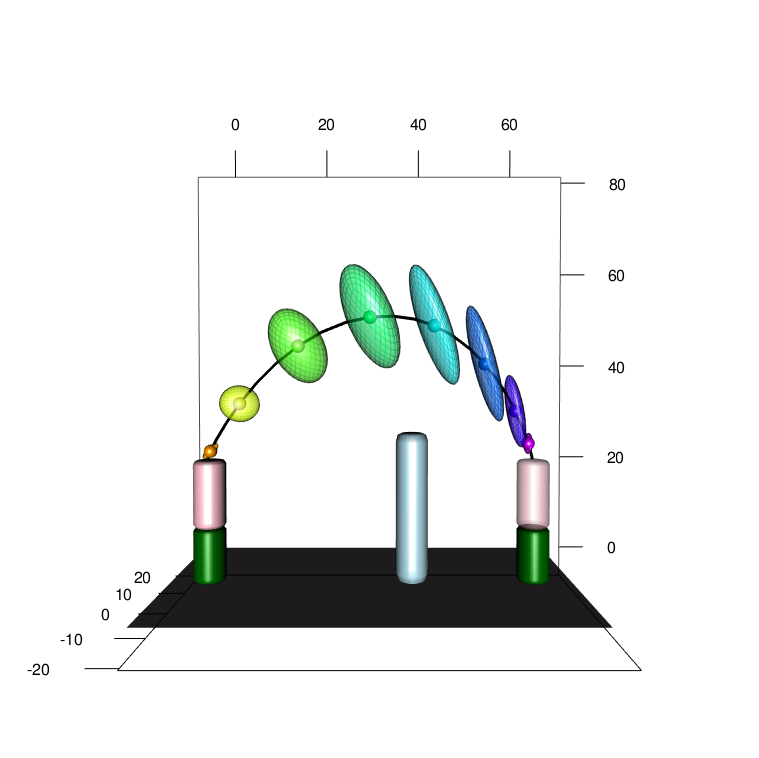}
\includegraphics[width = 0.3\textwidth , trim = 100 80 100 100, clip]{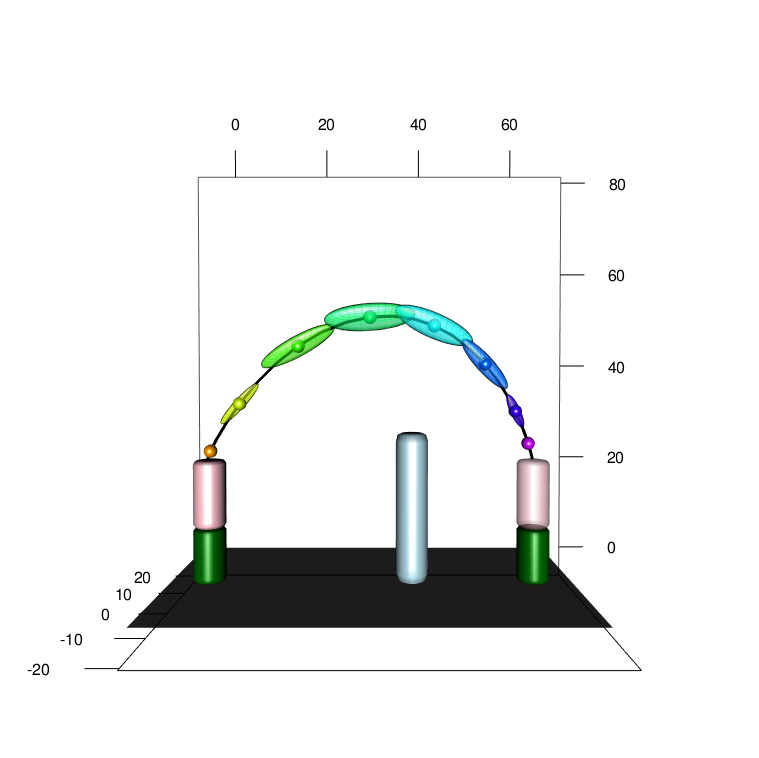}
\includegraphics[width = 0.3\textwidth , trim = 100 80 100 100, clip]{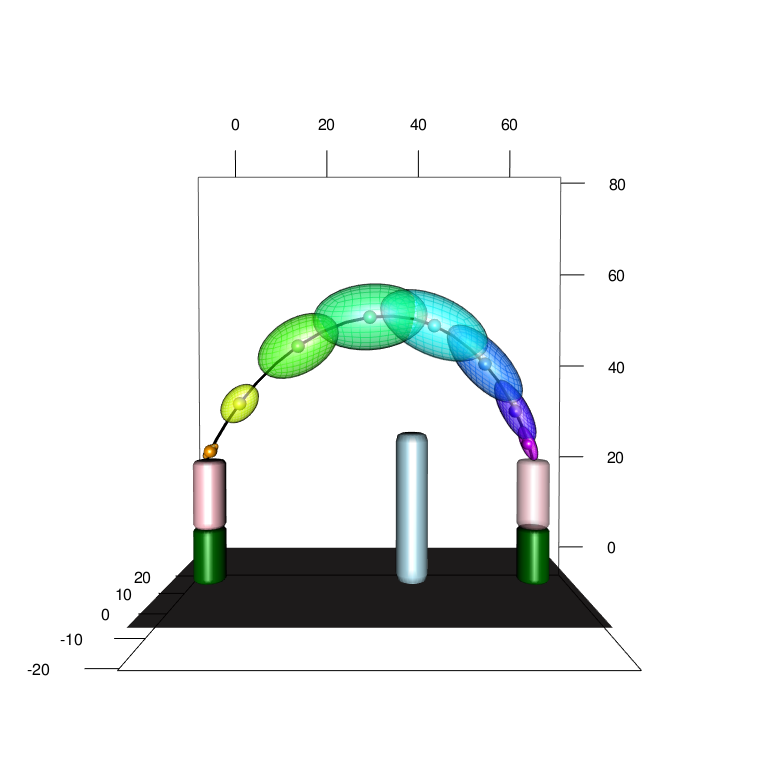}
\includegraphics[width = 0.3\textwidth , trim = 100 80 100 100, clip]{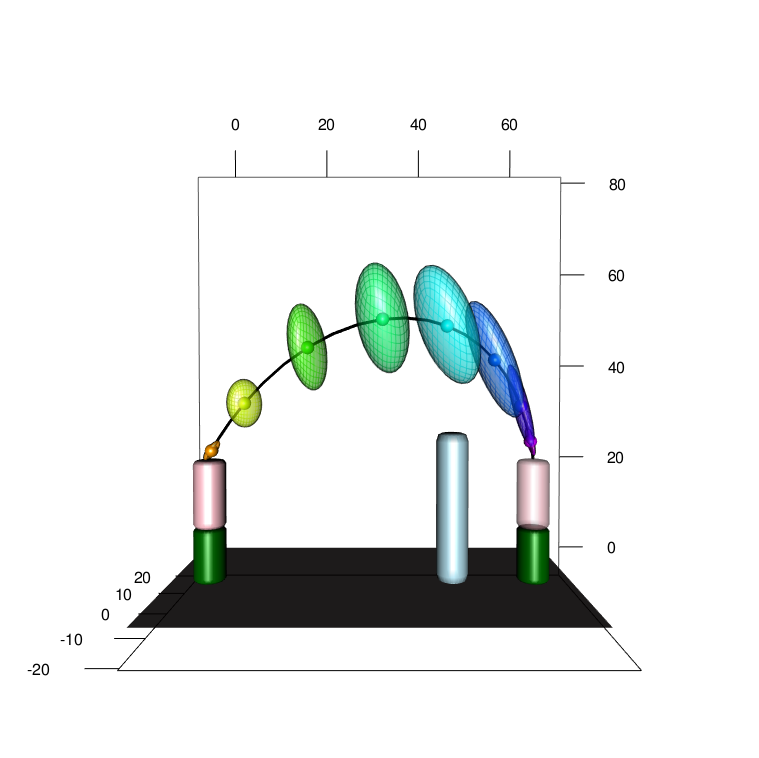}
\includegraphics[width = 0.3\textwidth , trim = 100 80 100 100, clip]{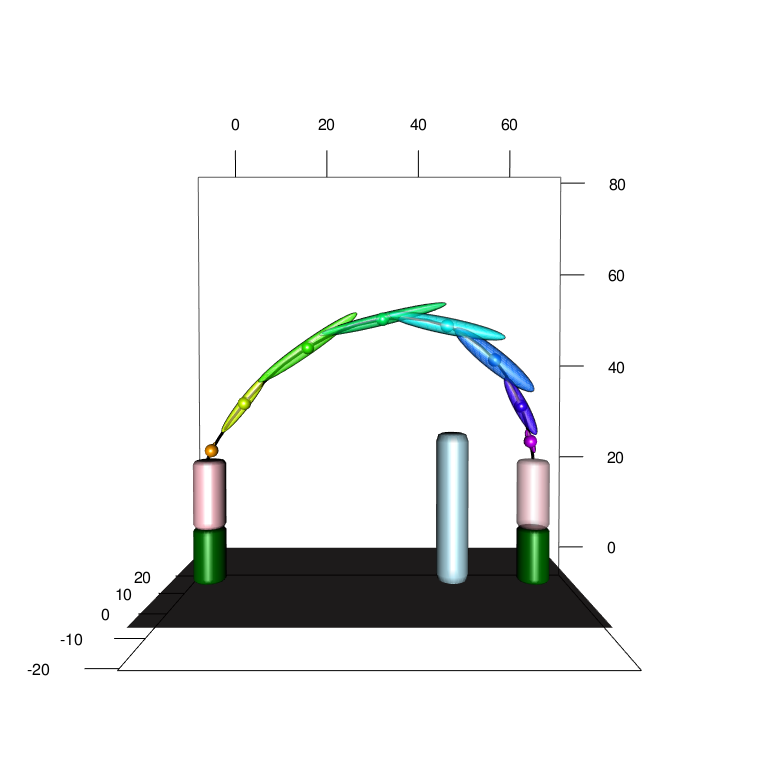}
\includegraphics[width = 0.3\textwidth , trim = 100 80 100 100, clip]{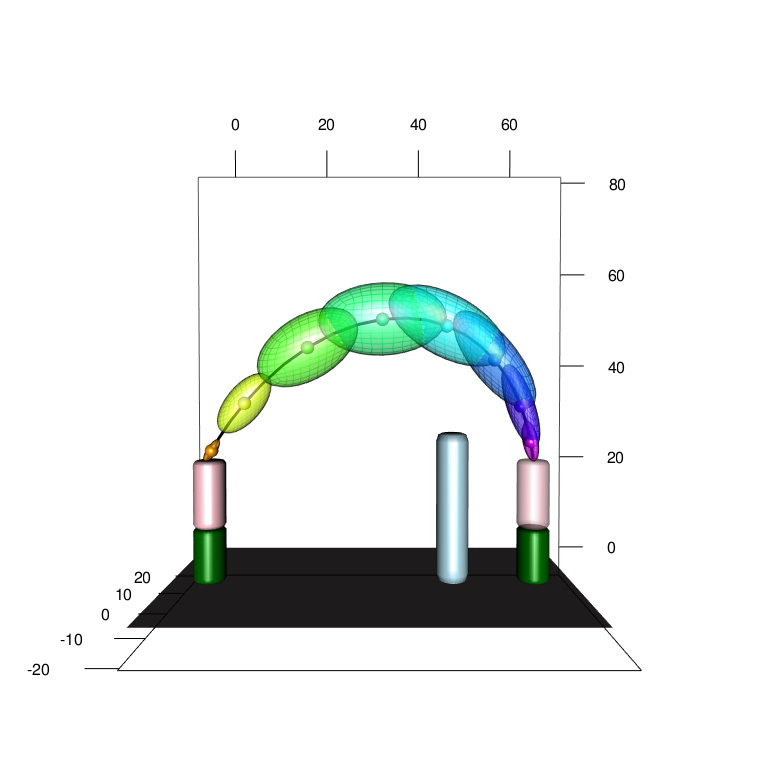}
\caption{Illustration of the experimental setup for the medium height obstacle at all obstacle distances (rows) with the mean trajectory plotted. Along the trajectory eight equidistant points (in percentual warped time) are marked, and at each point  95\% prediction ellipsoids are drawn. The three columns represent the random effects tied to participant, subjective reaction to height change and repetition, respectively.}\label{fig:variation1}
\end{figure}

\begin{figure}[!pth]
\centering
\includegraphics[width = 0.3\textwidth , trim = 30 210 90 210, clip]{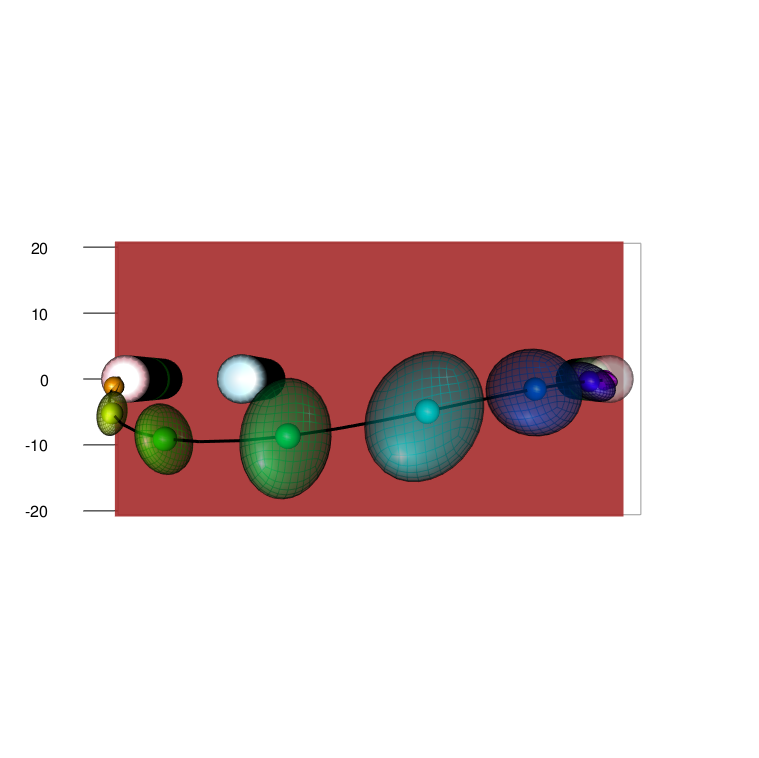}
\includegraphics[width = 0.3\textwidth , trim = 30 210 90 210, clip]{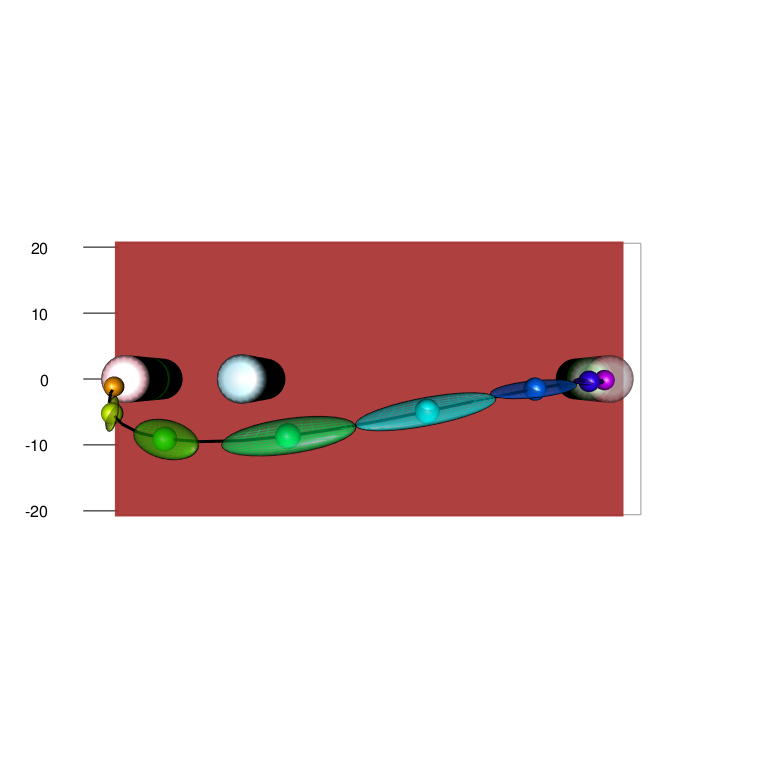}
\includegraphics[width = 0.3\textwidth , trim = 30 210 90 210, clip]{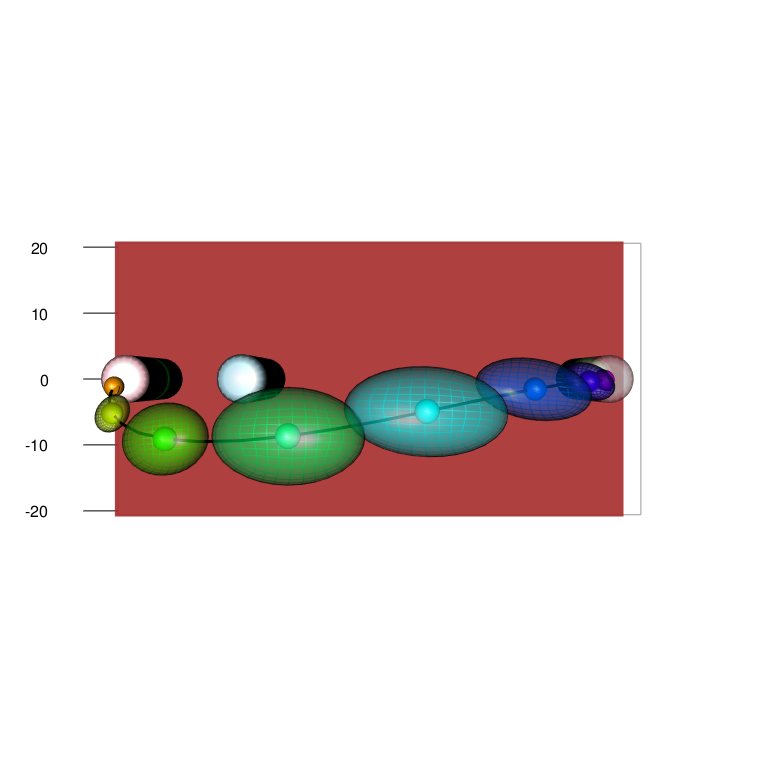}
\includegraphics[width = 0.3\textwidth , trim = 30 210 90 210, clip]{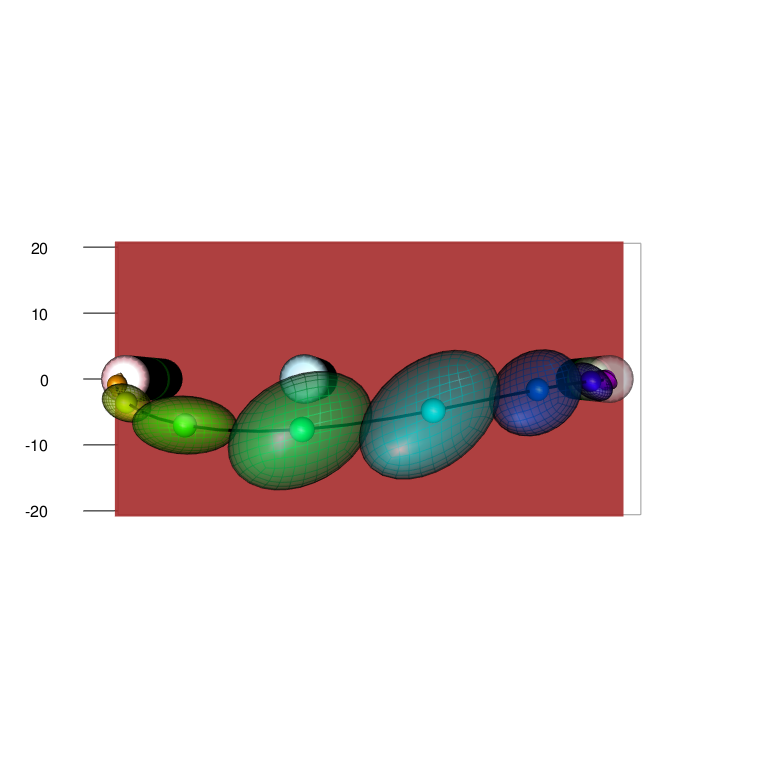}
\includegraphics[width = 0.3\textwidth , trim = 30 210 90 210, clip]{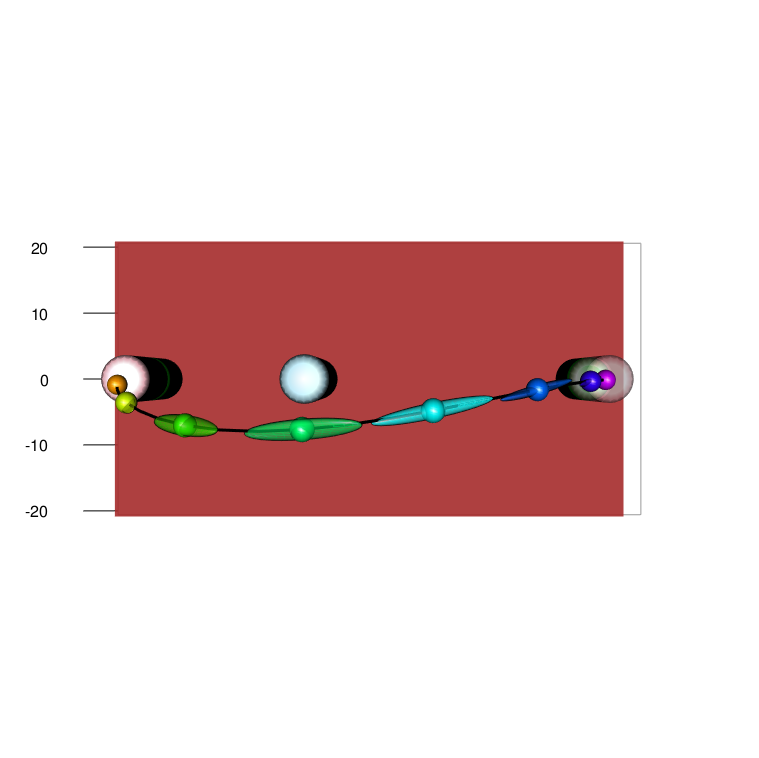}
\includegraphics[width = 0.3\textwidth , trim = 30 210 90 210, clip]{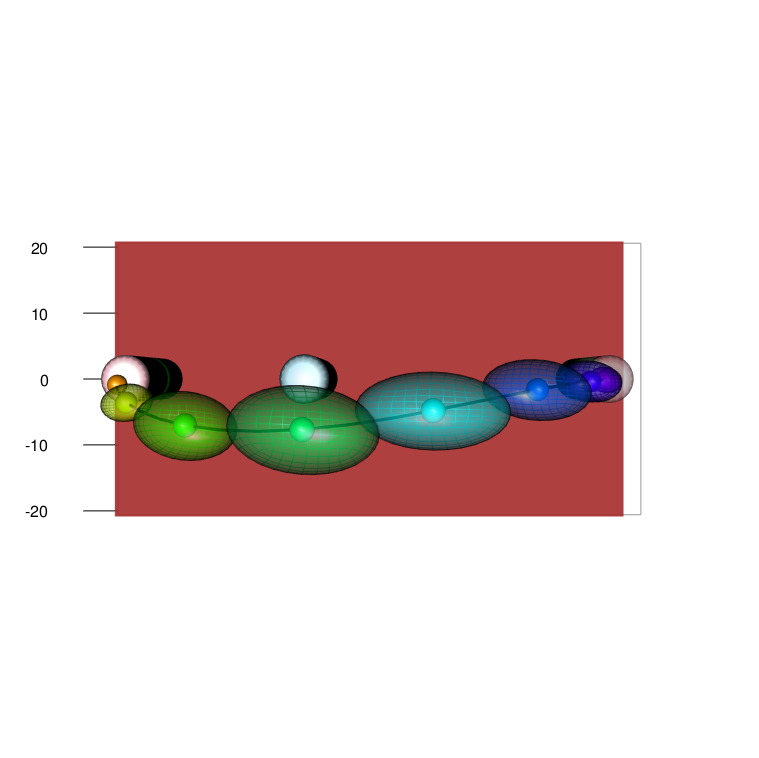}
\includegraphics[width = 0.3\textwidth , trim = 30 210 90 210, clip]{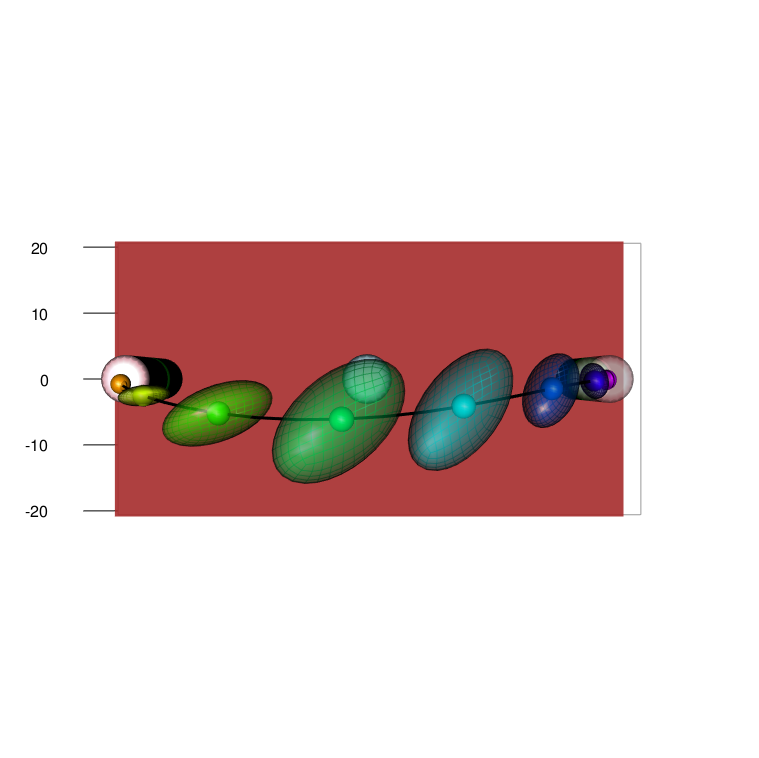}
\includegraphics[width = 0.3\textwidth , trim = 30 210 90 210, clip]{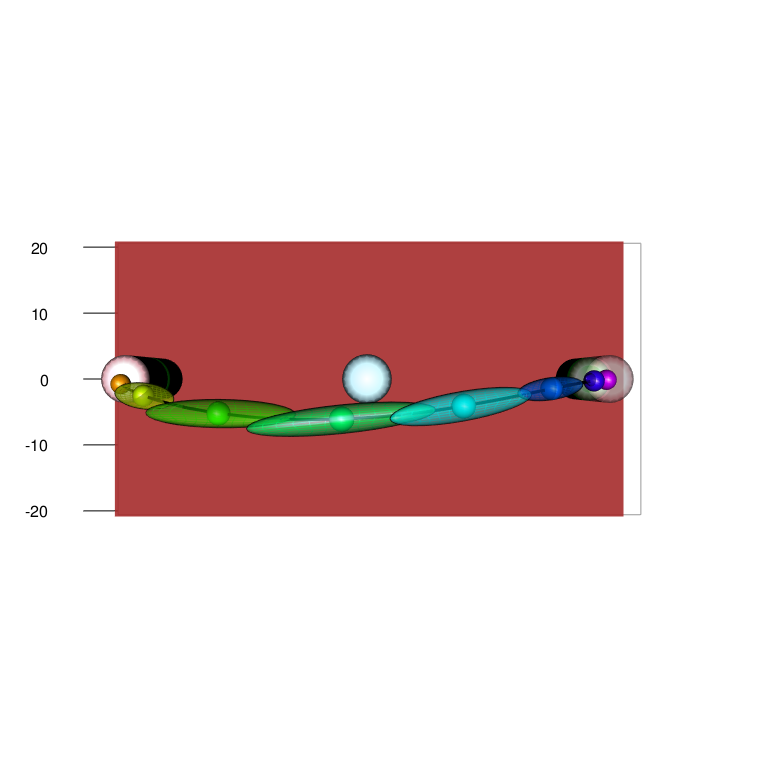}
\includegraphics[width = 0.3\textwidth , trim = 30 210 90 210, clip]{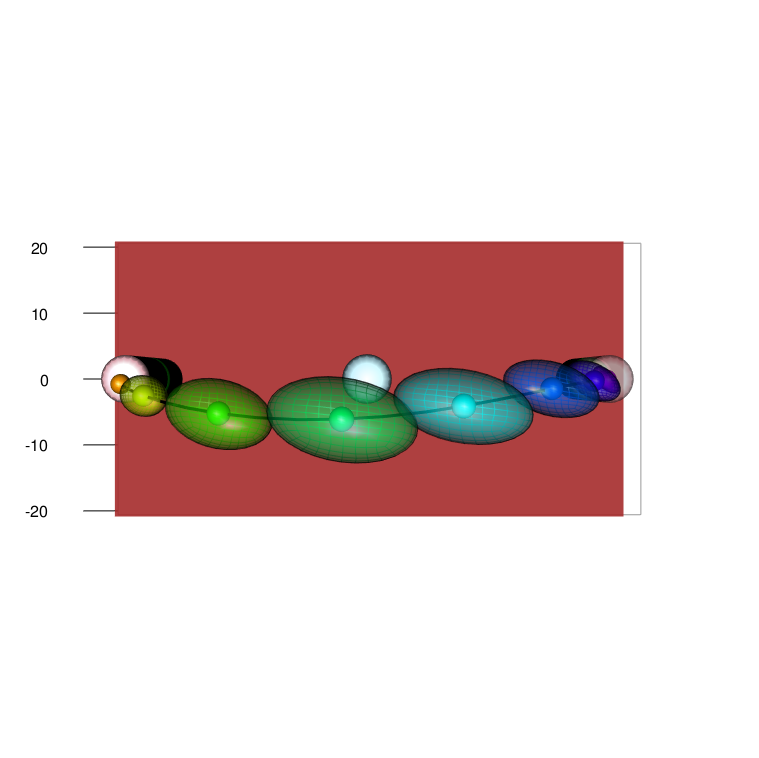}
\includegraphics[width = 0.3\textwidth , trim = 30 210 90 210, clip]{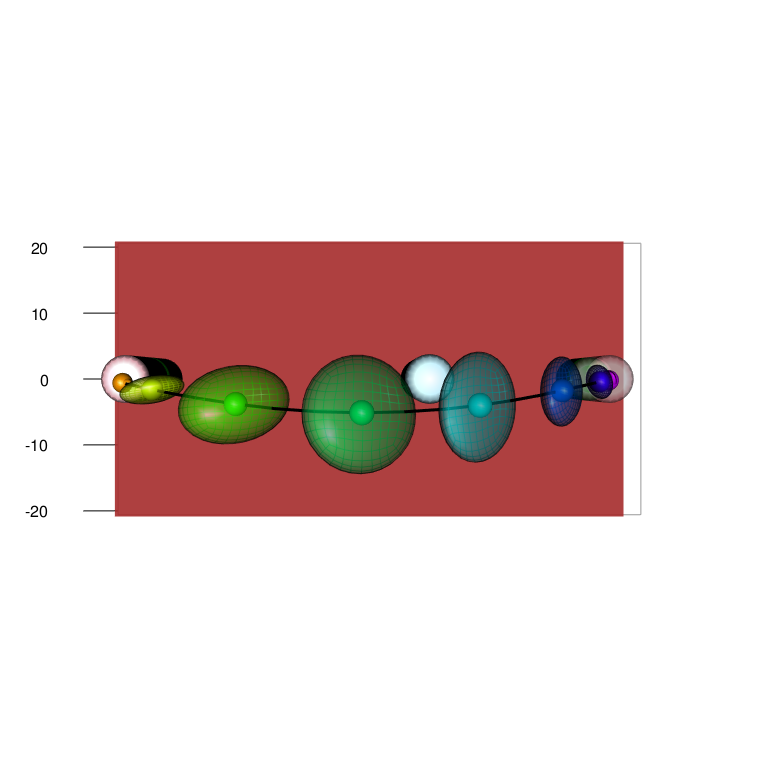}
\includegraphics[width = 0.3\textwidth , trim = 30 210 90 210, clip]{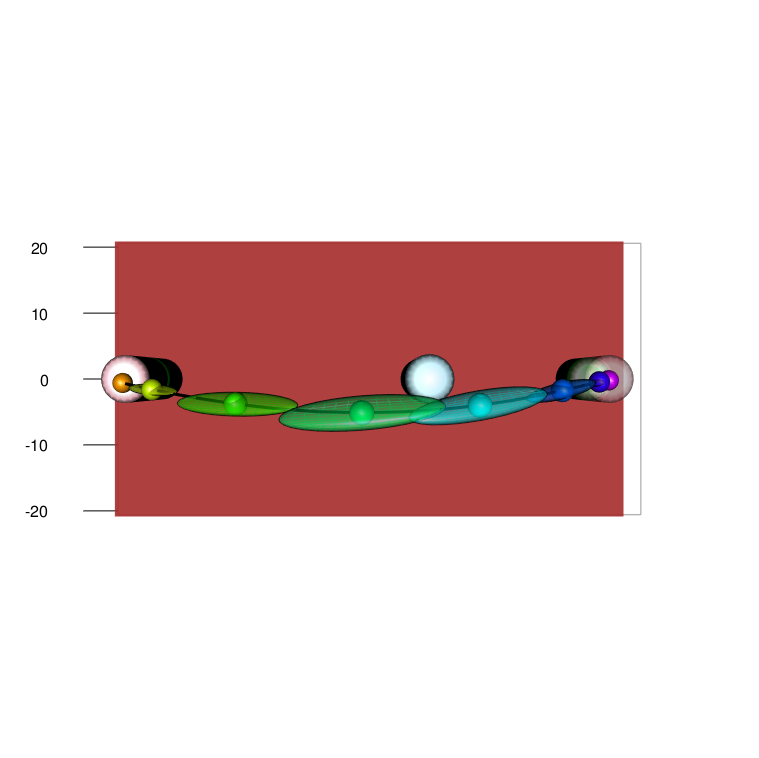}
\includegraphics[width = 0.3\textwidth , trim = 30 210 90 210, clip]{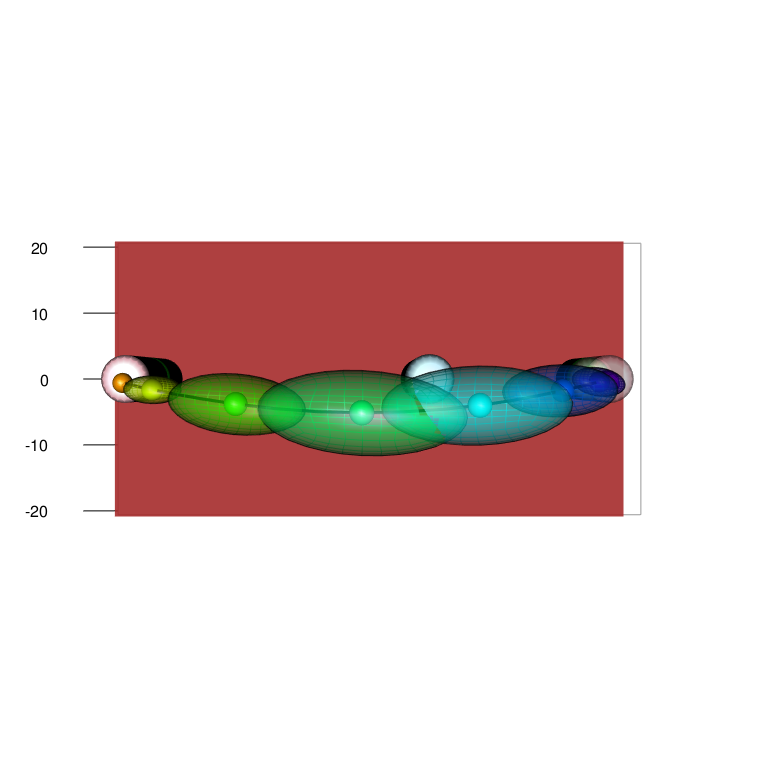}
\includegraphics[width = 0.3\textwidth , trim = 30 210 90 210, clip]{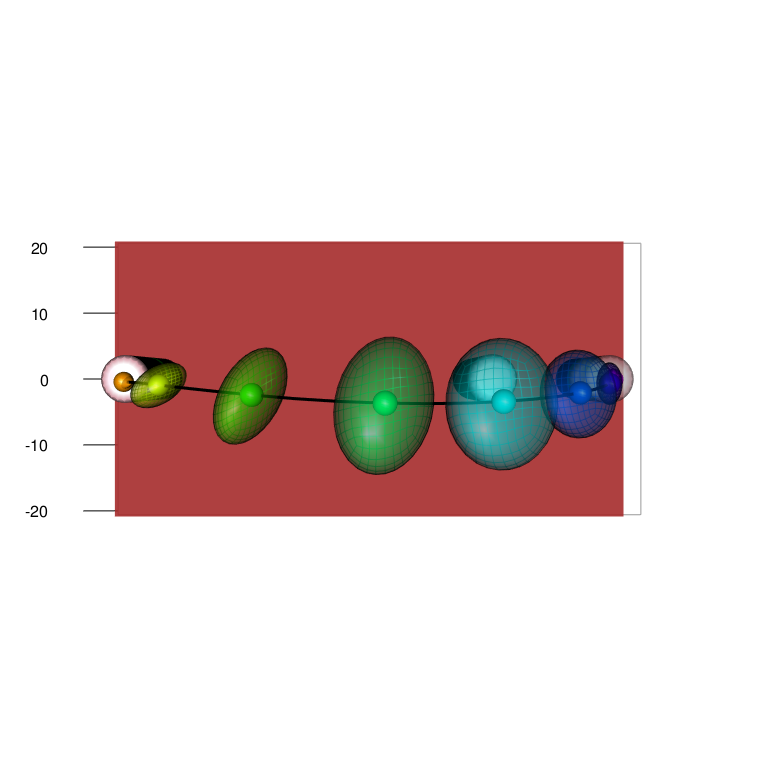}
\includegraphics[width = 0.3\textwidth , trim = 30 210 90 210, clip]{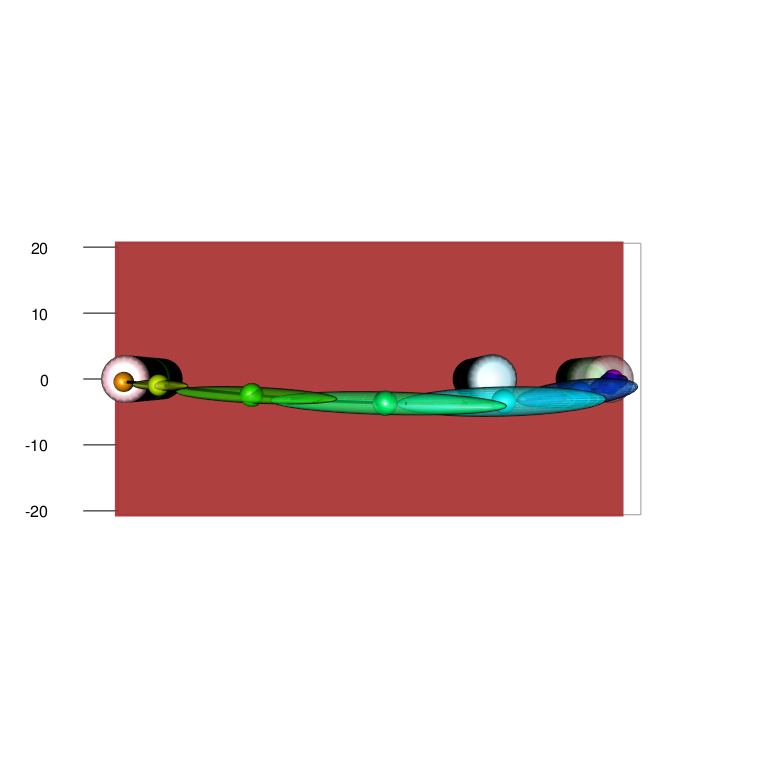}
\includegraphics[width = 0.3\textwidth , trim = 30 210 90 210, clip]{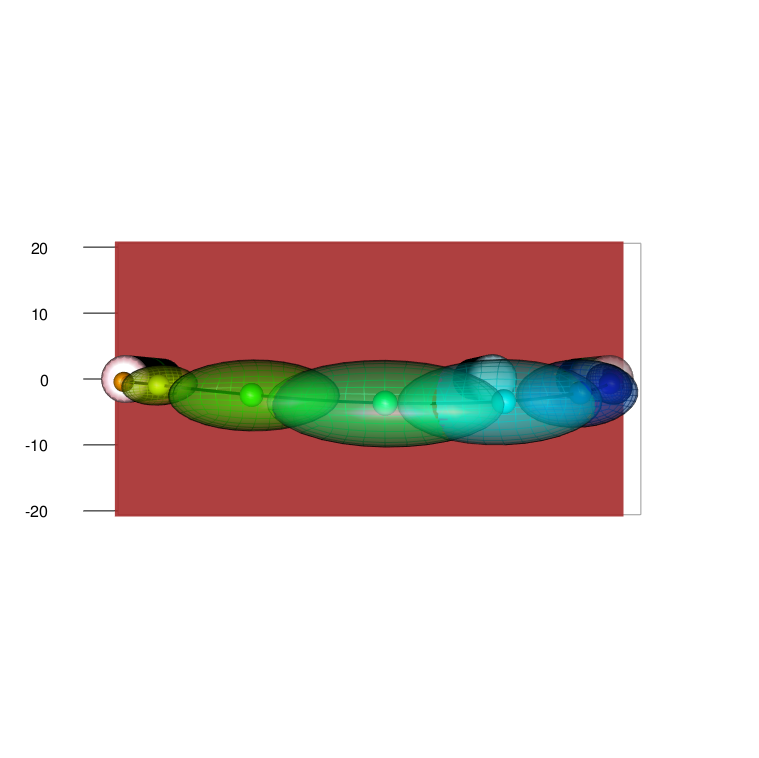}
\caption{Top-view of the experimental setup from the top for the medium height obstacle at all obstacle distances with the mean trajectory plotted. Along the trajectory eight equidistant points (in percentual warped time) are marked, and at each point  95\% prediction ellipsoids are drawn. The ordering is the same as in Figure~\ref{fig:variation1}.}\label{fig:variation2}
\end{figure}

\paragraph{Discovering the time-structure of variance along movement trajectory}
A strength of our approach to time warping is that we can estimate variability more reliably. In addition to observation noise, we model three sources of variation in the observed movement trajectories:  individual differences in the trajectory, individual differences caused by changing obstacle height, and variation from repetition to repetition. The variances described from these three sources are independent of obstacle height. In figures~\ref{fig:variation1}~and~\ref{fig:variation2}, 
two spatial representations of the mean movement path for the medium obstacle height are shown.
The five distances of the obstacle from the starting position are shown in the five rows of the figures. The three columns show variance originating from individual differences in the trajectory, individual differences caused by changing obstacle height, and variation from repetition to repetition (from left to right). 
Variance is illustrated at eight equidistant points in time along the mean path by ellipsoids that mark  95\% prediction for each level of variation. 

Note the asymmetry of the movements with respect to obstacle position,  both in terms of path and variation. This asymmetry  reflects the direction of the movement. 
Generally, variability is higher in the middle of the movement than early and late in the movement. Individual differences caused by  change in obstacle height (middle column) are small and lie primarily along the path. That is, individuals adapt the timing of the movement differently as height is varied. Individual differences in the movement path itself (left column) are largely differences in movement parameters: individuals differ in the maximal elevation and in the lateral positioning of their paths, not as much in the time structure of the movements. Variance from trial to trial (right column) is more evenly distributed, but is largest along the path reflecting variation in timing.  

\begin{figure}[!thp]
\centering
\includegraphics[scale = 0.5]{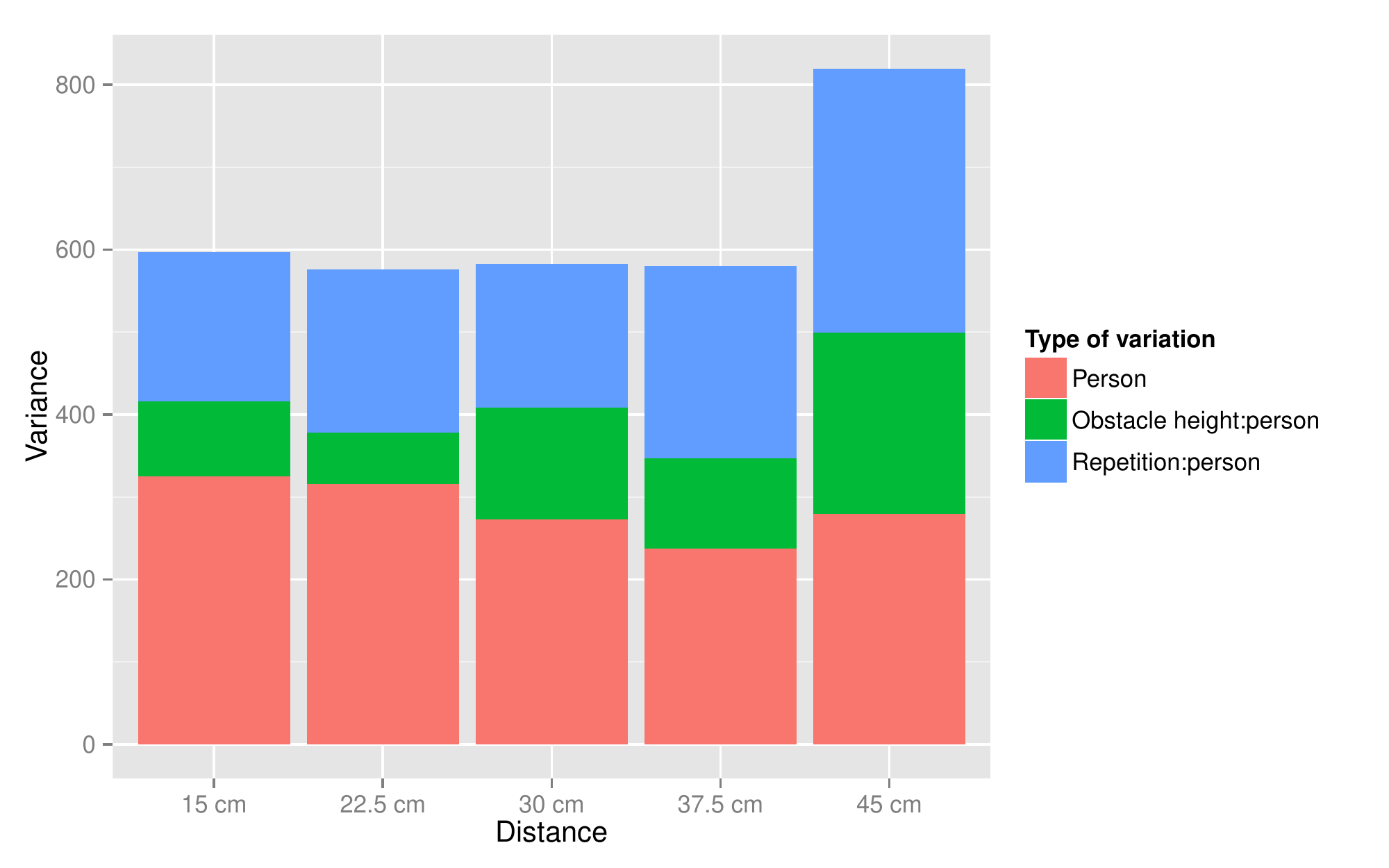}
\caption{Amount of variance explained by random effects.}\label{fig:variance}
\end{figure}

These descriptions are corroborated by the comparisons of the amounts of variance explained by the three effects in Figure~\ref{fig:variance}. The obstacle distance of $45$, in which the obstacle is close to the target lead to the largest variance in movement trajectory, with most of the increase over other conditions coming from repetition and individual differences caused by change in obstacle height. This suggests that this experimental condition is more difficult than the others, and perhaps much more so for the tall obstacle than for the small one.  Apart from this condition, we see that the largest source of variation are individual differences in movement trajectory. 
The second largest source of variation was repetition. Individual differences caused by change in obstacle height were systematically the smallest source of variance. 



\paragraph{Trajectory focal points}
\begin{figure}[!thp]
\centering
\includegraphics[width = 0.47\textwidth, trim = 60 50 50 0, clip]{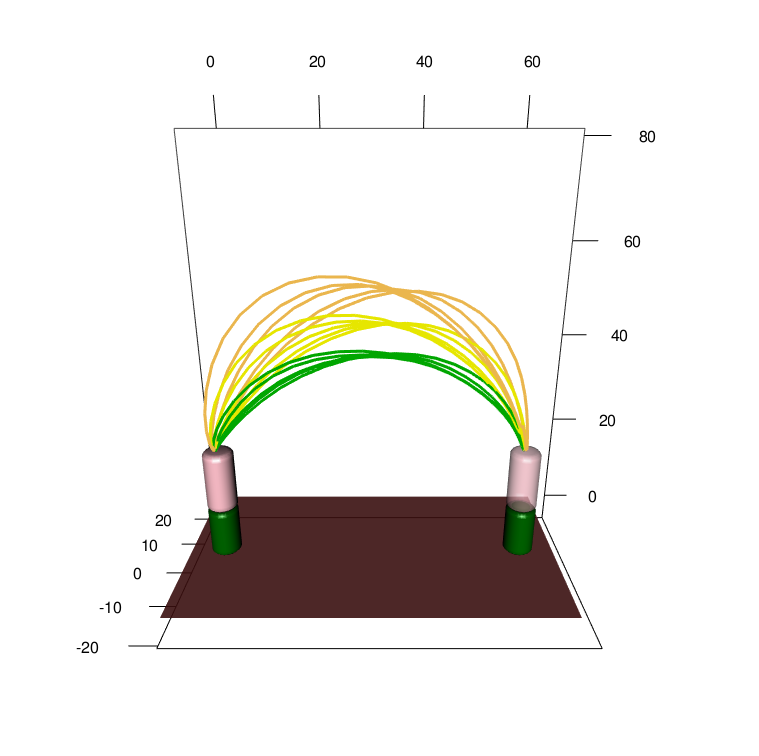}
\includegraphics[width = 0.47\textwidth, trim = 0 200 140 200, clip]{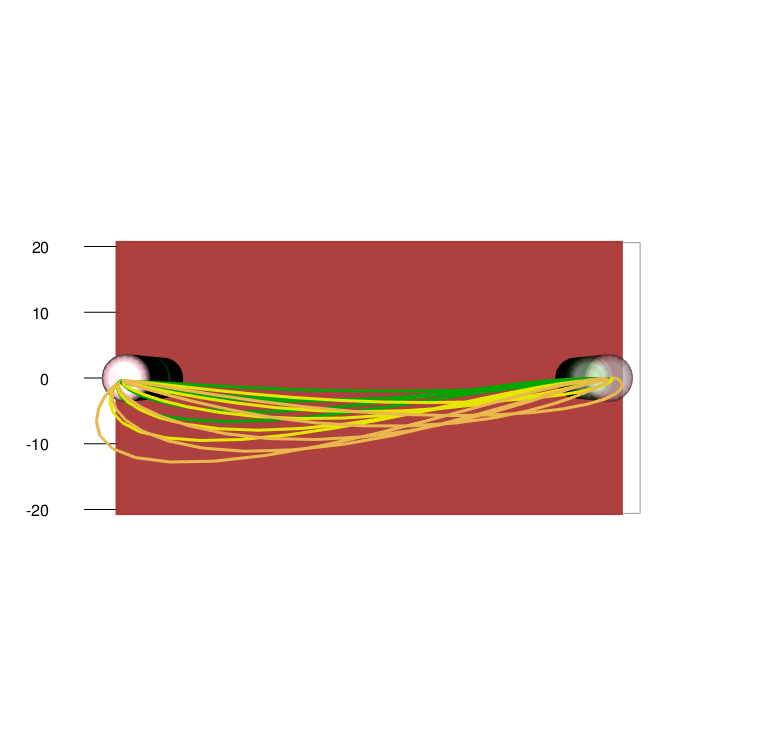}
\caption{Front and top view of the mean trajectories for the 15 obstacle avoidance setups. Green lines correspond to low obstacles, yellow to medium obstacles and orange to tall obstacles. }\label{fig:focal}
\end{figure}

A final demonstration of the strength of the method of analysis is illustrated in Figure~\ref{fig:focal}, which shows the mean paths for all obstacle distances and all obstacle heights. For each obstacle height,  the paths from different obstacle distances intersect both in the frontal and  the vertical plane. These focal points occur approximately at the same distance along the imagined line connecting start and end position. This pattern is clearly visible in the front view of the mean trajectories in Figure~\ref{fig:focal}. Due to the  limited variation of the path in the horizontal plane (Figure~\ref{fig:focal} top view), this effect is less clear in the horizontal plane. 
This pattern may reflect a scaling law, a form of invariance of an underlying path generation mechanism. 

\section*{Discussion} 
We have proposed a statistical framework for the modeling of human 
movement data. The hierarchical nonlinear mixed-effects model
systematically decomposes movements into a common effect that reflects the variation of movement variables with time during the movement, individual effects, that reflect individual differences, variation from trial to trial, as well as measurement noise. The model  amounts to a nonlinear time-warping approach that treats all sources of variances simultaneously. 

 We have outlined a method for performing maximum likelihood estimation of the model parameters, and demonstrated the approach by analyzing a set of human movement data on the basis of acceleration profiles in arm movements with obstacle avoidance. 
 The quality of the estimates was evaluated in a classification task, in which our model was better able to determine if a sample movement came from a particular participant compared to state-of-the-art template-based curve classification methods. These results indicate that the templates that emerge from our nonlinear warping procedure are both more consistent and richer in detail. 
 
We used the nonlinear time warping obtained from the acceleration profiles to analyze the spatial movement trajectories and their dependence on task conditions, here the dependence on obstacle height and obstacle placement along the path. We discovered that the warped movement path scales linearly with increasing obstacle height.  Furthermore, we separated the variation around the mean paths into three levels: individual differences of movement trajectory, individual differences caused by change in obstacle height, and trial to trial variability. This combination of models uncovered clear and coherent patterns in the structure of variance. Individual differences in trajectory and variance from trial to trial were the largest sources of variance, with individual differences being primarily at the level of movement parameters such as elevation and lateral extent of the movement while variance from trial to trial contained a larger amount of timing variance. We documented a remarkable property of the movement paths when obstacle distance along the path is varied at fixed obstacle height: all paths intersect at a single point in space. 

We believe that the approach we describe enhances the power of time series analysis as demonstrated in human movement data. The nonlinear time warping procedure makes it possible to obtain reliable estimates of variance along the movement trajectories and is strong in extracting individual differences. This advantage can be leveraged by combining the nonlinear time warping with factor analysis to extract systematic dependencies of movements on task conditions at the same time as tracking individual differences, both base-line and with respect to the dependence on task conditions, as well as variance across repetitions of the movement. 

{
Recent theoretical accounts have used the analysis of variance across repetitions of movements to uncover coordination among the many degrees of freedom of human movement systems \cite{LatashScholzSchoner2007}. Differences in variance between the subspace that keeps hypothesized relevant task variables invariant and the subspace within which such task variables vary support hypotheses about the task-dependent structure of the underlying control systems. Because variance is modulated in time differently across the two subspaces, a more principled decomposition of time dependence and variance from trial to trial would give such analyses new strength. Because this application requires the extension of the proposed method to multivariate time series, it is beyond the scope of this paper. Together with the considerable practical interest in identifying individual differences, these theoretical developments underscore that the method proposed here is timely and worth the methodological investment. }
\newpage 
\section*{Supporting Information}
{
\subsection*{A primer on model building for movement data}
\begin{figure}[!th]
\centering
\includegraphics[width = 0.47\textwidth]{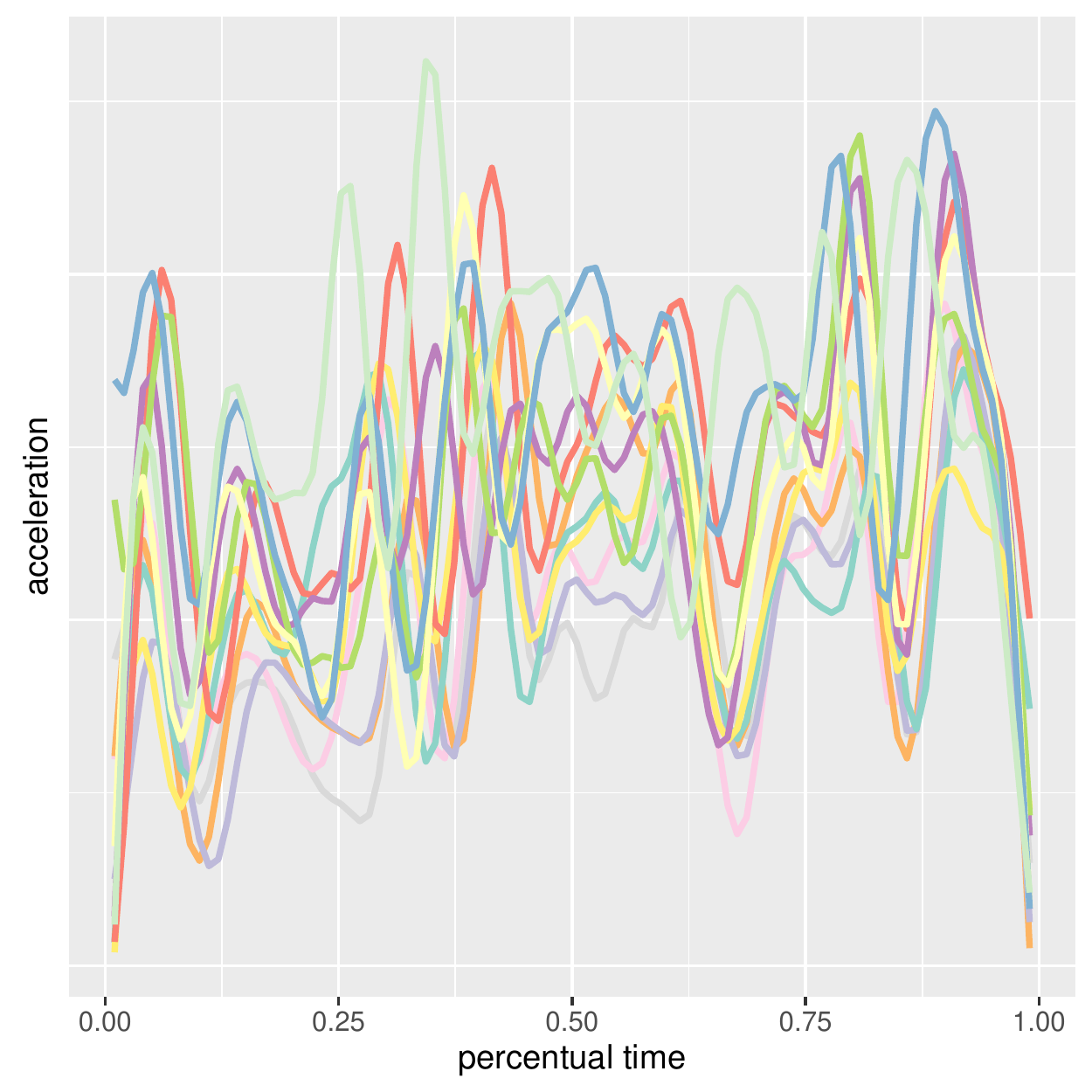}
\caption{Twelve acceleration functions corresponding to repetitions of a signature being written.}\label{fig:signa1}
\end{figure}
The two major types of variation in movement data are path variation (amplitude) and movement timing variation (time warping). From a statistical modeling perspective, it is natural to model these effects as realizations of random processes across repetitions of the task at hand, since this allows a data driven regularization of the predictions of these effects. We consider the class of models on the form
\begin{align}
y_i(t_k) = \theta(v(t_k, \bw_i)) + x_i(t_k) + \varepsilon_{ik}\label{eq:mod_primer}
\end{align}
where $\theta$ is the mean profile, $v$ is a warping function that depends on the random warping parameters $\bw_i\in\R^{n_w}$ that are assumed to be independent across $i$ and multivariate zero-mean Gaussian with covariance matrix $\sigma^2 C$, the $x_i$ terms are independent zero-mean Gaussian process with covariance function $\sigma^2 \mathcal{S}$, and the $\varepsilon_{ik}$-terms are independent zero-mean Gaussian noise with variance $\sigma^2$.

To use model~\eqref{eq:mod_primer} we need to choose the type of mean function $\theta$, the type of warping function $v$ as well as the covariance structures for $\bw_i$ and $x_i$. Below is a list of considerations of how to do these model choices based on the experiment at hand.
\begin{description}
\item[$\theta$:] A good allround choice is to model $\theta$ as a B-spline using a functional basis. For periodic movement sequences, a Fourier basis may sometimes be preferred. To choose the number of basis functions to use we need to consider the data at hand. For experiments with dense sampling and a clear systematic pattern in the trajectories the number of basis functions should be just be sufficiently high to model the mean pattern. If on the other hand the trajectories are sparsely observed in time or the common pattern is very unclear, one should choose a small number of basis functions to avoid local overfitting.
\item[$v$:] The behavior of the warping function should be driven by the random variables $\bw_i$. We will consider warping functions where $\bw_i$ models disparities from the identity mapping (corresponding to no warp), and $v$ is an interpolating function of these random disparities at a set of specified anchor points $\bt^{w}\in \R^{n_w}$. If one needs to predict derivatives such as velocity or acceleration of the observed profiles the interpolation should be smooth (e.g. a cubic spline). If no derivatives are needed, one should prefer simpler models such as linear interpolation as this reduces the nonlinear contribution of the derivative term in the linearization term $Z$ and thus reduces the complexity the estimation problem, see model~\eqref{model:linear}.
\item[$C$:] The covariance matrix of the random disparities $\bw_i$ should be chosen to respect the experimental setup. If movements are modeled in percentual time, and the beginning and end of the observed trajectories correspond to the same states across movement (e.g. beginning and end of movement), the model for $\bw_i$ should respect that. The simplest such model is to assume that $\bw_i$ is a Brownian bridge observed at the discrete anchor points $\bt^w$. For other types of data one may wish to include a random Gaussian time shift (add constant matrix to the covariance of the non-shift part of the model) or have an open end point which could be modeled using a Brownian motion model. If $n_w$ is low relative to the number of repetitions one may model $C$ as a completely free covariance matrix.
\item[$\mathcal{S}$:] A good allround choice for the covariance function of the amplitude variance is the Mat\'ern covariance function. The Mat\'ern covariance has three parameters, scale, range and smoothness. The scale parameter determines the variance of the process, the range parameter determines the strength of the correlation over time and the smoothness parameter determines the smoothness of the corresponding process. If one wants to simplify the optimization problem, one may fix the smoothness parameter at some value that represents sufficient smoothness, for example 2 corresponding to twice differentiable sample paths of the amplitude effect.  For experiments with fixed start and end values one may use a bridge process such as the Brownian bridge, however, it is often preferable to use a less specific model than an overly specific model. For additive effects such as $x_i$, even slight misspecification of a bridge covariance structure at the beginning and end of movement (where variance and covariance go from zero to non-zero) may result in considerable bias of the corresponding parameters. 
\end{description} 
Consider the handwriting signature data in percentual time in Figure~\ref{fig:signa1}. The data consist of 12 acceleration magnitude profiles, each with 98 observations, corresponding to repetitions of a signature being written by a participant. This data has previously been used as an example in \cite{Srivastava}.  A reproducible pattern across repetitions is in the nature of the task, and we also see a strong consistent pattern across the samples, but the curves are both misaligned and vary systematically in amplitude. Using the considerations above, we choose a B-spline basis with 40 interior knots to have sufficient flexibility to model the mean, we choose $v$ to be a piecewise linear interpolation of the disparities $\bw_i$ that we model as discretely observed Brownian bridges over $n_w=20$ equidistant anchor points. For the amplitude covariance $\mathcal{S}$ we choose a Mat\'ern covariance with unknown scale, range and smoothness. 

The alignment of the proposed model is displayed in Figure~\ref{fig:signa2}. We see a neat alignment of the samples and a mean function that represents the mean pattern well, with no indications of local overfitting. Similarly for the warping functions, we only see small systematic deviations from the identity warp, despite of the high number of anchor points. The maximum likelihood estimates for the variance parameters were as follows: the warp scale estimate was 14.1, the variance scale estimate for the amplitude effect was 54.4, the range parameter estimate was $8.2\cdot 10^{-3}$, the smoothness parameter was 6.2, and the noise variance $\sigma^2$ was estimated to be $1.4\cdot 10^{-4}$. This suggests that the systematic part of the amplitude variation explains more than 99.9\% of the amplitude variation, which fits well with the smooth functional samples.

To find the best among multiple models, one can compare different models using cross-vaildation, as was done in this paper, if such a setup is meaningful for the application at hand. In general one can do model selection based on the corrected conditional AIC of the linearized likelihoods \cite{greven2010behaviour}.

\begin{figure}[!ht]
\centering
\subfloat[Aligned samples and mean (dashed)]{\includegraphics[width = 0.47\textwidth]{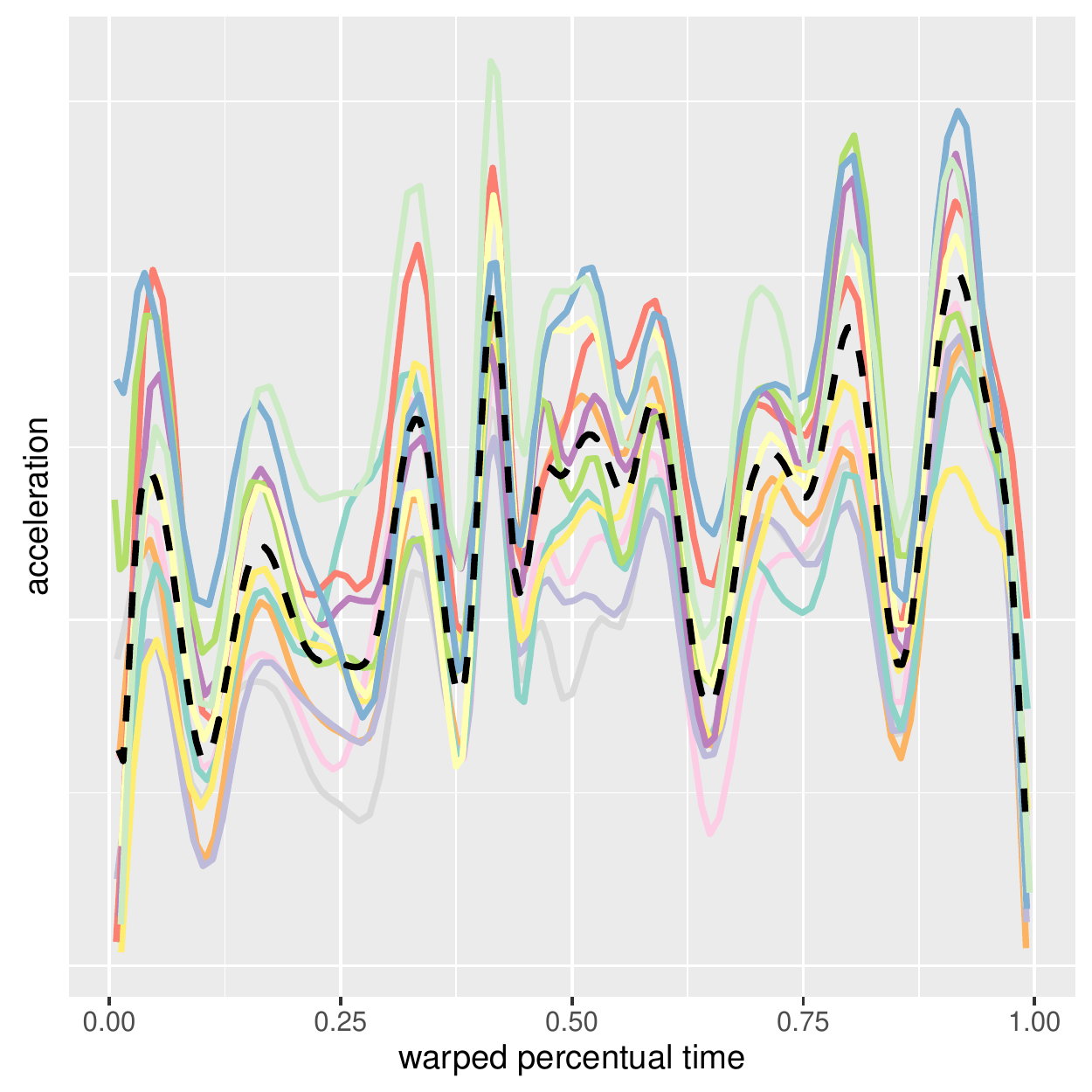}}
\subfloat[Warping functions compared to the identity (dashed)]{\includegraphics[width = 0.47\textwidth]{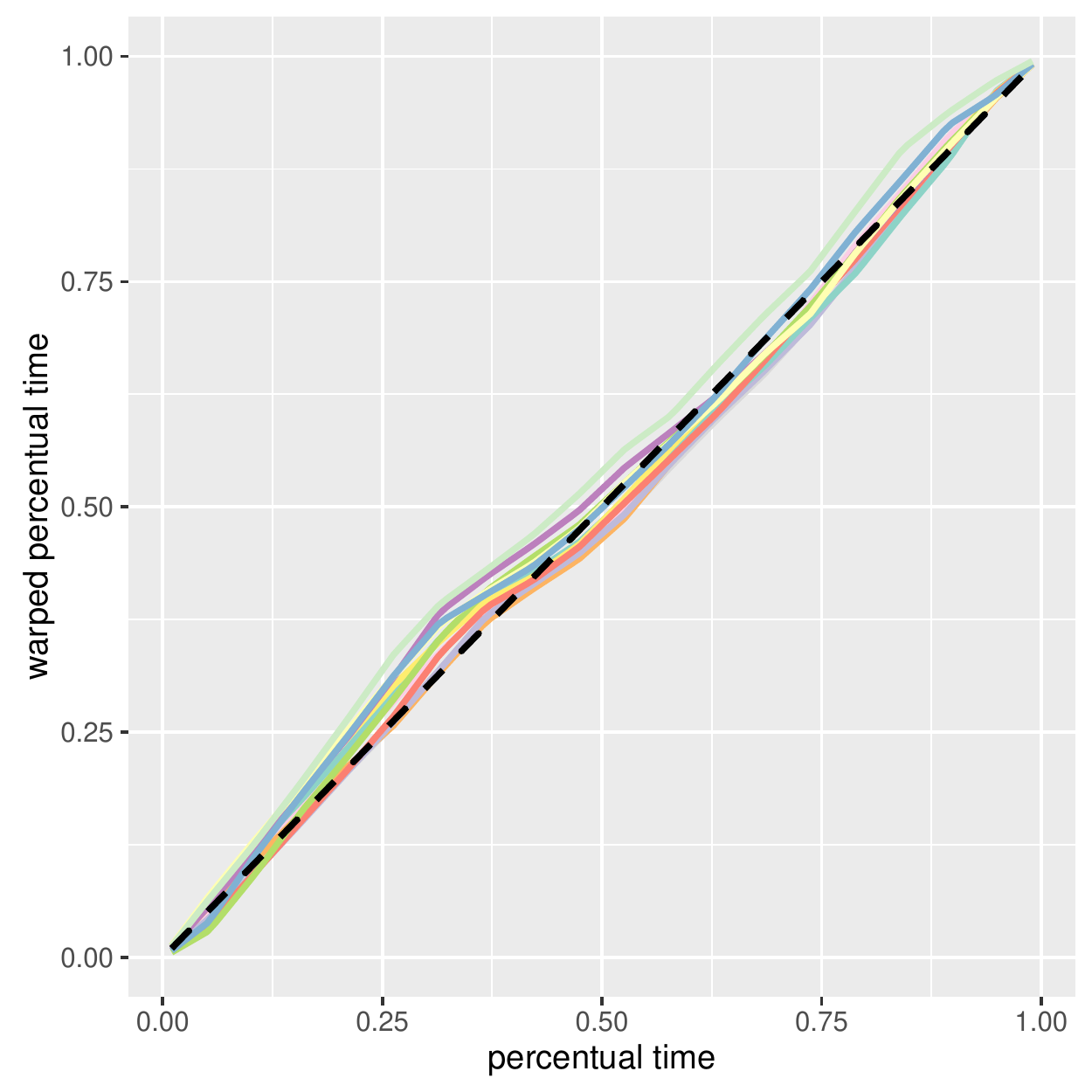}}
\caption{The aligned acceleration functions from Figure~\ref{fig:signa1} (a), along with the predicted warping functions (b). }\label{fig:signa2}
\end{figure}

\subsubsection*{R code for fitting the model to the signature data}
Suppose that \texttt{y} is a list containing the 12 acceleration trajectories and \texttt{t} is a list of the corresponding observation times. The model described above can be specified and fitted using the code given below. The methods in the \texttt{pavpop} R package are thoroughly documented with a wide array of examples in the package help pages and vignettes.
\begin{verbatim}
# Install and load pavpop R package
if (packageVersion("devtools") < 1.6) {
  install.packages("devtools")
}
devtools::install_github('larslau/pavpop')
library(pavpop)

# Set up basis function
kts <- seq(0, 1, length = 42)[2:41]
basis_fct <- make_basis_fct(kts = kts, intercept = TRUE, 
                            control = list(boundary = c(0, 1)))

# Set up warp function
tw <- seq(0, 1, length = 20)
warp_fct <- make_warp_fct('piecewise-linear', tw)

# Set up covariance functions and roughly initialize parameters
warp_cov_par <- c(tau = 10)
warp_cov <- make_cov_fct(Brownian, noise = FALSE, param = warp_cov_par, 
                         type = 'bridge')

amp_cov_par <- c(scale = 4, range = 1, smoothness = 2)
amp_cov <- make_cov_fct(Matern, noise = TRUE, param = amp_cov_par)

#
# Estimate in the model
#

# Rough bounds on parameters
lower <- c(1e-3, 1e-3, 1e-3, 1e-3)
upper <- c(1000, 10, 10, 10)

res <- pavpop(y, t, basis_fct, warp_fct, amp_cov, warp_cov, 
              iter = c(5, 20), homeomorphism = 'soft', 
              like_optim_control = list(lower = lower, upper = upper))
\end{verbatim}

\subsection*{Simulation study}
To evaluate the proposed algorithm for maximum likelihood estimation, we simulated data from the proposed model under the maximum likelihood estimates on the full data using a sampling setup identical to the central experiment ($d = 30.0$ cm, medium obstacle). We simulated 1000 outcomes and ran the estimation procedure as described in the section on modeling of effects and the algorithmic approach. The total runtime of the 1000 estimation procedures was approximately 6 hours on a 64-core machine. 

The densities of the integrated square estimation errors ($L^2$ error) for the estimated mean profiles (experiment and participant) are shown in the right panel of Figure~\ref{fig:fixed_simul}. For comparison, the experiment-specific and participant mean profiles have been estimated using ordinary least squares (OLS) estimation with the correctly specified spline model for the mean. The corresponding densities are shown in the left panel. We note that the densities are shown on squareroot scale to enable visual inspection of the differences. We see that the estimate for the experiment-specific mean profile is marginally more stable for the OLS estimation, but results are close-to perfect for both methods. For the participant-specific effects, however, we see that the proposed model that aligns samples within participant using a random warping function gives $L^2$ errors that are approximately an order of magnitude lower than simple OLS estimation. 

\begin{figure}[!ht]
\centering
\input{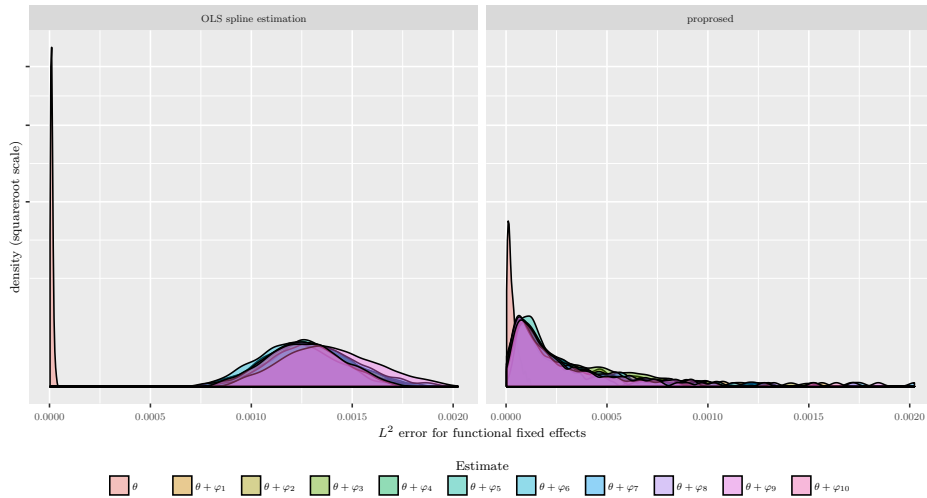}
\caption{Densities of the integrated square estimation errors ($L^2$ errors) for the common and participant-specific mean functions in the simulation study. The left panel shows results for ordinary least square (OLS) estimation and the right panel shows the results for the proposed model and estimation algorithm. Both models were fitted using the correctly specified spline model for the mean. Note that the density is displayed on squareroot scale. }\label{fig:fixed_simul}
\end{figure}

Figure~\ref{fig:warp_fixed_simul} displays densities of the differences between the maximum likelihood estimates for the true participant-specific warping parameters across participants. The estimates generally seem unbiased with small variance around the true warping parameters.

\begin{figure}[!ht]
\centering
\input{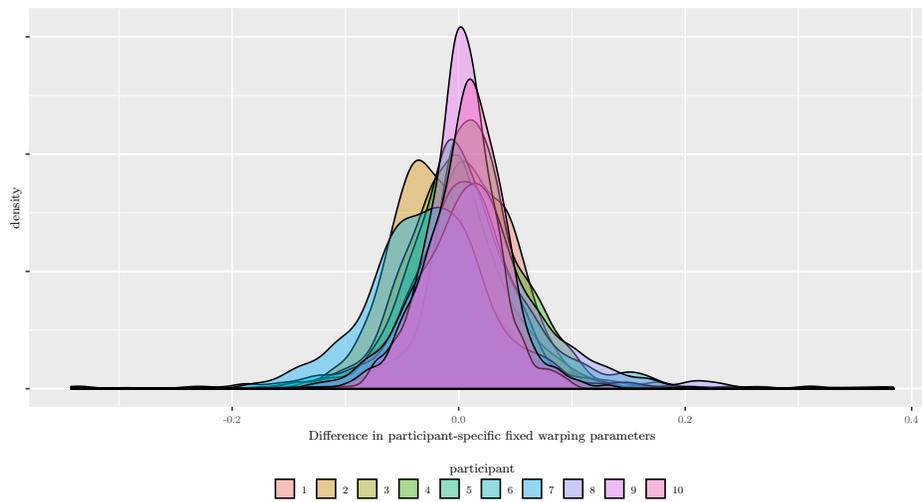}
\caption{Densities of the difference between the estimated and the true participant-specific warping parameters across participants.}\label{fig:warp_fixed_simul}
\end{figure}

Figure~\ref{fig:variances_simul} displays densities for the parameter estimates in the simulated experimental setups. We see that the estimators for the noise scale $\sigma$ and the scales for the warp parameters $\sigma \gamma$ both seem to have a small upward bias. The scale $\sigma\tau$ of the serially correlated effects and the range parameter $1/\alpha$ both seem to be estimated with very little or no bias. Slightly biased variance-parameter estimates are to be expected in likelihood-based inference \cite{Harville}, in particular in nonlinear models where bias-reducing estimation methods such as restricted maximum likelihood (REML), that are inherently linear, are not available. 

\begin{figure}[!ht]
\centering
\input{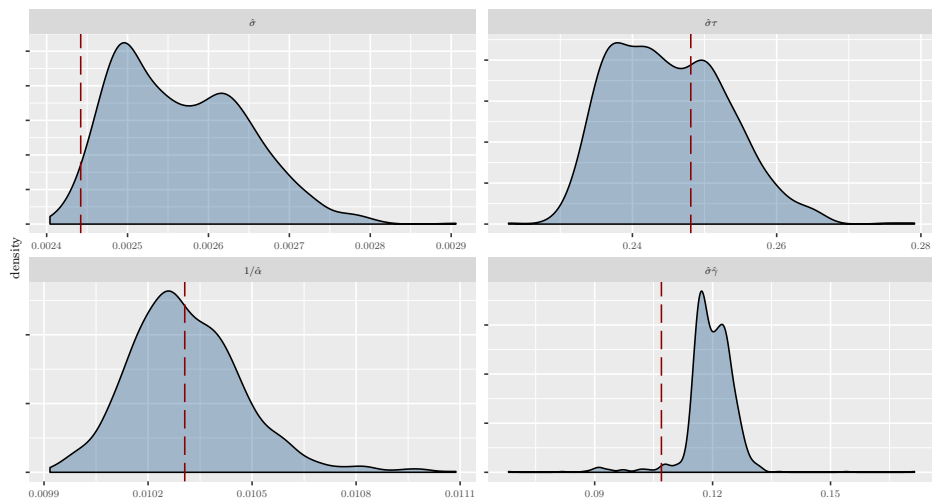}
\caption{Densities of the estimated variance parameters in the simulated experimental setups. Dashed red lines indicate the true values of the parameters. }\label{fig:variances_simul}
\end{figure}

}

\subsection*{Cross-validation grids for motion classification}
Cross-validation was done over:
\begin{description}
\item[MBM] number of bands $J$ in $ \{1,2,3,4,5,6\}$.
\item[DTW] {degrees of freedom for B-spline basis $ \{8, 13, 18, 23, 28, 33, 38\}$. For DTW$_{\text{p}}$ $\{8, 9, \dots, 18\}$.}
\item[FR]  number of principal components in $\{1,2,3,4,5\}$.
\item[FR$_{\textrm{E}}$]  number of principal components in $ \{1,2,3,4,5\}$ and weighting between phase and amplitude distance in $\{0.0, 0.5, \dots, 5.0\}$.
\end{description}

%
%
%
\bibliography{bib1.bib}

\end{document}